\documentclass[sigconf]{acmart}
\newcommand{\myparatight}[1]{\smallskip\noindent{\bf {#1}:}~}

\usepackage{pgf}
\usepackage{xcolor}
\usepackage{tikz}
\usetikzlibrary{tikzmark}

\usepackage{float}
\usepackage{stfloats}

\usepackage{newfloat}
\usepackage{listings}

\usepackage{bibentry}

\usepackage[skip=0pt]{caption}

\usepackage{amsmath,amsfonts}
\usepackage{xcolor}
\usepackage{multirow}
\usepackage{array} 
\usepackage{textcomp} 
\usepackage{float}
\usepackage{amsthm}

\usepackage{url}
\usepackage{verbatim}
\usepackage{graphicx} 
\usepackage{subcaption}
\usepackage{makecell}
\usepackage{bm}
 
\usepackage{algorithm}
\usepackage{algpseudocode}

\usepackage{xspace}

\usepackage{color, colortbl}

\definecolor{greyL}{RGB}{230,248,255}

\newcommand{\alg}{{\textsf{SafeFL}}\xspace}
\newcommand{\algFirst}{{\textsf{SafeFL-ML}}\xspace}
\newcommand{\algSecond}{{\textsf{SafeFL-CL}}\xspace}

\copyrightyear{2025}
\acmYear{2025}
\setcopyright{acmlicensed}\acmConference[ASIA CCS '25]{ACM Asia Conference on Computer and Communications Security}{August 25--29, 2025}{Hanoi, Vietnam}
\acmBooktitle{ACM Asia Conference on Computer and Communications Security (ASIA CCS '25), August 25--29, 2025, Hanoi, Vietnam}
\acmDOI{10.1145/3708821.3736194}
\acmISBN{979-8-4007-1410-8/2025/08}

\begin{document}

\captionsetup[subfigure]{skip=0pt} 
\captionsetup[subtable]{skip=0pt}

\title{Toward Malicious Clients Detection in Federated Learning}

\author{Zhihao Dou}
\authornote{Equal contribution. Zhihao Dou and Jiaqi Wang conducted this research while they were interns under the supervision of Minghong Fang.}
\affiliation{
	\institution{Duke University}
	\city{Durham}
        \country{USA}
}

\author{Jiaqi Wang}
\authornotemark[1]
\affiliation{
	\institution{Hainan Normal University}
	\city{Haikou}
        \country{China}
}

\author{Wei Sun}
\affiliation{
	\institution{Wichita State University}
	\city{Wichita}
        \country{USA}
}

\author{Zhuqing Liu}
\affiliation{
	\institution{University of North Texas}
	\city{Denton}
        \country{USA}
}

\author{Minghong Fang}
\affiliation{
	\institution{University of Louisville}
	\city{Louisville}
         \country{USA}
}

\begin{abstract}
Federated learning (FL) enables multiple clients to collaboratively train a global machine learning model without sharing their raw data. However, the decentralized nature of FL introduces vulnerabilities, particularly to poisoning attacks, where malicious clients manipulate their local models to disrupt the training process. While Byzantine-robust aggregation rules have been developed to mitigate such attacks, they remain inadequate against more advanced threats. In response, recent advancements have focused on FL detection techniques to identify potentially malicious participants. Unfortunately, these methods often misclassify numerous benign clients as threats or rely on unrealistic assumptions about the server's capabilities. In this paper, we propose a novel algorithm, \alg, specifically designed to accurately identify malicious clients in FL. The \alg approach involves the server collecting a series of global models to generate a synthetic dataset, which is then used to distinguish between malicious and benign models based on their behavior. Extensive testing demonstrates that \alg outperforms existing methods, offering superior efficiency and accuracy in detecting malicious clients.
\end{abstract}



\maketitle


\section{Introduction}

Federated learning (FL)~\cite{mcmahan2017communication} is a distributed machine learning approach that trains models on multiple decentralized clients, each holding its own local data, without sharing that data. 
This method primarily mitigates privacy concerns associated with centralized training methods.
In FL, a central server collects models from each participating client, maintaining data confidentiality while collectively enhancing the model.
The process unfolds in three main steps: first, the server distributes the current global model to clients or a selected group. Then, these clients adjust their local models using their specific local training data. Finally, they send their local models back to the server, where these models are combined according to a predefined aggregation rule to refine the global model. 
FL has been adopted in various real-world applications, including credit risk evaluation~\cite{webank}, speech recognition~\cite{paulik2021federated}, and next-word prediction~\cite{gboard}.

In recent years, much  research~\cite{wang2020tackling,karimireddy2020scaffold,zhu2021data,li2019convergence,chen2023enhancing,bonawitz2019towards,mohri2019agnostic,li2020federated,li2019fair} has focused on enhancing the efficiency of FL.
However, a significant obstacle to its widespread adoption lies in FL's inherent vulnerability to poisoning attacks carried out by malicious clients~\cite{2018How, baruch2019little, bhagoji2019analyzing, cao2022mpaf, 2019Local, xie2019dba, tolpegin2020data,zhang2024poisoning,yin2024poisoning}, due to its decentralized structure.
In these attacks, adversaries can undermine the integrity of the global model by tampering with their local training data or altering the models they send to the server, thereby contaminating the aggregated global model. These malicious actions typically manifest in two primary forms. The first, {\em untargeted attacks}~\cite{cao2022mpaf, 2019Local}, seeks to degrade the global model's performance on a wide range of test cases, aiming for a general reduction in accuracy across various scenarios. The second type, {\em targeted attacks}~\cite{2018How, baruch2019little, bhagoji2019analyzing, xie2019dba}, involves more strategic manipulation, where the attacker’s goal is to influence the global model in such a way that it produces specific, desired outputs for selected test cases, often with malicious intent. These attacks pose significant challenges to the robustness, reliability, and security of FL systems, making it imperative to develop effective defense mechanisms that can detect and neutralize such threats.

To mitigate poisoning attacks, several defensive mechanisms have been proposed in the literature~\cite{yin2018byzantine,blanchard2017machine,el2021collaborative,rajput2019detox,xie2019zeno,munoz2019byzantine,li2019rsa,naseri2022cerberus,kumari2023baybfed,rieger2022deepsight,karimireddy2020byzantine,li2021ditto,mozaffari2023every,ozdayi2021defending,wang2022flare,fang2025byzantine}. These defenses can generally be categorized into {\em detection-based} and {\em prevention-based} approaches. Detection-based defenses focus on identifying malicious clients in the FL system and subsequently removing them. 
For instance, in FLTrust~\cite{2021FLTrust}, the server possesses a small validation dataset that resembles the clients' training data. Using this validation data, the server generates a reference model, and a client is classified as benign if its local model aligns positively with this reference model.
Similarly, FLDetector~\cite{zhang2022fldetector} predicts a client’s model based on historical models, flagging clients as malicious if their actual and predicted models significantly diverge over multiple rounds. In contrast, prevention-based defenses aim to reduce the impact of malicious clients without removing them from the system. 
The Median~\cite{yin2018byzantine} method, for instance, computes the coordinate-wise median of local models to derive the global model. 
However, current defenses against poisoning attacks to FL face notable limitations. 
Existing detection-based methods struggle to accurately identify malicious clients in complex scenarios, often misclassifying benign clients as malicious, especially when both targeted and untargeted attacks are present. 
Furthermore, some methods like FLTrust rely on unrealistic assumptions, such as the server possessing a validation dataset that reflects the overall distribution of client data. 
On the other hand, prevention-based defenses cannot fully mitigate the effects of malicious clients, as these malicious clients remain within the FL system.

\myparatight{Our work}
In this study, we aim to fill this critical gap by presenting a novel detection-based defense mechanism, referred to as \alg, specifically designed to identify and mitigate the impact of malicious clients in FL systems. Our proposed \alg relies on the server maintaining its own unique dataset, enabling it to evaluate the integrity of local models submitted by participating clients. 
Given the inherent challenges associated with the server's ability to have complete knowledge of the data distribution across the diverse and heterogeneous devices of clients, we propose an innovative solution. Instead of relying on a dataset that mirrors the clients' data, the server generates a synthetic dataset derived from the trajectory of global models trained over multiple rounds. While this synthetic dataset does not replicate the actual distribution of client data, it is crafted to effectively distinguish between malicious and benign behaviors in models. By leveraging this dynamic dataset generation strategy, \alg ensures robust detection and defense against malicious activities, enhancing the overall security and reliability of FL systems.

After creating the synthetic dataset, the server leverages it to identify potential malicious clients by analyzing the distinct behavioral differences between malicious and benign local models when evaluated on this dataset. This approach operates under the assumption that malicious models, crafted with adversarial intent, will demonstrate significantly different performance compared to benign models. 
To implement this detection mechanism, we introduce two variations of our defense strategy, referred to as \algFirst and \algSecond, each utilizing a unique methodology for identifying malicious clients. The foundation of \algFirst lies in the observation that malicious local models tend to incur a higher loss on the synthetic dataset compared to benign models. Guided by this principle, the server evaluates the loss of each local model submitted by clients and calculates the median loss across all models. A client is classified as benign if the loss of its corresponding model falls below this median value; otherwise, it is flagged as potentially malicious. 
On the other hand, \algSecond employs a different strategy while still relying on the loss evaluation of local models against the synthetic dataset. Instead of utilizing a median-based threshold, the server applies a clustering algorithm to group models based on their loss values. This approach is grounded in the premise that the loss values of benign models are more likely to cluster closely together, reflecting their consistent and non-adversarial behavior. By categorizing models into clusters, \algSecond identifies outliers, which are indicative of malicious clients, with greater precision. 

We conduct an extensive evaluation of our proposed \alg using five diverse datasets, including large-scale benchmarks such as CIFAR-10~\cite{cifar10data}, STL-10~\cite{coates2011analysis}, and Tiny-ImageNet~\cite{deng2009imagenet}, as well as the FEMNIST~\cite{femnist} dataset, which is inherently heterogeneous.
These datasets span multiple domains to ensure a comprehensive assessment of our approach. Our evaluation also encompasses eleven poisoning attack scenarios and ten state-of-the-art FL defenses. Among these defenses are seven detection-based methods—FLAME~\cite{2021FLAME}, FLDetector~\cite{zhang2022fldetector}, FLTrust~\cite{2021FLTrust}, DeepSight~\cite{rieger2022deepsight}, BackdoorIndicator~\cite{li2024backdoorindicator}, FreqFed~\cite{fereidooni2024freqfed}, FedREDefense~\cite{xie2024fedredefense}—as well as three prevention-based strategies, namely Median~\cite{yin2018byzantine}, Trimmed mean~\cite{yin2018byzantine}, and Krum~\cite{blanchard2017machine}. 
Beyond these benchmarks, we explore a variety of practical settings in FL that reflect real-world challenges. These include scenarios where clients operate with highly non-independent and identically distributed training data, such as datasets limited to three classes per client. We also consider cases where clients utilize complex deep learning models, such as ResNet-20~\cite{2016Deep}, for local training. On the server side, we investigate the impact of employing different aggregation rules to combine the local models submitted by clients. This comprehensive evaluation highlights the robustness and adaptability of our method under diverse and challenging conditions.

The contributions of our work can be outlined as follows:
\begin{list}{\labelitemi}{\leftmargin=2em \itemindent=-0.3em \itemsep=.2em}
\item
We propose a novel detection framework, \alg, designed to identify malicious clients in FL.

\item 
We evaluate our proposed detection method on five datasets and against eleven distinct poisoning attacks, including the strong adaptive attack, and compare its performance with ten state-of-the-art FL defense baselines.

\item 
Extensive experiments show that our proposed \alg not only excels at identifying malicious clients in FL but also outperforms existing detection approaches.

\end{list}


\section{Background and Related Work} \label{sec:related}
\subsection{Background on federated learning (FL)}

A typical federated learning (FL) system consists of a central server and \( n \) distributed clients. 
Each client \( i \) has its own distinct database, referred to as the local training dataset \( D_i \), where \( i = 1, 2, \ldots, n \). 
The combined training dataset across all clients is represented as \( D \), where \( D = \bigcup_{i=1}^n D_i \). 
In an FL framework, these \( n \) clients collaborate under the coordination of the central server to train a shared global machine learning model. The primary goal of FL is to derive the optimal global model \( \mathbf{w}^* \), which is obtained by solving the following optimization problem
$
 \mathbf{w}^* = \arg \min_{\mathbf{w} \in \mathbb{R}^d} \sum_{i=1}^n \mathcal{L}(D_i, \mathbf{w}),
$
where $\mathcal{L}(D_i, \mathbf{w})$ represents the loss corresponding to the local training data of client $i$, based on the model parameter $\mathbf{w}$; $d$ is the dimension of $\mathbf{w}$.
FL tackles the above optimization problem through a decentralized methodology. The training procedure in round \( t \) is carried out in three sequential steps:

\begin{list}{\labelitemi}{\leftmargin=1.15em \itemindent=-0.08em \itemsep=.2em}

\item {\bf Step I (Global model synchronization):} The server transmits the current global model $\mathbf{w}^t$ to all clients or a portion of clients.

\item {\bf Step II (Local models updating):}
Upon receiving the global model $\mathbf{w}^t$ from the server, clients refine their local models using stochastic gradient descent (SGD). Specifically, client $i$ selects a mini-batch of training samples from its dataset $D_i$, computes a gradient $\mathbf{g}_{i}^{t}$ based on $\mathbf{w}^t$ and the sampled data, and then updates its local model as $\mathbf{w}_{i}^{t}=\mathbf{w}^{t}- \mu \cdot \mathbf{g}_{i}^t$, where $\mu$ is the learning rate. Finally, client $i$ sends its updated model $\mathbf{w}_{i}^{t}$ to the server.

\item {\bf Step III (Local models aggregation):}
After receiving the local models from the clients, the server applies a specified aggregation rule, denoted as $\text{AR}$, to combine the models. This is expressed as $\mathbf{w}^{t+1} = \text{AR}\{\mathbf{w}_1^{t}, \mathbf{w}_2^{t},...,\mathbf{w}_n^{t}\}$.

\end{list}

FL iteratively performs the outlined three steps until it meets the convergence criteria. When all participating clients are reliable and act without malicious intent, the server can adopt the straightforward FedAvg~\cite{mcmahan2017communication} algorithm to aggregate the local model updates. This approach updates the global model by averaging the local models received from the clients, calculated as \( \mathbf{w}^{t+1} = \frac{1}{n} \sum_{i=1}^{n} \mathbf{w}_{i}^{t} \).

\subsection{Poisoning Attacks to FL}

FL is inherently susceptible to poisoning attacks due to its decentralized structure. These attacks can be classified into two categories based on the attacker's objectives: untargeted attacks~\cite{cao2022mpaf,2019Local,shejwalkar2022back,shejwalkar2021manipulating,wang2025poisoning} and targeted attacks~\cite{2018How,baruch2019little,bhagoji2019analyzing,xie2019dba}. 
Untargeted poisoning attacks aim to degrade the overall performance of the global model across arbitrary test inputs. For instance, in the label-flipping attack~\cite{tolpegin2020data}, malicious clients alter the labels in their local training data to mislead the model. Fang et al.~\cite{2019Local} proposed a framework for untargeted FL attacks by formulating the problem as an optimization task. Their approach focuses on crafting malicious client updates that maximize the discrepancy between the aggregated global model updates before and after the attack. 
Conversely, targeted poisoning attacks, such as backdoor attacks, are designed to manipulate the global model to predict an attacker-specified label when provided with test inputs containing a specific, pre-determined trigger.

\subsection{Defenses against Poisoning Attacks to FL}
Various defenses~\cite{yin2018byzantine,blanchard2017machine,naseri2022cerberus,kumari2023baybfed,rieger2022deepsight,karimireddy2020byzantine,fang2025we,fang2024byzantine,fang2025provably,fang2022aflguard} have been proposed to counter poisoning attacks in FL. Some methods focus on identifying and excluding malicious clients from the FL system. For example, FLAME~\cite{2021FLAME} leverages the HDBSCAN~\cite{campello2013density} clustering algorithm to detect potentially malicious clients, while FLDetector~\cite{zhang2022fldetector} evaluates the consistency of a client’s local model to flag suspicious behavior. 
Other defense mechanisms aim to mitigate the impact of malicious clients without directly identifying them. For instance, Trimmed-mean~\cite{yin2018byzantine} and Median~\cite{yin2018byzantine} are coordinate-wise aggregation techniques that process each dimension of the local models independently. In Trimmed-mean, the values for each coordinate across clients' models are sorted, and the $k$ largest and $k$ smallest values are excluded. The average of the remaining $n-2k$ values is then calculated for each coordinate.

\myparatight{Limitations of existing defenses}Despite their advancements, existing FL defense mechanisms exhibit notable limitations. First, many defenses either result in the misclassification of a significant number of benign clients as malicious or fail to adequately reduce the influence of malicious clients, as their presence persists within the system. Second, some methods are based on unrealistic assumptions, such as the server requiring access to a clean dataset that accurately reflects the distribution of clients' training data.


\section{PROBLEM STATEMENT} \label{sec:problem}

\myparatight{Threat model} 
The threat model we employ follows the approach outlined in prior studies \cite{2018How,xie2019dba,baruch2019little,2019Local,shejwalkar2022back,shejwalkar2021manipulating}. 
To elaborate, the attacker controls a set of malicious clients, which may either be fake clients injected by the attacker or benign clients compromised by the attacker.
These malicious clients have the capability to transmit arbitrary local models to the server. The extent of the attacker's knowledge of the targeted FL system may vary. In cases of partial knowledge, the attacker possesses information solely about the local models and local training data on the malicious clients. In contrast, with full knowledge, the attacker possesses information about the local models on all clients and the aggregation rule employed by the server. This full knowledge attack represents the most severe scenario, and in this paper, we employ it to assess the efficacy of our proposed approach.

\myparatight{Defender’s knowledge and goal}%
Our goal is to develop a reliable FL detection method that identifies malicious clients based solely on their local model updates, without access to training data or prior knowledge of data distributions or attack strategies. The detection mechanism should ensure {\em robust learning integrity} by preserving benign clients in non-adversarial settings and avoiding their unintended removal. At the same time, it must achieve {\em effective threat mitigation} by detecting both targeted and non-targeted attacks during training, while maintaining high classification accuracy and minimizing the impact of malicious updates on the global model.


\section{Our \alg} 
\label{our_method}

\subsection{Overview}
In our proposed \alg, the server begins by collecting a trajectory of the global model and uses this information to generate a synthetic dataset. This synthetic dataset serves as a tool to identify potentially malicious clients by analyzing their behavior. 
Specifically, malicious local models often generate loss patterns that differ noticeably from those of benign models, facilitating their identification.

\subsection{Global Model Trajectory Collection}
\label{trajectory_collection}

In our proposed method, the server possesses its own distinct dataset. 
Upon receiving local models from clients, the server distinguishes between malicious and benign models by comparing their performance on this separate dataset. 
Ideally, we might assume the server having a small, clean training dataset, as posited in~\cite{2021FLTrust}, where it is assumed that both the server's dataset and the overall training dataset used by clients are drawn from the same distribution. 
However, in FL, this assumption does not hold in practice since clients' training data remain on their devices, making it challenging, if not impossible, for the server to have perfect knowledge of the distribution of clients' training data.

To address this challenge, our approach involves the server first gathering a trajectory of the global model and subsequently generating a synthetic dataset based on this collected trajectory. 
It is important to note that the generated synthetic dataset does not have to replicate the distribution of clients' training data. 
Instead, it merely needs to be a basis on which malicious and benign local models exhibit different performance.
Let $\{\mathbf{w}^1, \mathbf{w}^2, ..., \mathbf{w}^{\epsilon}\}$ represent the trajectory gathered by the server, where $\epsilon$ is the length of trajectory, and $\mathbf{w}^t$ denotes the global model at the $t$-th training round, $t=1,2,...,\epsilon$.
Note that we assume that the server gathers the first $\epsilon$ global models.
The primary challenge here is determining how to calculate each global model in the trajectory. The straightforward solution involves computing $\mathbf{w}^t$ by directly aggregating the $n$ received local models, that is, $\mathbf{w}^{t+1}=\text{AR}\{\mathbf{w}_1^{t}, \mathbf{w}_2^{t},...,\mathbf{w}_n^{t}\}$ for $t=1,2,...,\epsilon$. However, since some clients may act maliciously and send arbitrary local models to the server, the global model may be corrupted if potential malicious local models are not removed before aggregation. 
The server addresses this by excluding potentially malicious local models during the initial $\epsilon$ training rounds before constructing the global model trajectory.

Our primary insight is that, during poisoning attacks on FL, malicious clients typically alter either the directions and/or magnitudes of their local models. 
By leveraging this understanding, the server categorizes the received clients' local models into several clusters. The local models within the largest clusters are considered benign, underpinning the belief that the majority of clients in FL are benign. 
Moreover, local models from benign clients tend to cluster together.
Let $\text{Cluster}()$ represent the clustering technique employed by the server, such as K-means algorithm. 
Let $\mathcal{H}^t$ denote the largest cluster formed when the server categorizes the $n$ local models received into several clusters at training round $t$, where $t=1,2,...,\epsilon$.
Thus, we can express this as:
\begin{align}
\mathcal{H}^t = \text{Cluster}(\mathbf{w}_1^{t}, \mathbf{w}_2^{t},...,\mathbf{w}_n^{t}).
\end{align}

After that, the server computes $\mathbf{w}^t$ by aggregating the local models within cluster $\mathcal{H}^t$ as
$
\mathbf{w}^{t+1} = \text{AR}\{ \mathbf{w}_i^{t}, i \in \mathcal{H}^t\}.
$
This process enables the server to gather the trajectory $\{\mathbf{w}^1, \mathbf{w}^2, ..., \mathbf{w}^{\epsilon}\}$ over the initial $\epsilon$ training rounds.

\subsection{Synthetic Data Generation}

With the collection of the global model trajectory $\{\mathbf{w}^1, \mathbf{w}^2, ..., \mathbf{w}^{\epsilon}\}$, the server can now leverage this information to generate a synthetic dataset, drawing on insights from recent studies in dataset condensation~\cite{wang2022cafe,cazenavette2022dataset,zhao2020dataset,liu2023slimmable,zhao2023dataset,kim2022dataset,pi2023dynafed}. 
This trajectory represents the sequence of global models generated over the first $\epsilon$ rounds of training. By analyzing this sequence, the server can effectively create synthetic data that mimics the underlying patterns learned by the model over time.
To better understand how this process works, let's consider a network denoted by $f$, which represents the model architecture used to generate the synthetic dataset. The primary goal here is to generate a synthetic dataset, represented as $D_{\text{syn}} = \{\mathbf{X}, \mathbf{Y}\}$. In this dataset, $\mathbf{X}$ represents the input features and $\mathbf{Y}$ corresponds to the labels. The objective is to ensure that when we train the network $f$ on this synthetic data, the results are comparable to those obtained when the network is trained on the full, real dataset $D$, which contains data from all participating clients in the FL process. Essentially, we want the synthetic data to be a good approximation of the real data in terms of its ability to train the network effectively.

To generate this synthetic data, consider two global models: $\mathbf{w}^{\alpha}$ and $\mathbf{w}^{\alpha+\Delta}$. Here, $\mathbf{w}^{\alpha}$ represents a global model from the collected trajectory $\{\mathbf{w}^1, \mathbf{w}^2, ..., \mathbf{w}^{\epsilon}\}$, and $\mathbf{w}^{\alpha+\Delta}$ is another global model in the same trajectory, where $\Delta$ is a step parameter that defines the number of rounds between the two models. By training the network $f$ for $\Delta$ steps starting from the model $\mathbf{w}^{\alpha}$ using the synthetic dataset $D_{\text{syn}}$, we aim to obtain a model that closely resembles $\mathbf{w}^{\alpha+\Delta}$. The idea is that the synthetic data should enable the network to transition from $\mathbf{w}^{\alpha}$ to $\mathbf{w}^{\alpha+\Delta}$, just as it would if trained on real data $D$.
This synthetic data generation can be formulated as an optimization problem, where the goal is to minimize the difference between the model obtained by training on the synthetic data and the target model $\mathbf{w}^{\alpha+\Delta}$. Formally, the problem is expressed as follows:
\begin{equation}
\label{sys_gen}
\begin{gathered}
    \min _{\mathbf{X},\mathbf{Y}} \|\hat{\mathbf{w}} - \mathbf{w}^{\alpha+\Delta}\|_2^2 \ ,
\\
\text{s.t.} \ \hat{\mathbf{w}}= f(\mathbf{X},\mathbf{Y}, \mathbf{w}^{\alpha}, \Delta),
\end{gathered}
\end{equation}
where $f(\mathbf{X},\mathbf{Y}, \mathbf{w}^{\alpha}, \Delta)$ represents training the network $f$ on the synthetic dataset $D_{\text{syn}} = \{\mathbf{X}, \mathbf{Y}\}$ for $\Delta$ steps, starting from the model $\mathbf{w}^{\alpha}$. 
The objective is to find the synthetic data $\mathbf{X}$ and $\mathbf{Y}$ that minimize the squared Euclidean distance $\|\hat{\mathbf{w}} - \mathbf{w}^{\alpha+\Delta}\|_2^2$, ensuring the resulting model $\hat{\mathbf{w}}$ closely aligns with the global model $\mathbf{w}^{\alpha+\Delta}$.

We can apply the gradient descent method to iteratively solve Problem~(\ref{sys_gen}). 
The synthetic dataset generation algorithm (SynGen) is outlined in Algorithm~\ref{sys_gen_alg}. 
Specifically, in each iteration, we begin by randomly and uniformly selecting $\alpha$ from the sequence \(\{1, 2, \ldots, \epsilon-\Delta\}\). 
Then, we retrieve \(\mathbf{w}^{\alpha}\) and \(\mathbf{w}^{\alpha+\Delta}\) from the trajectory \(\{\mathbf{w}^1, \mathbf{w}^2, \ldots, \mathbf{w}^{\epsilon}\}\). 
Note that the sequence \(\{1, 2, \ldots, \epsilon-\Delta\}\) starts at 1 and ends at $\epsilon-\Delta$ to ensure that both \(\mathbf{w}^{\alpha}\) and \(\mathbf{w}^{\alpha+\Delta}\) are within the trajectory.
Following this, the server obtains $\hat{\mathbf{w}}$ by training the network for $\Delta$ steps (Line~\ref{sys_gen_alg_train_h_step} in Algorithm~\ref{sys_gen_alg}). 
The server then calculates the gradient of $\|\hat{\mathbf{w}} - \mathbf{w}^{\alpha+\Delta}\|_2^2$ with respect to $\mathbf{X}$ and $\mathbf{Y}$, and proceeds to update the features and labels of the synthetic dataset using gradient descent (Lines~\ref{sys_gen_alg_train_grad}-\ref{sys_gen_alg_train_gd}).
At the conclusion of Algorithm~\ref{sys_gen_alg}, we obtain the synthetic dataset $D_{\text{syn}}$.
Note that in Algorithm~\ref{sys_gen_alg}, \(\mathbf{X}^{\kappa}\) and \(\mathbf{Y}^{\kappa}\) represent the features and labels of the synthetic dataset at iteration \(\kappa\), respectively.

\begin{algorithm}[t]
\small
		\caption{SynGen.}
		\label{local_training}
		\begin{algorithmic}[1]
        \renewcommand{\algorithmicensure}{\textbf{Input:}}
        \Ensure Global model trajectory $\{\mathbf{w}^1, \mathbf{w}^2, ..., \mathbf{w}^{\epsilon}\}$;  training iterations $\Psi$; network $f$; learning rate $\gamma$; parameter $\Delta$.
        \renewcommand{\algorithmicensure}{\textbf{Output:}}
         \Ensure $D_{\text{syn}}$. 
         \State Initialize $\mathbf{X}^1$ and $\mathbf{Y}^1$.
		\For {$\kappa = 1, 2, \cdots, \Psi$}
             \State Randomly and uniformly select the $\alpha$ from the sequence \(\{1, 2, \ldots, \epsilon-\Delta\}\).
             \State Retrieve \(\mathbf{w}^{\alpha}\) and \(\mathbf{w}^{\alpha+\Delta}\) from the model trajectory \(\{\mathbf{w}^1, \mathbf{w}^2, \ldots, \mathbf{w}^{\epsilon}\}\).
             \State Train the network \( f \) on the current synthetic dataset for \(\Delta\) iterations to obtain the updated model \(\hat{\mathbf{w}}\).
             \label{sys_gen_alg_train_h_step}
             \State Evaluate the squared Euclidean distance $\|\hat{\mathbf{w}} - \mathbf{w}^{\alpha+\Delta}\|_2^2$, then derive the gradients $\nabla_{\mathbf{X}^{\kappa}}\|\hat{\mathbf{w}} - \mathbf{w}^{\alpha+\Delta}\|_2^2$ and $\nabla_{\mathbf{Y}^{\kappa}}\|\hat{\mathbf{w}} - \mathbf{w}^{\alpha+\Delta}\|_2^2$.
             \label{sys_gen_alg_train_grad}
              \State Update the features and labels by applying gradient descent: $\mathbf{X}^{\kappa+1}=\mathbf{X}^{\kappa} - \gamma \nabla_{\mathbf{X}^{\kappa}}\|\hat{\mathbf{w}} - \mathbf{w}^{\alpha+\Delta}\|_2^2$, $\mathbf{Y}^{\kappa+1}=\mathbf{Y}^{\kappa} - \gamma \nabla_{\mathbf{Y}^{\kappa}}\|\hat{\mathbf{w}} - \mathbf{w}^{\alpha+\Delta}\|_2^2$.
             \label{sys_gen_alg_train_gd}
		\EndFor
		\end{algorithmic}
  \label{sys_gen_alg}
\end{algorithm}

\subsection{Malicious Clients Detection Using $D_{\text{syn}}$}
Once we have acquired the synthetic dataset $D_{\text{syn}}$, we can utilize it to detect potential malicious clients. 
In the following, we detail two variants of our defense strategy, named \algFirst and \algSecond, each employing a distinct approach to detect malicious clients.

\begin{algorithm}[t!]
    \caption{\alg.}
    \label{our_alg}
    \begin{algorithmic}[1]
        \renewcommand{\algorithmicrequire}{\textbf{Input:}}
        \renewcommand{\algorithmicensure}{\textbf{Output:}}
        \Require The $n$ clients, each with local training datasets $D_i$ for $i = 1, 2, \dots, n$; the total number of global training rounds $T$; learning rate $\mu$; aggregation rule AR; clustering algorithm $\text{Cluster}()$; network $f$; and parameters $\epsilon$, $\Psi$, $\gamma$, and $\Delta$.
        \Ensure Global model $\mathbf{w}^T$. 
        \State Initialize $\mathbf{w}^1$.
        \State $\mathcal{S} \leftarrow \emptyset$.
        \State $\mathcal{S} \leftarrow \mathcal{S} \cup \{\mathbf{w}^{1}\}$.
        \For{$t=1,2,\cdots,T$}
            \State // Step I (Global model synchronization).
            \label{step_I_FL}
            \State Server distributes the current global model $\mathbf{w}^t$ to all clients.
            \State // Step II (Local models updating).
            \For {each client $i=1,2,\cdots,n$ in parallel}
                \State Client \( i \) updates its local model \(\mathbf{w}_i^t\) using \(\mathbf{w}^t\) and \(D_i\).
                \State Send $\mathbf{w}_{i}^{t}$ to the server.
            \EndFor
             \label{step_II_FL}
            \State // Step III (Aggregation and global model updating).
            \label{step_III_FL}
            \tikz[overlay, remember picture] \coordinate (start1);
            \State // Global model trajectory collection.
            \If {$t < \epsilon$}
                \State $\mathcal{H}^t = \text{Cluster}(\mathbf{w}_1^{t}, \mathbf{w}_2^{t},...,\mathbf{w}_n^{t})$.
                \label{cluester_one}
                \State $\mathbf{w}^{t+1} = \text{AR}\{ \mathbf{w}_i^{t}, i \in \mathcal{H}^t\}$.
                \State $\mathcal{S} \leftarrow \mathcal{S} \cup \{\mathbf{w}^{t+1}\}$.
            \EndIf
            \tikz[overlay, remember picture] \coordinate (end1);
            \tikz[overlay, remember picture] \coordinate (start2);
            \State // Synthetic data generation.
            \label{alg_Synthetic}
            \If {$t = \epsilon$}
                \State $D_{\text{syn}}=\text{SynGen}(\mathcal{S},\Psi,f,\gamma,\Delta)$.
            \EndIf
            \label{end_alg_Synthetic}
            \tikz[overlay, remember picture] \coordinate (end2);
            \tikz[overlay, remember picture] \coordinate (start3);
            \State // Malicious clients detection using $D_{\text{syn}}$.
            \label{alg_Malicious_detection}
            \If {$t \ge \epsilon$}
                \State The server computes the loss \(l_i^t\) by applying client \(i\)'s local model \(\mathbf{w}_i^t\) on the synthetic dataset \(D_{\text{syn}}\), for \(i = 1, 2, \ldots, n\).
                \If {\algFirst is used}
                    \State Calculate \(r_i^t\) for each client using Eq.~(\ref{adap}).
                    \State $\mathbf{w}^{t+1}=\sum_{i=1}^n r_i^t  \mathbf{w}_i^{t}$. 
                \ElsIf {\algSecond is used}
                    \State $\mathcal{Q}^t = \text{Cluster}(l_1^{t}, l_2^{t}, \ldots, l_n^{t})$.
                    \label{cluester_two}
                    \State \(\mathbf{w}^{t+1} = \text{AR}\{\mathbf{w}_i^t: i \in \mathcal{Q}^t\}\).
                \EndIf
            \EndIf
             \label{end_step_III_FL}
            \tikz[overlay, remember picture] \coordinate (end3);
        \EndFor
    \end{algorithmic}
    \vspace{-0.9em} 
    \begin{tikzpicture}[overlay, remember picture]
        \fill[blue!20, rounded corners, fill opacity=0.2] 
            ([xshift=-23.0em, yshift=-0.3em] start1) rectangle ([xshift=21.2em, yshift=-0.2em] end1);
        \fill[green!20, rounded corners, fill opacity=0.2] 
            ([xshift=-4.6em, yshift=-0.3em] start2) rectangle ([xshift=21.0em, yshift=-0.2em] end2);
        \fill[red!20, rounded corners, fill opacity=0.2] 
            ([xshift=-4.8em, yshift=-0.3em] start3) rectangle ([xshift=21.0em, yshift=-0.2em] end3);
    \end{tikzpicture}
\end{algorithm}

\myparatight{\algFirst}In this section, we introduce the \textsf{SafeFL-MedianLoss} (\algFirst), which is the first variant of our approach.
In FL, malicious clients often aim to maximize their attack impact by manipulating the directions and/or magnitudes of their local models. 
The fundamental idea behind \algFirst is that malicious local models typically result in a larger loss when evaluated on the synthetic dataset $D_{\text{syn}}$ compared to the loss observed with benign local models.
Using this observation, the server calculates the loss for every received local model, and then computes the median of these $n$ losses, where $n$ is the total number of clients.
In particular, at the training round $t$, one has that:
\begin{align}
l_{\text{Med}}^{t} =\text{Median}\{l_1^{t},l_2^{t},...,l_n^{t}\}, 
\end{align}
where $l_i^{t}$ represents the loss when the server employs client $i$'s local model $\mathbf{w}_i^t$ to compute the loss on the synthetic dataset $D_{\text{syn}}$ during training round $t$.

In \algFirst, client $i$ is identified as benign if its loss $l_i^{t}$ is smaller than $l_{\text{Med}}^{t}$, where $i = 1, 2, \ldots, n$. 
Furthermore, concerning clients identified as benign, higher losses suggest poorer alignment with the training data distribution, while lower losses reflect better fitting performance. Clients with lower losses are considered more reliable.
Thus, at training round $t$, we can derive the weight of client $i$ using the following formula:
\begin{align}
r_i^t=\left\{\begin{array}{ll}
\frac{(l^t_i)^{-1}}{\sum_{j=1}^n(l^t_j)^{-1}}, & \text { if } l_{i}^{t} \leq  l_{\text{Med}}^{t}, \\
0, & \text { otherwise, }
\end{array}\right.
\label{adap}
\end{align}

Then, the server can aggregate the local models using the weighted average approach as $\mathbf{w}^{t+1}=\sum_{i=1}^n r_i^t  \mathbf{w}_i^{t}$.

\myparatight{\algSecond}Our median loss selection method, \algFirst, identifies half of the clients as suspicious in each round, regardless of whether their losses closely match or significantly deviate from the median. While this approach ensures consistency in detecting a fixed proportion of suspicious clients, it risks misclassifying benign clients with losses slightly above the median as malicious. Such misclassification can lead to the exclusion of valuable local training data, as a substantial number of benign local models are left out during aggregation, potentially degrading model performance. To address these shortcomings, we propose integrating a clustering-based approach for identifying potential malicious clients based on their computed loss values.

In our \textsf{SafeFL-ClusterLoss} (\algSecond) method, the server also evaluates the loss of each received local model on the synthetic dataset \(D_{\text{syn}}\). However, instead of relying on the median loss for classification, the server employs a clustering algorithm to group the loss values. This approach is motivated by the observation that losses derived from benign local models are more likely to form a coherent cluster. 
Let \(\mathcal{Q}^t\) denote the largest cluster obtained when the server partitions the \(n\) losses into multiple clusters during training round \(t\). Formally, one has that:
\begin{align}
    \mathcal{Q}^t = \text{Cluster}(l_1^{t}, l_2^{t}, \ldots,l_n^{t}).
\end{align}

Using this cluster, the server computes the global model \(\mathbf{w}^{t+1}\) as \(\mathbf{w}^{t+1} ={\text{AR}\{\mathbf{w}_i^t: i \in \mathcal{Q}^t\}}\), where \(\text{AR}\) denotes the aggregation rule applied to the local models within the largest cluster. This clustering-based method ensures that aggregation focuses on benign clients, thereby improving the robustness and effectiveness of the model.

Algorithm~\ref{our_alg} provides an overview of the complete process for our proposed method, \alg. 
Lines~\ref{step_I_FL}-\ref{step_II_FL} outline the first two steps of standard FL. In Lines~\ref{step_III_FL}-\ref{end_step_III_FL}, \alg identifies potential malicious clients and updates the global model using contributions from the remaining clients. 
Our \alg consists of three main components: ``global model trajectory collection'', ``synthetic data generation'', and ``malicious clients detection using $D_{\text{syn}}$''. 
Specifically, during the first $\epsilon$ training rounds, the server collects global model trajectories, forming a set denoted as $\mathcal{S}$.
At training round $\epsilon$, the server utilizes these collected trajectories to generate a synthetic dataset $D_{\text{syn}}$.
It is important to note that $D_{\text{syn}}$ is constructed only once at round $\epsilon$, as detailed in Lines~\ref{alg_Synthetic}-\ref{end_alg_Synthetic} of Algorithm~\ref{our_alg}.
Once the synthetic dataset $D_{\text{syn}}$ is generated, the server uses it to detect malicious clients, as described in Lines~\ref{alg_Malicious_detection}-\ref{end_step_III_FL}.


\section{Experiments} \label{sec:exp}

\subsection{Experimental Setup}

\subsubsection{Datasets}
In our experiments, we incorporate the following five datasets: CIFAR-10~\cite{cifar10data}, MNIST~\cite{mnist}, FEMNIST~\cite{femnist}, STL-10~\cite{coates2011analysis}, and Tiny-ImageNet~\cite{deng2009imagenet}. 
\textcolor{black}{
The details of these datasets are shown in Appendix~\ref{sec:setting_app_dataset}.
}

\subsubsection{Poisoning attacks to FL} 
We consider four single-method poisoning attacks (Trim attack~\cite{2019Local}, Scaling attack~\cite{2018How}, Distributed Backdoor Attack (DBA)~\cite{xie2019dba}, and Adaptive attack~\cite{shejwalkar2021manipulating}), along with two hybrid poisoning strategies (Trim+DBA, and Scaling+DBA attacks) to evaluate the effectiveness of our proposed detection method. In the case of single-method attacks, every malicious client uses the same strategy to craft their local models. For instance, in a Trim attack scenario, all involved malicious clients adopt the Trim attack technique to formulate the models they submit to the server. Conversely, in hybrid attacks, different malicious clients might utilize varying attack strategies.
\textcolor{black}{ 
See Appendix~\ref{sec:setting_app_attack} for a detailed description of these attacks.
}
Note that we also consider \textcolor{black}{five} more advanced attacks is Section~\ref{sec:discussion_limitation}.

\subsubsection{Defenses against Poisoning Attacks to FL} 
In this paper, we compare our \alg against seven detection-based approaches (FLAME~\cite{2021FLAME}, FLDetector~\cite{zhang2022fldetector}, FLTrust~\cite{2021FLTrust}, DeepSight~\cite{rieger2022deepsight}, BackdoorIndicator~\cite{li2024backdoorindicator}, FreqFed~\cite{fereidooni2024freqfed}, FedREDefense~\cite{xie2024fedredefense}) and three prevention-based methods (such as Median~\cite{yin2018byzantine}, Trimmed mean (TrMean)~\cite{yin2018byzantine}, and Krum~\cite{blanchard2017machine}).
\textcolor{black}{ 
Refer to Appendix~\ref{sec:setting_app_defense} for comprehensive details of these defenses.
}

\subsubsection{Non-IID setting}
FL is characterized by the non-independent and identically distributed (Non-IID) nature of training data across clients. To simulate this, we follow the approach from~\cite{2019Local}. In a dataset with $M$ classes, clients are divided into $M$ groups. Each training sample with label $g$ is assigned to group $g$ with probability $q$ and to other groups with a probability of $\frac{1-q}{M-1}$. The parameter $q$ determines the degree of Non-IID distribution; when $q = \frac{1}{M}$, the data is IID, otherwise, it is Non-IID. For CIFAR-10, MNIST, STL-10, and Tiny-ImageNet, we set $q = 0.5$, while FEMNIST remains unchanged due to its inherently Non-IID distribution.

\subsubsection{Evaluation metrics}
We evaluate using five metrics: three for detection—detection accuracy (DACC), false positive rate (FPR), and false negative rate (FNR)—and two for the final global model—testing accuracy (TACC) and attack success rate (ASR). 
For detection-based methods, we assess both detection performance and final model accuracy, while for prevention-based methods, we focus solely on final model accuracy. Detection methods, including baselines and our \alg, identify malicious clients during each training round, with results averaged over all rounds.
It is important to note that we detect malicious clients in every round. When clients are identified as malicious, the server ignores their local models for that particular round, rather than permanently removing them from the system. This approach is taken because malicious clients may choose to attack in certain rounds and refrain from attacking in others. Permanently removing clients upon detection could result in the exclusion of some benign clients.

\myparatight{a)~Detection accuracy (DACC)}
DACC measures the percentage of clients correctly classified, ensuring benign clients are identified as benign and malicious clients are recognized as malicious.

\myparatight{b)~False positive rate (FPR)}
FPR represents the ratio of benign clients incorrectly predicted as malicious.

\myparatight{c)~False negative rate (FNR)}
FNR denotes the proportion of malicious clients erroneously classified as benign.

\myparatight{d)~Testing accuracy (TACC)}
TACC represents the proportion of test samples accurately predicted by the final global model.

\myparatight{e)~Attack success rate (ASR)}
ASR represents the proportion of targeted test samples that are classified into the specific label designated by the attacker.

Higher DACC indicates stronger detection performance, while lower FPR and FNR indicate better detection capabilities. For all defenses, higher TACC and lower ASR reflect a more robust model.

\subsubsection{Parameter settings}

\begin{table*}[htbp]
  \centering
  \footnotesize
  \caption{Detection performance of various detection-based methods is assessed using DACC (\(\uparrow\)), FPR (\(\downarrow\)), and FNR (\(\downarrow\)) metrics. Here, \(\uparrow\) denotes better detection performance with higher values, and \(\downarrow\) denotes better performance with lower values.}
\begin{tabular}{|c|c|ccc|ccc|ccc|ccc|ccc|}
\hline
\multirow{2}{*}{{Attack}} & \multirow{2}{*}{{Defense}} & \multicolumn{3}{c|}{{CIFAR-10}} & \multicolumn{3}{c|}{{MNIST}} & \multicolumn{3}{c|}{{FEMNIST}} & \multicolumn{3}{c|}{{STL-10}} & \multicolumn{3}{c|}{{Tiny-ImageNet}} \\ \cline{3-17} 
                                 &                                     & DACC        & FPR        & FNR        & DACC       & FPR        & FNR       & DACC        & FPR        & FNR        & DACC       & FPR        & FNR        & DACC          & FPR          & FNR          \\ \hline
\multirow{9}{*}{No attack}              & FLAME                               & 0.57        & NA         & NA         & 0.57       & NA         & NA        & 0.61        & NA         & NA         & 0.65       & NA         & NA         & 0.56          & NA           & NA           \\
                                 & FLDetector                          & 0.65        & NA         & NA         & 0.52       & NA         & NA        & 0.63        & NA         & NA         & 0.59       & NA         & NA         & 0.68          & NA           & NA           \\
                                 & FLTrust                             & 0.82        & NA         & NA         & 0.77       & NA         & NA        & 0.89        & NA         & NA         & 0.75       & NA         & NA         & 0.83          & NA           & NA           \\
                                 & DeepSight                           & 1.00        & NA         & NA         & 0.97       & NA         & NA        & 0.93        & NA         & NA         & 0.92       & NA         & NA         & 0.98          & NA           & NA           \\
                                 & BackdoorIndicator                           & 0.95        & NA         & NA         & 0.93       & NA         & NA        & 0.97        & NA         & NA         & 0.92       & NA         & NA         & 0.86          & NA           & NA           \\
\hline
\end{tabular}
  \label{tab:detection}%
\end{table*}%

By default, our experiments involve 100 clients for CIFAR-10, MNIST, and STL-10 datasets, 300 clients for FEMNIST, and 400 clients for Tiny-ImageNet.
In the default setup, 30\% of clients are considered malicious.
For the MNIST and FEMNIST datasets, we have employed a four-layer Convolutional neural network (CNN), as detailed in Table~\ref{cnn} in Appendix, as the model architecture. 
In the case of CIFAR-10, STL-10 and Tiny-ImageNet datasets, we have adopted the widely recognized ResNet-20 architecture~\cite{2016Deep} as the architecture.
%
%
The batch size is set as 64 for CIFAR-10, MNIST, FEMNIST, and Tiny-ImageNet datasets, and 32 for the STL-10 dataset. The total training rounds are set at 1500 for CIFAR-10, 1000 for STL-10, MNIST, and FEMNIST datasets, and 2000 for Tiny-ImageNet. In CIFAR-10 and Tiny-ImageNet, the initial learning rate is 0.15 for the first 1000 training rounds, subsequently reduced by a factor of 0.5 every 250 rounds thereafter. For STL-10, MNIST, and FEMNIST, the learning rate is initially set to 0.10 for the first 500 rounds and then reduced by a factor of 0.5 every 250 rounds after that.
\textcolor{black}{
By default, we assume that all clients participate in each training round (i.e., a selection rate of 100\%).
}
We consider the worst-case setting where the attacker performs attacks in every training round.

For our \alg, we use 
 K-means clustering algorithm~\cite{hartigan1979algorithm} when generating the synthetic data, and we use the Mean-shift clustering algorithm~\cite{cheng1995mean} to group the losses in \algSecond. 
The global model trajectory length (\(\epsilon\)) for the synthetic dataset is set to 25 for MNIST, CIFAR-10, and FEMNIST, and 30 for STL-10 and Tiny-ImageNet. The parameter \(\gamma\) is consistently set to 0.1 across all datasets during the synthetic dataset generation. The value of \(\Psi\) is configured as 5000 for CIFAR-10, STL-10, and MNIST, 8500 for FEMNIST, and 10,000 for Tiny-ImageNet. 
Following~\cite{pi2023dynafed}, the server uses the same network \( f \) to generate the synthetic dataset as the clients use for local training.
Furthermore, the $\Delta$ is fixed at 15 for all datasets. For each dataset, the synthetic dataset size is set to 100.
In our \alg, once the server identifies the malicious clients, it aggregates the remaining local models using the FedAvg method, meaning the aggregation rule AR in \alg is configured as FedAvg.

\subsection{Experimental results}
\label{results}

\begin{table*}[t]
  \centering
    \footnotesize
  \caption{Performance of final global models obtained through various detection-based methods, where TACC (\(\uparrow\)) and ASR (\(\downarrow\)) metrics are considered. \(\uparrow\) denotes better performance with higher values, and \(\downarrow\) denotes better performance with lower values.}
\begin{tabular}{|c|c|cc|cc|cc|cc|cc|}
\hline
\multirow{2}{*}{{Attack}} & \multirow{2}{*}{{Defense}} & \multicolumn{2}{c|}{{CIFAR-10}}                                  & \multicolumn{2}{c|}{{MNIST}}                                    & \multicolumn{2}{c|}{{FEMNIST}}                                  & \multicolumn{2}{c|}{{STL-10}}                                   & \multicolumn{2}{c|}{{Tiny-ImageNet}} \\ \cline{3-12} 
                                 &                                     & TACC                              & ASR                                & TACC                              & ASR                                & TACC                              & ASR                                & TACC                              & ASR                                & TACC                 & ASR                  \\ \hline
\multirow{9}{*}{{\makecell {No attack}}}              & FLAME                               & 0.72                              & NA                                 & 0.91                              & NA                                 & 0.60                              & NA                                 & 0.47                              & NA                                 & 0.46                 & NA                   \\
                                 & FLDetector                          & 0.77                              & NA                                 & 0.96                              & NA                                 & 0.63                              & NA                                 & 0.45                              & NA                                 & 0.41                 & NA                   \\
\hline
\end{tabular}
\label{tab:tacc}
\end{table*}

\begin{table}[]
    \scriptsize
     \footnotesize
  \addtolength{\tabcolsep}{-4.205pt}  
  \caption{The performance of the final global model obtained through various prevention-based methods and our approach, where TACC (\(\uparrow\)) and ASR (\(\downarrow\)) metrics are considered. Here, \(\uparrow\) denotes better performance with higher values, and \(\downarrow\) denotes better performance with lower values.}
\begin{tabular}{|c|c|cc|cc|cc|cc|cc|}
\hline
\multirow{2}{*}{{Attack}} & \multirow{2}{*}{{Defense}} & \multicolumn{2}{c|}{{CIFAR-10}} & \multicolumn{2}{c|}{{MNIST}} & \multicolumn{2}{c|}{{FEMNIST}} & \multicolumn{2}{c|}{{STL-10}} & \multicolumn{2}{c|}{{\makecell {Tiny-\\ImageNet}}} \\ \cline{3-12} 
 &     & TACC     & ASR      & TACC    & ASR  & TACC  & ASR  & TACC    & ASR              & TACC  & ASR  \\ \hline
\multirow{5}{*}{{\makecell {No\\ attack}}}              & Median                              & 0.71              & NA                & 0.96             & NA               & 0.63              & NA                & 0.47              & NA               & 0.45                 & NA                   \\
                                 
                                 & TrMean                       & 0.70              & NA                & 0.94             & NA               & 0.62              & NA                & 0.49              & NA               & 0.44                 & NA                   \\
    & Krum                                & 0.65              & NA                & 0.84             & NA               & 0.59              & NA                & 0.44              & NA               & 0.39                 & NA                   \\
                                 & \cellcolor{greyL}\algFirst                           & \cellcolor{greyL}0.82              & \cellcolor{greyL}NA                & \cellcolor{greyL}0.98    & \cellcolor{greyL}NA               & \cellcolor{greyL}0.67              & \cellcolor{greyL}NA                & \cellcolor{greyL}0.53              & \cellcolor{greyL}NA               & \cellcolor{greyL}0.51                 & \cellcolor{greyL}NA                   \\
\hline
\end{tabular}
\label{tab:prevention}
   \vspace{-15pt}
\end{table}

\begin{table*}[htbp]
        \footnotesize
  \addtolength{\tabcolsep}{-1.625pt}  
\caption{The impact of the malicious client ratio, total client number, and Non-IID degree is analyzed using the CIFAR-10 dataset. DACC values are reported for the Trim attack, Scaling attack, and DBA attack. The results of Trim+DBA attack, Scaling+DBA attack, and Adaptive attack are shown in Table~\ref{para_impact_app} in Appendix. ``BDIndicator'' refers to the ``BackdoorIndicator'' method.}
\subfloat[Impact of fraction of malicious clients.]
{\tabcolsep 0.1cm
\begin{tabular}{|c|c|ccccc|}
\hline
\multirow{2}{*}{Attack}          & \multirow{2}{*}{Defense} & \multicolumn{5}{c|}{Malicious client ratio} \\ \cline{3-7} 
                                 &                          & 0\%      & 10\%     & 20\%     & 30\%    & 40\%    \\ \hline
\multirow{9}{*}{{\makecell {Trim\\ attack}}}        & FLAME        & 0.57                    & 0.62                     & 0.73                     & 0.78                     & 0.81                      \\
                             & FLDetector   & 0.65                    & 0.96                     & 0.95                     & 0.96                     & 0.94                      \\
                             & FLTrust      & 0.80                    & 0.90                     & 0.91                     & 0.85                     & 0.87                      \\
                             & DeepSight    & 1.00                    & 0.89                     & 0.88                     & 0.88                     & 0.87                      \\
                             & BDIndicator    & 0.95                    & 0.73                     & 0.75                     & 0.73                     & 0.76                      \\
                             & FreqFed      & 0.81                    & 0.84                     & 0.86                     & 0.89                     & 0.87                      \\
                             & FedREDefense & 0.78                    & 0.85                     & 0.89                     & 0.85                     & 0.83                      \\
                             & \cellcolor{greyL}\algFirst    & \cellcolor{greyL}0.94                    & \cellcolor{greyL}0.94                     & \cellcolor{greyL}0.94                     & \cellcolor{greyL}0.90                     & \cellcolor{greyL}0.97                      \\
                             & \cellcolor{greyL}\algSecond    & \cellcolor{greyL}1.00                    & \cellcolor{greyL}0.99                     & \cellcolor{greyL}1.00                     & \cellcolor{greyL}1.00                     & \cellcolor{greyL}0.97                      \\ \hline
\multirow{9}{*}{{\makecell {Scaling\\ attack}}}     & FLAME        & 0.57                    & 0.69                     & 0.78                     & 0.84                     & 0.89                      \\
                             & FLDetector   & 0.65                    & 0.96                     & 0.98                     & 1.00                     & 0.75                      \\
                             & FLTrust      & 0.80                    & 0.72                     & 0.70                     & 0.75                     & 0.82                      \\
                             & DeepSight    & 1.00                    & 0.82                     & 0.85                     & 0.88                     & 0.92                      \\
                             & BDIndicator    & 0.95                    & 0.95                     & 0.97                     & 0.94                     & 0.93                      \\
                             & FreqFed      & 0.81                    & 0.84                     & 0.86                     & 0.84                     & 0.87                      \\
                             & FedREDefense & 0.78                    & 0.84                     & 0.87                     & 0.87                     & 0.86                      \\
                             & \cellcolor{greyL}\algFirst    & \cellcolor{greyL}0.94                    & \cellcolor{greyL}0.97                     & \cellcolor{greyL}0.94                     & \cellcolor{greyL}0.91                     & \cellcolor{greyL}0.99                      \\
                             & \cellcolor{greyL}\algSecond    & \cellcolor{greyL}1.00                    & \cellcolor{greyL}0.99                     & \cellcolor{greyL}1.00                     & \cellcolor{greyL}1.00                     & \cellcolor{greyL}1.00                      \\ \hline
\multirow{9}{*}{{\makecell {DBA\\ attack}}}         & FLAME        & 0.57                    & 0.70                     & 0.74                     & 0.82                     & 0.84                      \\
                             & FLDetector   & 0.65                    & 0.91                     & 0.88                     & 0.89                     & 0.92                      \\
                             & FLTrust      & 0.80                    & 0.84                     & 0.87                     & 0.80                     & 0.82                      \\
                             & DeepSight    & 1.00                    & 0.85                     & 0.87                     & 0.88                     & 0.90                      \\
                             & BDIndicator    & 0.95                    & 0.95                     & 0.94                     & 1.00                     & 0.97                      \\
                             & FreqFed      & 0.81                    & 0.87                     & 0.89                     & 0.89                     & 0.89                      \\
                             & FedREDefense & 0.78                    & 0.72                     & 0.77                     & 0.75                     & 0.75                      \\
                             & \cellcolor{greyL}\algFirst    & \cellcolor{greyL}0.94                    & \cellcolor{greyL}0.93                     & \cellcolor{greyL}0.97                     & \cellcolor{greyL}0.94                     & \cellcolor{greyL}0.95                      \\
                             & \cellcolor{greyL}\algSecond    & \cellcolor{greyL}1.00                    & \cellcolor{greyL}0.97                     & \cellcolor{greyL}0.97                     & \cellcolor{greyL}1.00                     & \cellcolor{greyL}1.00                      \\ \hline
\end{tabular}
\label{fraction_mali}
}
\subfloat[Impact of total number of clients.]
{
\begin{tabular}{|c|c|ccccc|}
\hline
\multirow{2}{*}{Attack}          & \multirow{2}{*}{Defense} & \multicolumn{5}{c|}{Total client number} \\ \cline{3-7} 
                                 &                          & 60     & 80     & 100    & 120   & 150   \\ \hline
\multirow{9}{*}{{\makecell {Trim\\ attack}}}        & FLAME        & 0.74                   & 0.79                   & 0.78                    & 0.82                    & 0.83                     \\
                             & FLDetector   & 0.90                   & 0.95                   & 0.96                    & 0.97                    & 0.94                     \\
                             & FLTrust      & 0.79                   & 0.84                   & 0.85                    & 0.85                    & 0.86                     \\
                             & DeepSight    & 0.85                   & 0.87                   & 0.88                    & 0.87                    & 0.85                     \\
                             & BDIndicator    & 0.71                   & 0.76                   & 0.73                    & 0.74                    & 0.73                     \\
                             & FreqFed      & 0.85                   & 0.85                   & 0.89                    & 0.87                    & 0.89                     \\
                             & FedREDefense & 0.88                   & 0.88                   & 0.85                    & 0.86                    & 0.87                     \\
                             & \cellcolor{greyL}\algFirst    & \cellcolor{greyL}0.94                   & \cellcolor{greyL}0.96                   & \cellcolor{greyL}0.90                    & \cellcolor{greyL}0.94                    & \cellcolor{greyL}0.92                     \\
                             & \cellcolor{greyL}\algSecond    & \cellcolor{greyL}1.00                   & \cellcolor{greyL}0.97                   & \cellcolor{greyL}1.00                    & \cellcolor{greyL}0.95                    & \cellcolor{greyL}0.99                     \\ \hline
\multirow{9}{*}{{\makecell {Scaling\\ attack}}}     & FLAME        & 0.83                   & 0.82                   & 0.84                    & 0.80                    & 0.84                     \\
                             & FLDetector   & 0.90                   & 0.93                   & 1.00                    & 0.95                    & 0.92                     \\
                             & FLTrust      & 0.74                   & 0.75                   & 0.75                    & 0.77                    & 0.74                     \\
                             & DeepSight    & 0.87                   & 0.85                   & 0.88                    & 0.86                    & 0.83                     \\
                             & BDIndicator    & 0.95                   & 0.92                   & 0.94                    & 0.97                    & 0.90                     \\
                             & FreqFed      & 0.85                   & 0.86                   & 0.84                    & 0.84                    & 0.85                     \\
                             & FedREDefense & 0.84                   & 0.85                   & 0.87                    & 0.87                    & 0.83                     \\
                             & \cellcolor{greyL}\algFirst    & \cellcolor{greyL}0.94                   & \cellcolor{greyL}0.97                   & \cellcolor{greyL}0.94                    & \cellcolor{greyL}0.91                    & \cellcolor{greyL}0.99                     \\
                             & \cellcolor{greyL}\algSecond    & \cellcolor{greyL}1.00                   & \cellcolor{greyL}0.99                   & \cellcolor{greyL}1.00                    & \cellcolor{greyL}1.00                    & \cellcolor{greyL}1.00                     \\ \hline
\multirow{9}{*}{{\makecell {DBA\\ attack}}}         & FLAME        & 0.83                   & 0.85                   & 0.82                    & 0.81                    & 0.84                     \\
                             & FLDetector   & 0.87                   & 0.85                   & 0.89                    & 0.88                    & 0.88                     \\
                             & FLTrust      & 0.80                   & 0.81                   & 0.80                    & 0.82                    & 0.76                     \\
                             & DeepSight    & 0.83                   & 0.84                   & 0.88                    & 0.87                    & 0.84                     \\
                             & BDIndicator    & 0.97                   & 0.92                   & 1.00                    & 0.94                    & 0.95                     \\
                             & FreqFed      & 0.85                   & 0.86                   & 0.89                    & 0.89                    & 0.84                     \\
                             & FedREDefense & 0.75                   & 0.77                   & 0.75                    & 0.77                    & 0.76                     \\
                             & \cellcolor{greyL}\algFirst    & \cellcolor{greyL}0.95                   & \cellcolor{greyL}0.94                   & \cellcolor{greyL}0.94                    & \cellcolor{greyL}0.97                    & \cellcolor{greyL}0.94                     \\
                             & \cellcolor{greyL}\algSecond    & \cellcolor{greyL}1.00                   & \cellcolor{greyL}0.97                   & \cellcolor{greyL}1.00                    & \cellcolor{greyL}0.99                    & \cellcolor{greyL}0.98                     \\ \hline
\end{tabular}
\label{total_number_clients}
}
\subfloat[Impact of degree of Non-IID.]
{
\begin{tabular}{|c|c|ccccc|}
\hline
\multirow{2}{*}{Attack}          & \multirow{2}{*}{Defense} & \multicolumn{5}{c|}{Non-IID degree} \\ \cline{3-7} 
                                 &                          & 0.1      & 0.3      & 0.5      & 0.7     & 0.9     \\ \hline
\multirow{9}{*}{{\makecell {Trim\\ attack}}}        & FLAME        & 0.71                    & 0.76                    & 0.78                    & 0.79                    & 0.77                     \\
                             & FLDetector   & 0.89                    & 0.87                    & 0.96                    & 0.98                    & 1.00                     \\
                             & FLTrust      & 0.81                    & 0.85                    & 0.85                    & 0.82                    & 0.89                     \\
                             & DeepSight    & 0.82                    & 0.84                    & 0.88                    & 0.87                    & 0.94                     \\
                             & BDIndicator    & 0.71                    & 0.69                    & 0.73                    & 0.77                    & 0.79                     \\
                             & FreqFed      & 0.80                    & 0.82                    & 0.89                    & 0.89                    & 0.89                     \\
                             & FedREDefense & 0.89                    & 0.83                    & 0.85                    & 0.78                    & 0.86                     \\
                             & \cellcolor{greyL}\algFirst    & \cellcolor{greyL}0.93                    & \cellcolor{greyL}0.91                    & \cellcolor{greyL}0.90                    & \cellcolor{greyL}0.93                    & \cellcolor{greyL}0.95                     \\
                             & \cellcolor{greyL}\algSecond    & \cellcolor{greyL}0.95                    & \cellcolor{greyL}0.99                    & \cellcolor{greyL}1.00                    & \cellcolor{greyL}0.97                    & \cellcolor{greyL}0.97                     \\ \hline
\multirow{9}{*}{{\makecell {Scaling\\ attack}}}     & FLAME        & 0.82                    & 0.84                    & 0.84                    & 0.81                    & 0.90                     \\
                             & FLDetector   & 0.86                    & 0.95                    & 1.00                    & 0.97                    & 0.92                     \\
                             & FLTrust      & 0.77                    & 0.76                    & 0.75                    & 0.75                    & 0.78                     \\
                             & DeepSight    & 0.72                    & 0.75                    & 0.88                    & 0.82                    & 0.84                     \\
                             & BDIndicator    & 0.88                    & 0.92                    & 0.94                    & 0.96                    & 0.86                     \\
                             & FreqFed      & 0.80                    & 0.85                    & 0.87                    & 0.89                    & 0.89                     \\
                             & FedREDefense & 0.89                    & 0.83                    & 0.91                    & 0.78                    & 0.86                     \\
                             & \cellcolor{greyL}\algFirst    & \cellcolor{greyL}0.91                    & \cellcolor{greyL}0.93                    & \cellcolor{greyL}0.90                    & \cellcolor{greyL}0.95                    & \cellcolor{greyL}0.95                     \\
                             & \cellcolor{greyL}\algSecond    & \cellcolor{greyL}0.93                    & \cellcolor{greyL}0.94                    & \cellcolor{greyL}1.00                    & \cellcolor{greyL}1.00                    & \cellcolor{greyL}1.00                     \\ \hline
\multirow{9}{*}{{\makecell {DBA\\ attack}}}         & FLAME        & 0.75                    & 0.79                    & 0.82                    & 0.84                    & 0.83                     \\
                             & FLDetector   & 0.82                    & 0.85                    & 0.89                    & 0.90                    & 0.92                     \\
                             & FLTrust      & 0.82                    & 0.83                    & 0.80                    & 0.81                    & 0.82                     \\
                             & DeepSight    & 0.84                    & 0.85                    & 0.88                    & 0.87                    & 0.94                     \\
                             & BDIndicator    & 0.95                    & 0.97                    & 1.00                    & 0.97                    & 0.98                     \\
                             & FreqFed      & 0.86                    & 0.88                    & 0.89                    & 0.89                    & 0.91                     \\
                             & FedREDefense & 0.72                    & 0.77                    & 0.75                    & 0.76                    & 0.79                     \\
                             & \cellcolor{greyL}\algFirst    & \cellcolor{greyL}0.91                    & \cellcolor{greyL}0.91                    & \cellcolor{greyL}0.94                    & \cellcolor{greyL}0.97                    & \cellcolor{greyL}0.99                     \\
                             & \cellcolor{greyL}\algSecond    & \cellcolor{greyL}0.93                    & \cellcolor{greyL}0.95                    & \cellcolor{greyL}1.00                    & \cellcolor{greyL}1.00                    & \cellcolor{greyL}0.98                     \\ \hline
\end{tabular}
\label{non_iid_impact}
}
\vspace{-2pt}
\end{table*}

\myparatight{\alg is effective}
In Table \ref{tab:detection}, we present the detection performance of our \alg and other detection-based approaches. ``No attack'' means all clients are benign (there are no malicious clients in the system). ``NA'' means not applicable.
We observe that our proposed \alg method demonstrates remarkable detection efficacy. 
First, when all clients are benign, our proposed \alg ensures maximum preservation of benign clients and prevents their unintended exclusion. This demonstrates that \alg achieves the objective of ``Robust learning integrity''.
For instance, on the CIFAR-10 dataset, under no attack, \algSecond achieves a perfect DACC of 1.00, indicating its capability to maintain high detection accuracy without attack. 
\algFirst also can recognize the most benign clients. 
However, other detection-based approaches like FLAME and FLDetector can recognize only half of the benign clients. 

Second, in the presence of malicious clients, our proposed \alg effectively detects the majority of them while minimizing false detections of benign clients.
For instance, under the hybrid attack strategy, the Trim+DBA attack poses a significant challenge, yet \algSecond achieves a robust DACC of 0.96 on CIFAR-10, showcasing its resilience. In contrast, other baselines struggle significantly under this attack. On CIFAR-10, their DACC scores are limited to 0.75, 0.78, and 0.76, with corresponding FPR of 0.45, 0.39, and 0.27 for DeepSight, BackdoorIndicator, and FedREDefense, respectively, indicating a tendency to misclassify many benign clients as malicious.
Similarly, the Scaling+DBA attack continues to mislead FLDetector, FLTrust, and FreqFed, resulting in DACC values no larger than 0.87 on both CIFAR-10 and MNIST datasets. 
In contrast, \algFirst demonstrates robust performance, achieving DACC values of 0.95 and 0.97 on CIFAR-10 and MNIST, respectively, closely approaching the DACC values of 0.98 and 1.00 achieved by \algSecond.
In addition, other baseline methods fail to detect malicious clients effectively when the FL training process is applied to STL-10 and Tiny-ImageNet under the Scaling+DBA attack.

Table~\ref{tab:tacc} presents the TACC and ASR of the final global model obtained using various detection methods. 
It is important to note that the ASR metric is relevant only for targeted attacks, including the Scaling attack, DBA attack, Trim+DBA attack, and Scaling+DBA attack.
From the table, we observe that \algSecond effectively defends against diverse attack types, achieving high DACC while maintaining elevated TACC and low ASR across multiple datasets. 
In contrast, other prevention-based methods fail to sustain high task accuracy. For example, on the MNIST dataset, the TACC of DeepSight under no attack is 0.98 but drops significantly to 0.84 under the Adaptive attack. This indicates that the global model trained using DeepSight lacks accuracy. Conversely, global models trained with \alg under different attack scenarios remain almost as accurate as those trained in the absence of attacks. For example, on the CIFAR-10 dataset, the TACC of \algFirst remains consistently at 0.82 under no attack, and it can be maintained at 0.80 under the strong Trim attack.
Table~\ref{tab:prevention} highlights the TACC and ASR of the final global models produced by various prevention-based defense mechanisms. These defenses are generally ineffective in mitigating the impact of malicious clients. For instance, with the TrMean method on CIFAR-10, the TACC drops from 0.70 in the absence of attacks to 0.23 under the Trim attack, rendering the resulting global model highly inaccurate.
In summary, compared to other detection-based and prevention-based methods, the final global model trained using our \alg demonstrates superior accuracy, thereby achieving the ``Effective Threat Mitigation'' objective.

\textcolor{black}{
Figures~\ref{exp:safefl-ml-cifar}–\ref{exp:safefl-ml-Tiny} in Appendix show the loss values of local models for benign and malicious clients, evaluated on the synthetic dataset using \algFirst, under six attacks across five datasets at the 750th training round. Similarly, Figures~\ref{exp:safefl-cl-cifar}–\ref{exp:safefl-cl-STL10} in Appendix present the corresponding results for \algSecond under the same settings.
Note that, by default, our experimental setup assumes that the first 30\% of clients are malicious.
As shown in Figures~\ref{exp:safefl-ml-cifar}–\ref{exp:safefl-cl-STL10},
}
the loss values of malicious local models are significantly higher than those of benign models. This observation reinforces the motivation behind our method: malicious local models often exhibit distinct loss patterns compared to benign ones, making their detection feasible.

\myparatight{Impact of the fraction of malicious clients} Table~\ref{fraction_mali} presents the DACC of various detection-based methods as the fraction of malicious clients ranges from 0\% to 40\%, considering the Trim attack, Scaling attack, DBA attack, and the CIFAR-10 dataset. 
The results for Trim+DBA attack, Scaling+DBA attack, and Adaptive attack are shown in Table~\ref{fraction_mali_app} in Appendix.
Note that for all the ablation study experiments, unless otherwise specified, only the DACC values are reported.
As illustrated in Table~\ref{fraction_mali} and Table~\ref{fraction_mali_app}, FLAME exhibits substantial variations in response to changes in the fraction of malicious clients. On the other hand, as the proportion of malicious clients increases from 0\% to 40\%, the DACC of \algFirst remains consistently at least 0.90 under both Trim and Trim+DBA. Similarly, \algSecond consistently achieves a DACC of at least 0.95 across all attack scenarios, demonstrating superior stability and performance compared to the other methods.

\myparatight{Impact of the total number of clients} Table \ref{total_number_clients} shows the DACC under Trim attack, Scaling attack, DBA attack for different detection-based methods with different total numbers of clients. 
The results of the Trim+DBA attack, Scaling+DBA attack, and Adaptive attack are presented in Table~\ref{total_number_clients_app} in the Appendix.
The fraction of malicious clients is still set to 30\% by default. As the number of total clients increases, all method performance generally remains stable. 
For our \algSecond under Trim attack, which achieves its highest DACC at 60 and 100 total clients. Additionally, \algSecond maintains a significant advantage, with almost all of its DACC values outperforming those of the other methods. 
However, FLAME and FLTrust consistently exhibit a DACC no larger than 0.86 under the Trim attack, regardless of the total number of clients.

\myparatight{Impact of degree of Non-IID} Table~\ref{non_iid_impact} presents the DACC results for different detection-based defense methods across Non-IID levels ranging from 0.1 to 0.9, consider the Trim attack, Scaling attack, DBA attack. 
Table~\ref{non_iid_impact_app} in the Appendix presents the results for the Trim+DBA attack, Scaling+DBA attack, and Adaptive attack.
According to Table~\ref{non_iid_impact} and Table~\ref{non_iid_impact_app}, FLDetector is significantly affected by the degree of Non-IID.
The reason is that when the clients' training data are highly heterogeneous, the server in FLDetector faces difficulty in predicting clients' local models using their historical information.
Regardless of the degree of Non-IID, \algSecond consistently maintains the highest DACC among all methods. Under the Trim attack, when the Non-IID value is 0.1, \algSecond outperforms FLAME by 0.24.

\myparatight{Impact of the selection rate} 
\textcolor{black}{
Under our default configuration, all clients are assumed to participate in every training round. In this section, we explore a more practical scenario in which the server randomly selects only a fraction of clients to engage in each round. In this setting, a malicious client can carry out an attack only if it is selected. Table~\ref{exp_selection_rate} in Appendix presents the detection performance of various methods on the CIFAR-10 dataset under different client selection rates. As shown, our proposed detection method remains effective at identifying malicious clients even when only a subset of clients participate in each round.
}

\begin{figure*}[!t]
  \centering
    \begin{subfigure}{0.163\textwidth}
    \includegraphics[width=\textwidth]{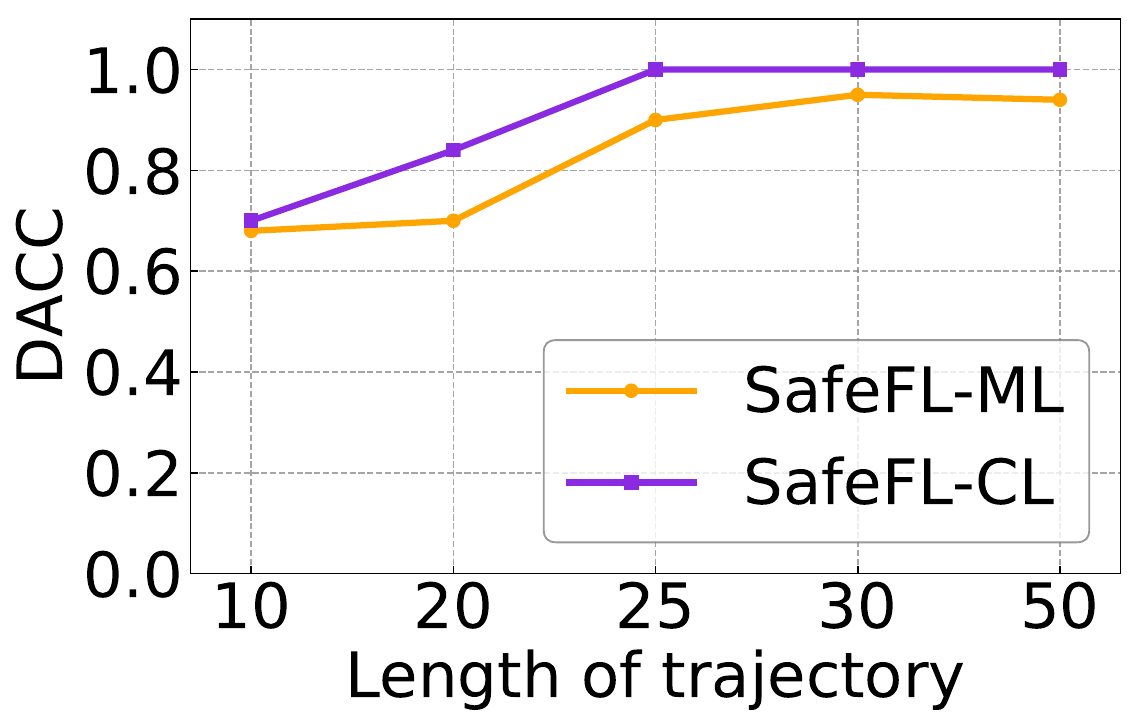}
    \caption{Trim attack}
  \end{subfigure}
    \begin{subfigure}{0.163\textwidth}
    \includegraphics[width=\textwidth]{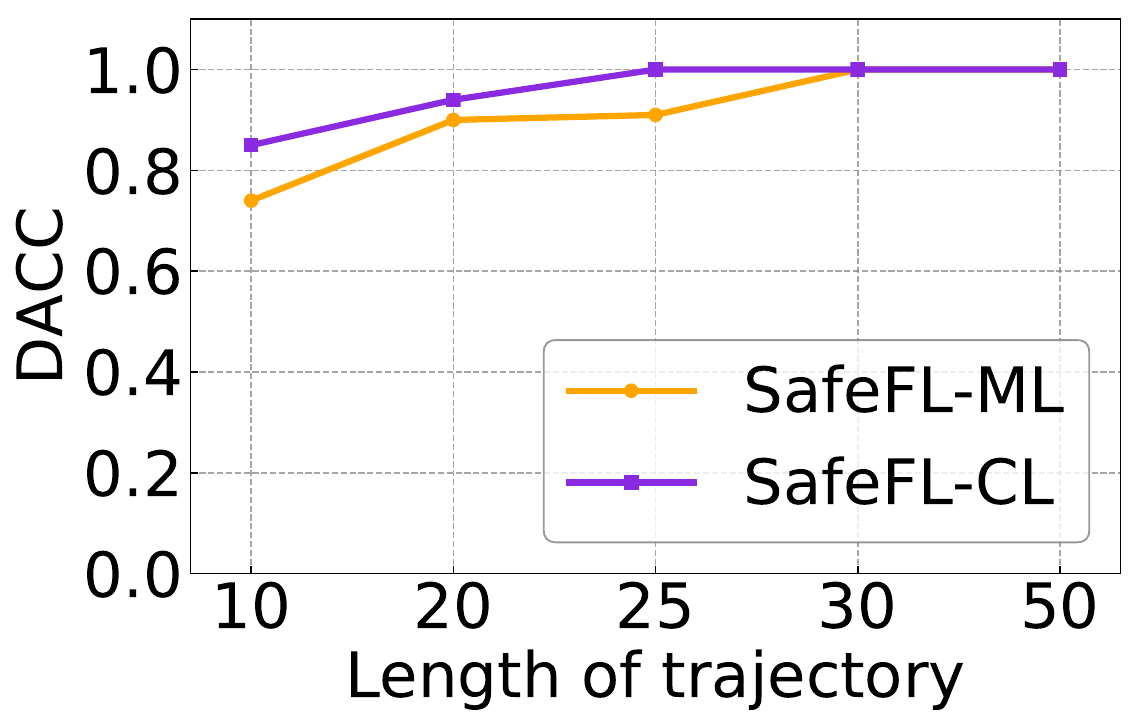}
    \caption{Scaling attack}
  \end{subfigure}
  \begin{subfigure}{0.163\textwidth}
    \includegraphics[width=\textwidth]{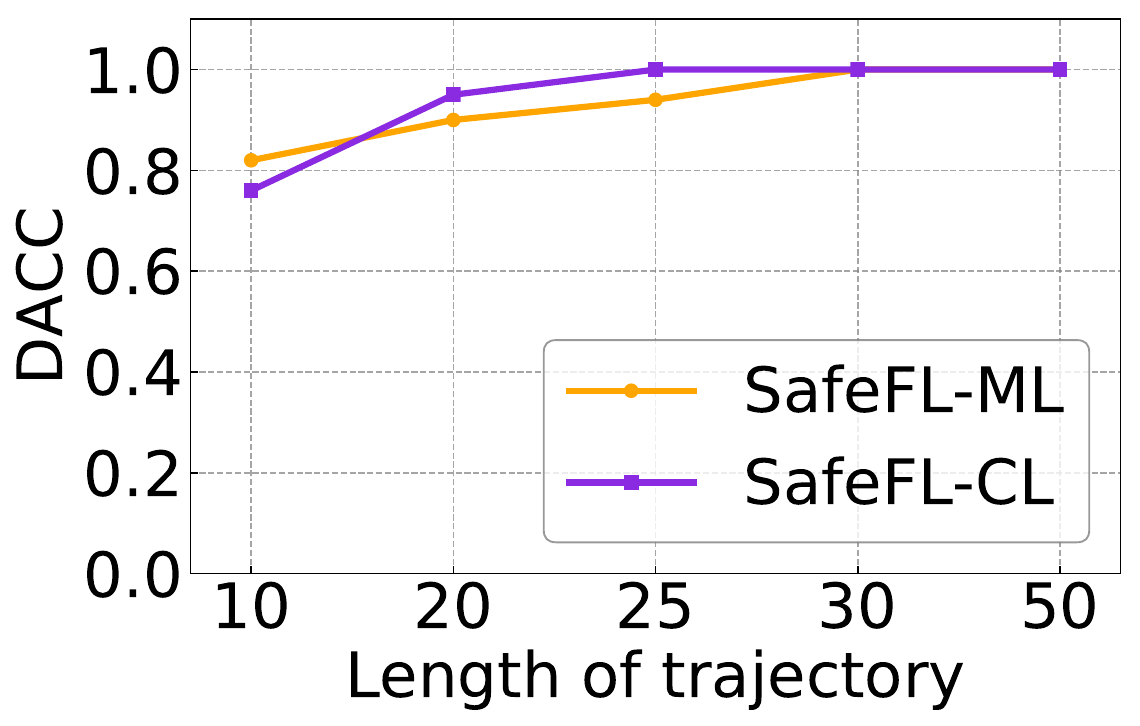}
    \caption{DBA attack}
  \end{subfigure}
  \begin{subfigure}{0.163\textwidth}
    \includegraphics[width=\textwidth]{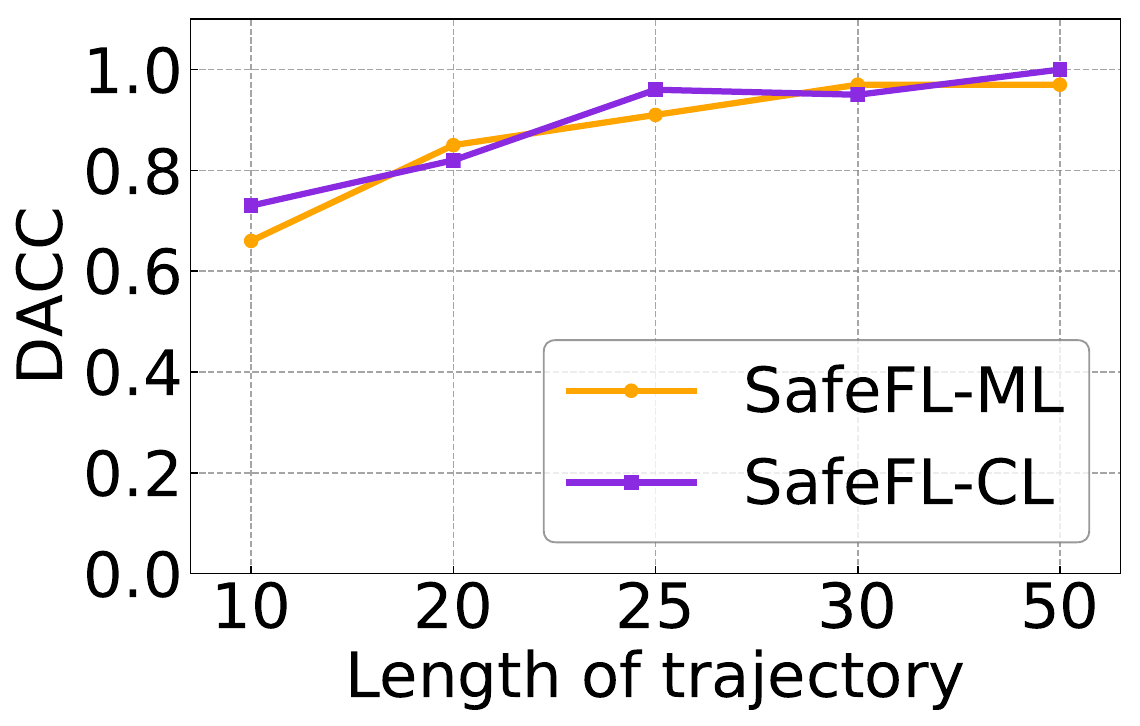}
    \caption{Trim+DBA attack}
  \end{subfigure}
    \begin{subfigure}{0.163\textwidth}
    \includegraphics[width=\textwidth]{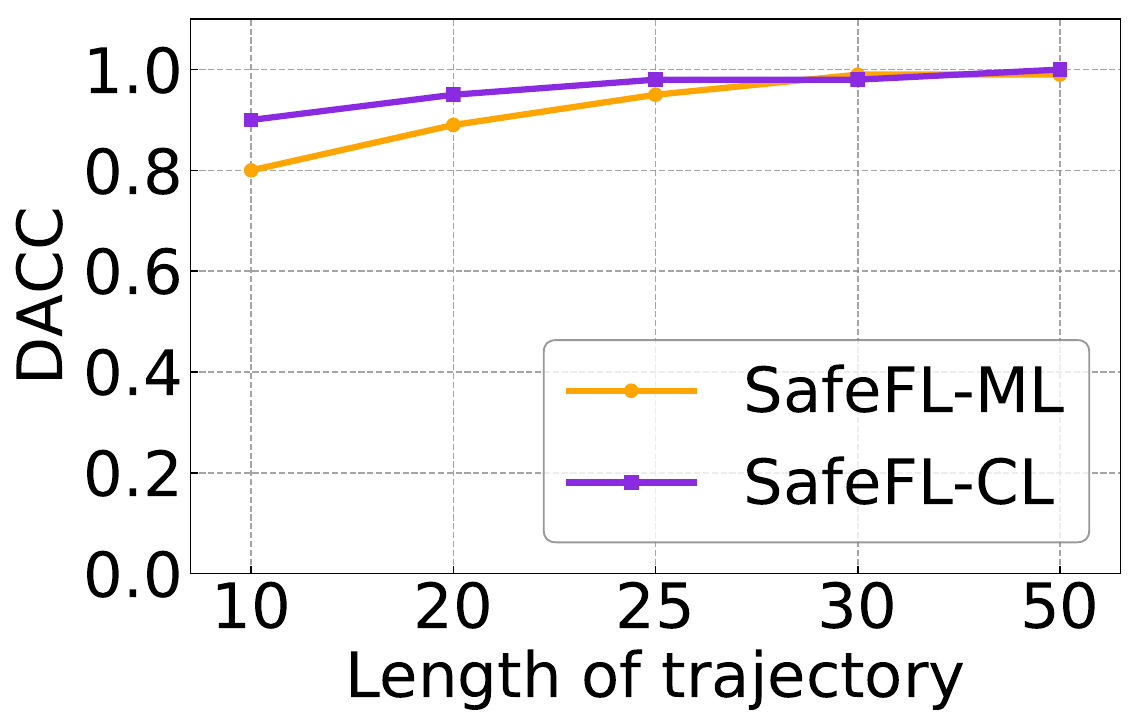}
    \caption{Scaling+DBA attack}
  \end{subfigure}
    \begin{subfigure}{0.163\textwidth}
    \includegraphics[width=\textwidth]{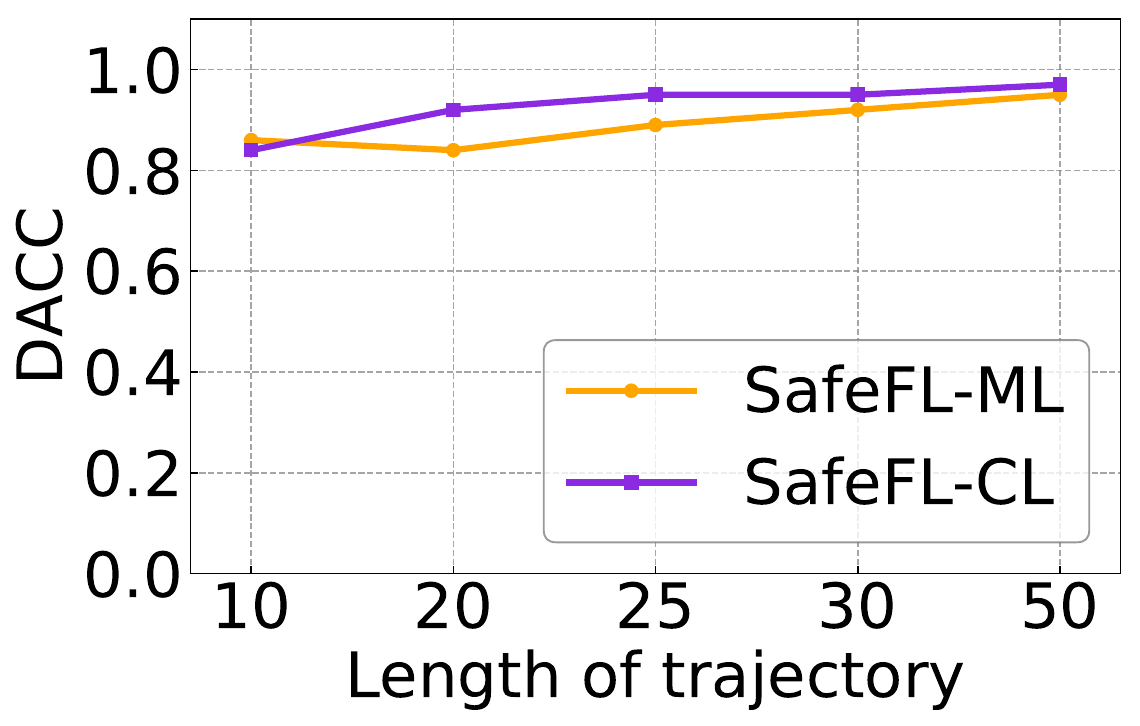}
    \caption{Adaptive attack}
  \end{subfigure}
  \caption{Impact of length of trajectory, where CIFAR-10 dataset is considered.}
  \label{iteration_full}
  \vspace{-8pt}
\end{figure*}

\myparatight{Impact of trajectory length} Figure~\ref{iteration_full} illustrates the impact of trajectory length, defined as the number of global models used to generate synthetic data or the value of $\epsilon$, on the detection of malicious clients in \alg.
Figure~\ref{iteration_full} reveals a positive correlation between trajectory length and DACC for both \algFirst and \algSecond. A longer trajectory length consistently enhances performance and improves DACC for \alg. However, when the trajectory length reaches 25, the improvement in DACC begins to plateau.

\begin{figure*}[t!]
  \centering
     \begin{subfigure}{0.163\textwidth}
    \includegraphics[width=\textwidth]{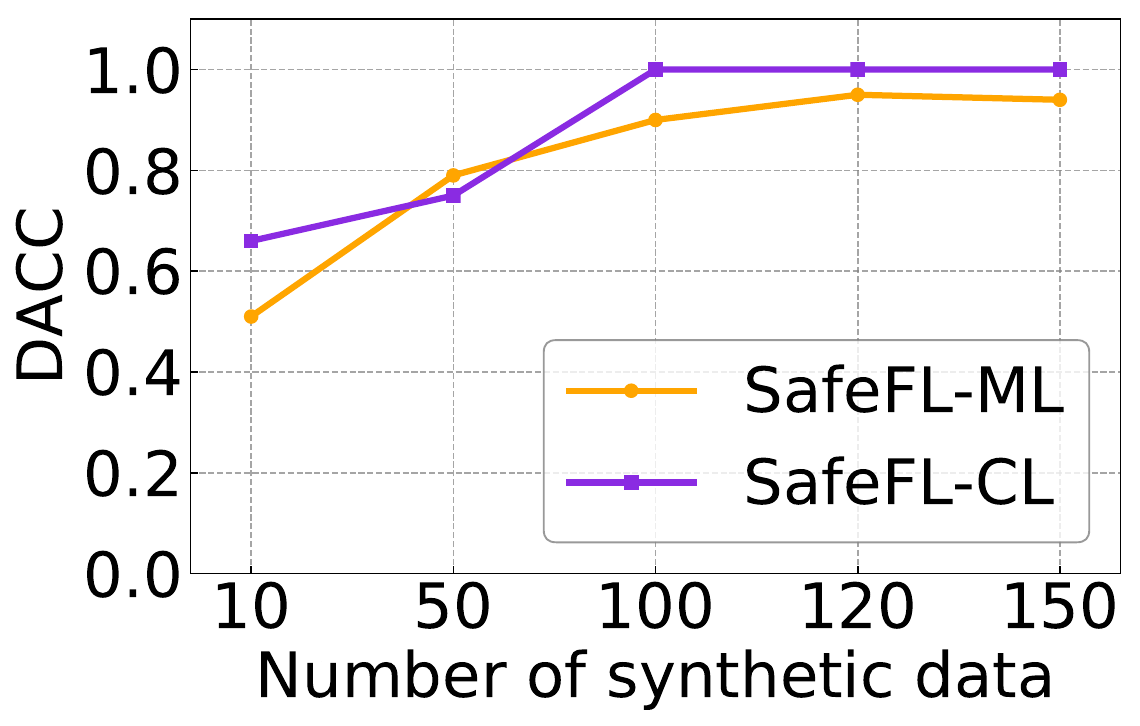}
    \caption{Trim attack}
  \end{subfigure}
    \begin{subfigure}{0.163\textwidth}
    \includegraphics[width=\textwidth]{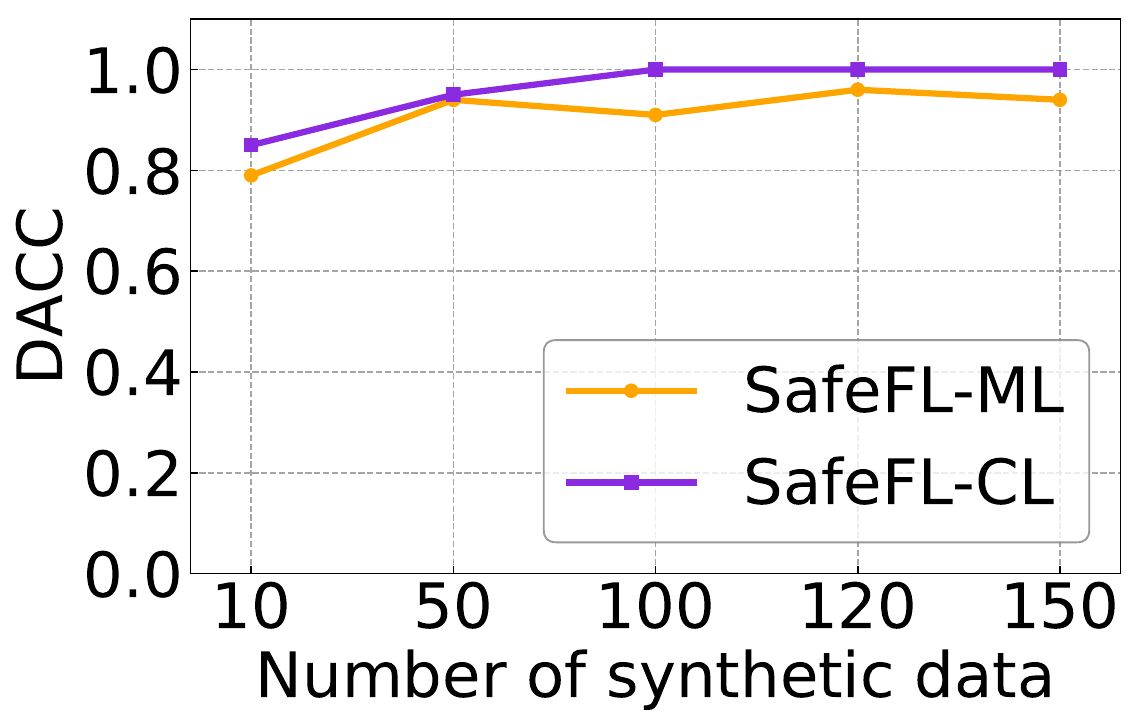}
    \caption{Scaling attack}
  \end{subfigure}
  \begin{subfigure}{0.163\textwidth}
    \includegraphics[width=\textwidth]{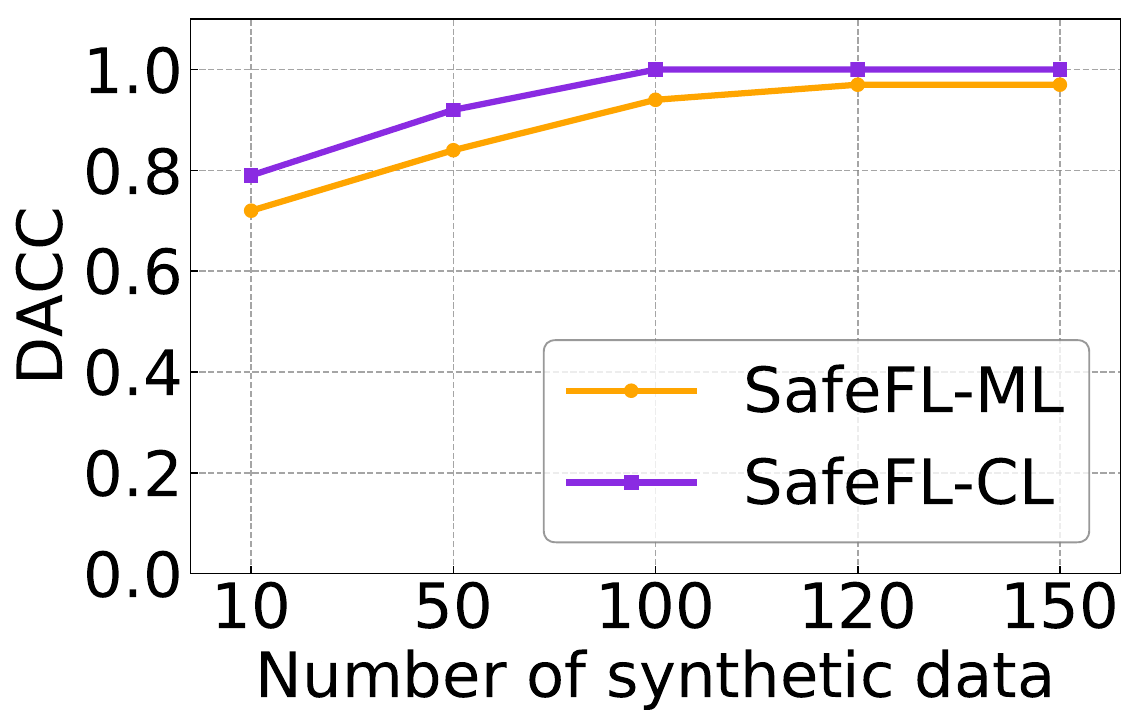}
    \caption{DBA attack}
  \end{subfigure}
  \begin{subfigure}{0.163\textwidth}
    \includegraphics[width=\textwidth]{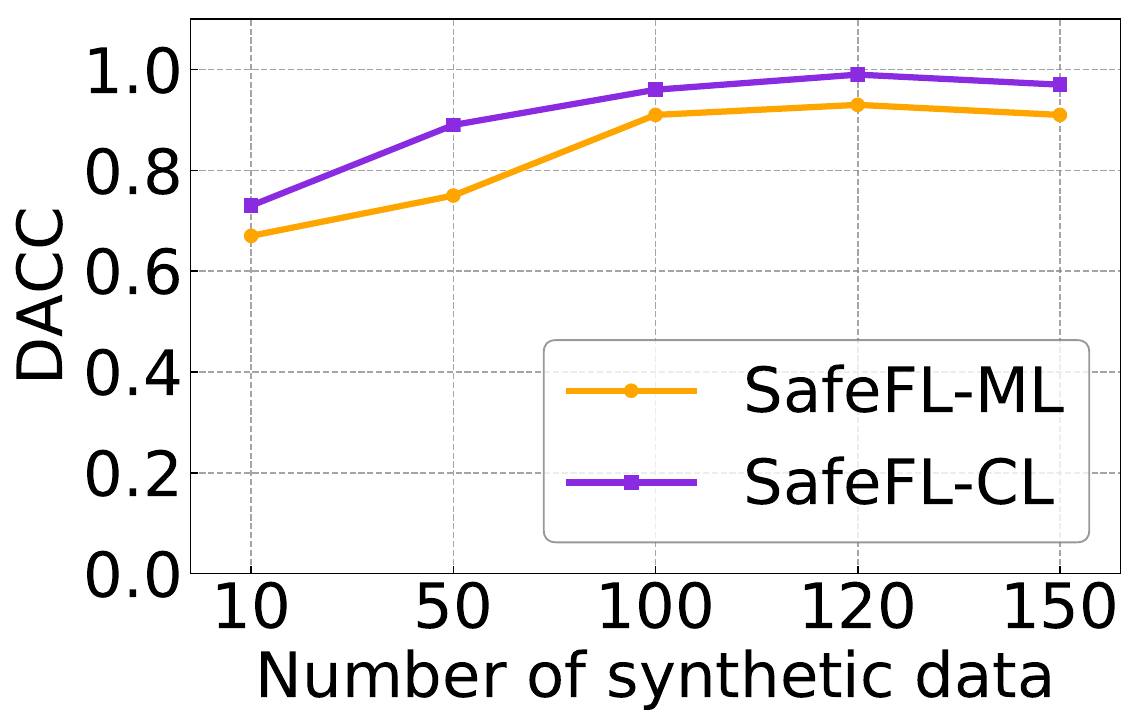}
    \caption{Trim+DBA attack}
  \end{subfigure}
   \begin{subfigure}{0.163\textwidth}
    \includegraphics[width=\textwidth]{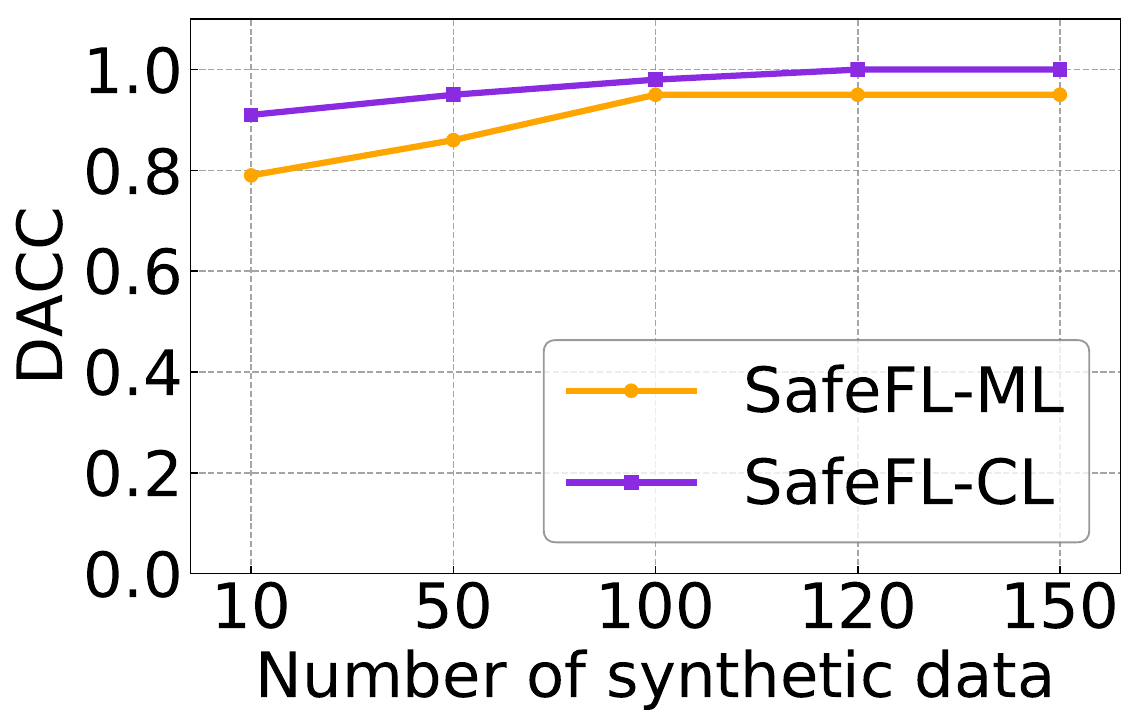}
    \caption{Scaling+DBA attack}
  \end{subfigure}
  \begin{subfigure}{0.163\textwidth}
    \includegraphics[width=\textwidth]{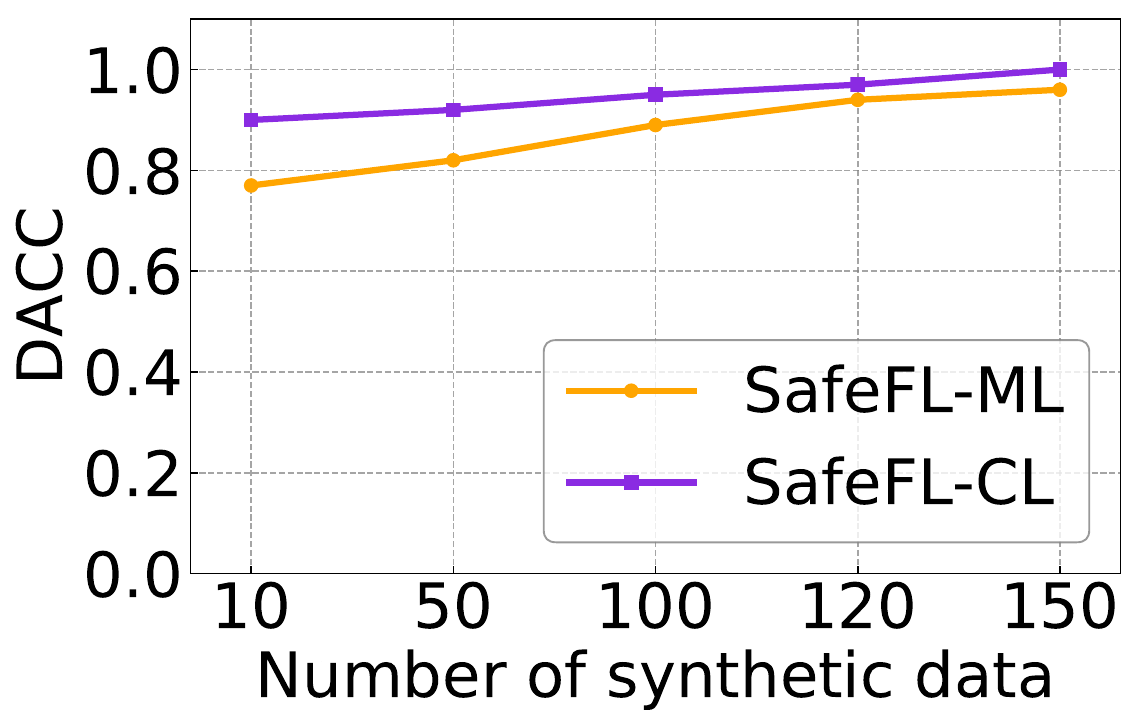}
    \caption{Adaptive attack}
  \end{subfigure}
  \caption{Impact of the number of synthetic data, where CIFAR-10 dataset is considered.}
  \label{server_full_gen}
  \vspace{-8pt}
\end{figure*}

\myparatight{Impact of number of synthetic data} Figure~\ref{server_full_gen} depicts the effect of the number of synthetic data on \alg's detection performance. The number of synthetic data refers to the total number of examples in the synthetic dataset.
Similar to the effect of trajectory length, there is a positive correlation between the amount of synthetic data and DACC. Notably, a larger volume of synthetic data more significantly enhances \algSecond's DACC. For instance, under the Trim+DBA attack, increasing the synthetic data from 10 to 150 raises \algSecond's DACC from 0.72 to 0.98, whereas \algFirst's DACC increases from 0.63 to 0.92.

\myparatight{Different variants of \algSecond} \algSecond performs clustering twice, as indicated in Line~\ref{cluester_one} and Line~\ref{cluester_two} of Algorithm~\ref{our_alg}.
We refer to these two clustering instances as A \& B, where A corresponds to the clustering algorithm used in Line~\ref{cluester_one}, and B pertains to the clustering algorithm applied to the losses (see Line~\ref{cluester_two}). Table~\ref{clus} examines the performance of various variants of our \algSecond.
In the different variants, we apply various clustering algorithms, such as K-means~\cite{hartigan1979algorithm}, Mean-shift~\cite{cheng1995mean} or DBSCAN~\cite{bozdemir2021privacy}, to both A and B.
As shown in Table~\ref{clus}, the ``Mean-shift \& Mean-shift'' variant exhibits poor detection performance against the Trim attack, whereas the ``Kmeans \& Mean-shift'' variant (which corresponds to our proposed \algSecond) achieves the best detection results.


\section{Discussion and Limitations} 
\label{sec:discussion_limitation}

\begin{table*}[htbp]
  \centering
   \footnotesize
\caption{Different variants of \algSecond on CIFAR-10, with DACC values reported.}
    \begin{tabular}{|c|c|c|c|c|}
\hline
          & K-means \& K-means   & Mean-shift \& Mean-shift & K-means \& DBSCAN       & \cellcolor{greyL} K-means \& Mean-shift (\algSecond) \\
\hline
    Trim attack     &   0.98    &   0.79    &  0.70     & \cellcolor{greyL} 1.00 \\
\hline
    Scaling attack     &  0.79     &  0.83     & 0.87      & \cellcolor{greyL} 1.00 \\
\hline
    DBA attack     &   0.94    &   0.72    &   0.75    &\cellcolor{greyL} 1.00 \\
\hline
    Trim+DBA attack     & 0.92      & 0.75      &  0.72     & \cellcolor{greyL} 0.96\\
\hline
    Scaling+DBA attack     &   0.83    & 0.75      & 0.80      &\cellcolor{greyL} 0.98 \\
\hline
    Adaptive attack     &  0.92     & 0.79      &   0.72    &\cellcolor{greyL} 0.95 \\
\hline
    \end{tabular}%
\label{clus}
   \vspace{-10pt}
\end{table*}%

\begin{table*}[]
\footnotesize
\caption{Detection results on CIFAR-10, with each client having three classes of training data and DACC values reported.}
\centering
\footnotesize
\begin{tabular}{|c|c|c|c|c|c|c|c|c|c|}
\hline
Attack      & FLAME & FLDetector & FLTrust & DeepSight & BackdoorIndicator & FreqFed & FedREDefense & \cellcolor{greyL}\algFirst & \cellcolor{greyL}\algSecond \\ \hline
Trim attack & 0.75  & 0.92       & 0.85    & 0.78      & 0.65        & 0.84    & 0.82         & \cellcolor{greyL}0.90      & \cellcolor{greyL}0.97      \\ \hline
Scaling attack     & 0.80  & 0.84       & 0.82    & 0.87      & 0.89        & 0.87    & 0.78         & \cellcolor{greyL}0.88                       & \cellcolor{greyL}0.96                        \\ \hline
DBA attack  & 0.81  & 0.72       & 0.79    & 0.85      & 0.74        & 0.72    & 0.84         & \cellcolor{greyL}0.93      & \cellcolor{greyL}0.99      \\ \hline
Trim+DBA attack   & 0.79  & 0.85       & 0.73    & 0.79      & 0.72        & 0.79    & 0.81         & \cellcolor{greyL}0.87      & \cellcolor{greyL}0.96      \\ \hline
Scaling+DBA attack & 0.80  & 0.78       & 0.72    & 0.82      & 0.79        & 0.80    & 0.81         & \cellcolor{greyL}0.91                       & \cellcolor{greyL}0.99                        \\ \hline
Adaptive attack    & 0.67  & 0.74       & 0.70    & 0.77      & 0.72        & 0.77    & 0.80         & \cellcolor{greyL}0.89                       & \cellcolor{greyL}0.92                        \\ \hline
\end{tabular}
\label{three_class}
   \vspace{-2pt}
\end{table*}

\begin{table}[]
\footnotesize
\caption{DACC of \alg with different aggregation rules on CIFAR-10.}
\begin{tabular}{|cll|cll|c|c|c|}
\hline
\multicolumn{3}{|c|}{Attack}                       & \multicolumn{3}{c|}{Defense}    & Median & TrMean & Krum \\ \hline
\multicolumn{3}{|c|}{\multirow{2}{*}{Trim attack}}        & \multicolumn{3}{c|}{\algFirst}  & 0.95   & 0.96   & 0.93 \\
\multicolumn{3}{|c|}{}                             & \multicolumn{3}{c|}{\algSecond} & 1.00   & 0.99   & 0.98 \\ \hline

\multicolumn{3}{|c|}{\multirow{2}{*}{Scaling attack}}        & \multicolumn{3}{c|}{\algFirst}  & 0.97   & 0.93   & 0.88 \\
\multicolumn{3}{|c|}{}                             & \multicolumn{3}{c|}{\algSecond} & 1.00   & 0.99   & 0.98 \\ \hline

\multicolumn{3}{|c|}{\multirow{2}{*}{DBA attack}}         & \multicolumn{3}{c|}{\algFirst}  & 0.92   & 0.92   & 0.95 \\
\multicolumn{3}{|c|}{}                             & \multicolumn{3}{c|}{\algSecond} & 0.94   & 1.00   & 1.00 \\ \hline
\multicolumn{3}{|c|}{\multirow{2}{*}{Trim+DBA attack}}    & \multicolumn{3}{c|}{\algFirst}  & 0.87   & 0.91   & 0.91 \\
\multicolumn{3}{|c|}{}                             & \multicolumn{3}{c|}{\algSecond} & 1.00   & 0.98   & 0.98 \\ \hline

\multicolumn{3}{|c|}{\multirow{2}{*}{Scaling+DBA attack}}        & \multicolumn{3}{c|}{\algFirst}   & 0.92   & 0.94   & 0.92 \\
\multicolumn{3}{|c|}{}                             & \multicolumn{3}{c|}{\algSecond} & 0.98   & 0.96   & 0.98 \\ \hline

\multicolumn{3}{|c|}{\multirow{2}{*}{Adaptive attack}}        & \multicolumn{3}{c|}{\algFirst}  & 0.88   & 0.87   & 0.85 \\
\multicolumn{3}{|c|}{}                             & \multicolumn{3}{c|}{\algSecond} & 0.91   & 0.90   & 0.94 \\ \hline

\end{tabular}
\label{dacc_our_prevent}
   \vspace{-10pt}
\end{table}

\myparatight{More extreme Non-IID distribution} 
This section examines a more extreme Non-IID scenario, as detailed in \cite{mcmahan2017communication}. The training data distribution among clients is purely label-based, with each client receiving data from only three specific classes. For example, Client A's training dataset consists solely of labels from 0 to 2, whereas Client B's dataset is restricted to labels from 3 to 5.
Under such condition, detection accuracy results of various methods are shown in Table~\ref{three_class}. Our method still significantly outperforms existing approaches, demonstrating the effectiveness of our \alg under extreme Non-IID setting.

\myparatight{More untargeted attacks} In this section, we use extra sophisticated untargeted attacks to further examine our method's detection ability. We implement the experiments on CIFAR-10.
Table~\ref{three_attacks} shows the detection results of different methods for the Label flipping attack~\cite{tolpegin2020data}, and ``A little is enough'' (LIE) attack~\cite{baruch2019little}.
Note that in the Label Flipping attack, the attacker alters the labels of training examples on malicious clients. 
In the LIE attack, the attacker strategically introduces small perturbations to the local models of malicious clients to avoid detection.
Table~\ref{three_attacks} in Appendix shows that our \alg, particularly \algSecond, achieves a perfect DACC of 1.00 against both advanced attacks. In contrast, BackdoorIndicator performs poorly under the Label flipping attack.

\myparatight{More backdoor attacks}
To further assess the robustness of our proposed detection methods, we evaluate them against three advanced backdoor attacks: Neurotoxin attack~\cite{zhang2022neurotoxin}, Irreversible backdoor attack~\cite{nguyen2023iba}, and Clean-label backdoor attack~\cite{zeng2023narcissus}. Table~\ref{exp:three_attack_detection} in Appendix reports the detection performance of our methods alongside all baseline detection approaches across five datasets, while Table~\ref{exp:three_attack_tacc} in Appendix presents the performance of the final global models trained using each detection method. As shown in both tables, baseline methods struggle to effectively identify and mitigate these sophisticated attacks. For example, FLTrust yields a high false positive rate (FPR) of 0.49 under the Irreversible backdoor attack on CIFAR-10 dataset. In contrast, our methods maintain strong detection capabilities and remain robust even in the presence of these advanced threats.

\myparatight{More evaluation metrics}
To provide a more comprehensive assessment of global model performance, we also incorporate three additional metrics: precision, recall, and F1-score. For these metrics, higher values indicate better detection effectiveness. Table~\ref{three_n_index} in Appendix presents the precision, recall, and F1-score of various detection methods under six standard attacks (Trim, Scaling, DBA, Trim+DBA, Scaling+DBA, and Adaptive attacks), while Table~\ref{exp:three_attack_detection_index} in Appendix shows the corresponding results for three advanced backdoor attacks (Neurotoxin, Irreversible backdoor, and Clean-label backdoor attacks). As shown in both tables, our proposed detection methods consistently achieve high precision, recall, and F1-scores across different attacks and datasets.

\myparatight{Computational overhead and storage usage of different methods}
In our methods, the server constructs a synthetic dataset based on the trajectory of global models accumulated over multiple training rounds. While this step enhances detection capabilities, it may introduce additional computational and storage overhead. Figure~\ref{time_comsume_fig} in Appendix presents the total running time across all datasets under the Trim attack, with similar patterns observed for other attack types. For our methods, the reported runtime includes the time required for synthetic data generation, clustering, and loss-based filtering. As shown, our methods incur only a modest increase in computational cost compared to FedAvg, whereas FedREDefense exhibits the highest computational overhead among all methods. Table~\ref{exp:memory} in Appendix summarizes the additional server-side storage requirements for FLDetector, FedREDefense, \algFirst, and \algSecond. Note that FedAvg and other baseline detection methods do not require any extra server storage. 
As shown in Table~\ref{exp:memory}, our methods require no more than 22.40 GB of additional storage, which is considered reasonable for high-capacity servers such as those commonly deployed in modern data centers.

\myparatight{\alg uses different aggregation rules}
By default, our proposed \alg utilizes the FedAvg rule to aggregate the detected benign local models. As observed in the experimental results in Section~\ref{results}, using the FedAvg rule is sufficient, as \alg effectively detects the majority of malicious clients. In this section, we examine the scenario where \alg employs prevention-based methods such as Median, TrMean, or Krum to aggregate the detected benign local models. The results, presented in Table~\ref{dacc_our_prevent}, indicate that \alg maintains strong detection performance even when prevention-based methods are used. For example, when \algSecond applies the Median method for aggregation and Trim attack is considered, it achieves a DACC of 1.00.

\myparatight{Discussion on the threat model for the ratio of malicious clients}
Table~\ref{fraction_mali} and Table~\ref{fraction_mali_app} demonstrate that our proposed \alg framework effectively detects malicious clients even when 40\% of the clients are compromised. Fractions higher than 40\%, such as 45\% or 50\%, were not included in our experiments, as such scenarios are considered impractical. As highlighted in~\cite{shejwalkar2022back}, achieving such a high proportion of malicious clients is unlikely in real-world FL environments. 
The decentralized nature of FL systems makes it difficult or even impossible for the attacker to control such a significant fraction of participating clients for malicious activities.

\myparatight{Discussion of biases introduced by the synthetic dataset}
In this section, we examine whether the synthetic dataset generated by our methods introduces bias into the global model. Ideally, the final model should perform equitably across groups defined by sensitive attributes, such as sex. To evaluate fairness, we use two standard metrics: Equalized odds~\cite{hardt2016equality}, which measures bias conditioned on the true label, and Demographic parity~\cite{dwork2012fairness}, which requires predictions to be independent of the sensitive attribute.
Our analysis is conducted on the CIFAR-10 dataset under two highly Non-IID scenarios. In Setting I, each client has access to samples from only three specific classes (note that further reducing this to one or two classes prevents convergence even without attacks). In Setting II, we simulate extreme data heterogeneity by setting the Non-IID degree to 0.1, as suggested in~\cite{2019Local}. Since CIFAR-10 lacks predefined sensitive attributes, we define a proxy sensitive attribute by grouping labels based on parity (odd vs. even).
Table~\ref{exp:bias} in Appendix presents the Equalized odds and Demographic parity scores of our methods under various attack conditions on CIFAR-10, with lower values indicating greater fairness. Detection performance under Setting I is reported in Table~\ref{three_class}, while results under Setting II are provided in Table~\ref{non_iid_impact} and Table~\ref{non_iid_impact_app}.
Table~\ref{exp:bias} reveals two key findings: (1) Highly Non-IID data induces bias in the global model even without attacks, consistent with prior work~\cite{chang2023bias,badar2024fairtrade} (e.g., FedAvg scores 0.79 in Equalized Odds under Setting I). (2) Our methods maintain similar fairness to FedAvg in the no-attack case, introducing no extra bias.

\myparatight{Potential challenges introduced by \alg}
In our proposed method, the server collects and stores multiple global models, which are then utilized to generate the synthetic dataset. This approach enhances the framework's ability to identify malicious clients effectively. However, it also introduces potential privacy concerns, as the storage and usage of multiple global models may inadvertently expose sensitive information about the clients’ data.
Although addressing privacy is not the primary focus of this paper, these concerns can be effectively alleviated by leveraging well-established privacy-preserving techniques. 
As an example, incorporating differential privacy~\cite{abadi2016deep} allows for the protection of individual client data while still supporting the generation of synthetic datasets for detection purposes. In particular, each client applies differential privacy by adding noise to its local model before transmitting it to the server. In our experiments, the noise is sampled from a Gaussian distribution $N(0, \varrho)$, where $\varrho$ denotes the noise level. Table~\ref{exp:noise_level} in the Appendix reports the impact of varying noise levels on the CIFAR-10 dataset, with results measured by DACC and TACC. As shown in the table, under a non-adversarial setting, excessive noise can negatively affect the global model’s testing accuracy (TACC), even when using the FedAvg aggregation rule. For instance, when the noise level is set to 2, the TACC of FedAvg drops to 0.70, compared to 0.85 in the absence of noise. This highlights a trade-off: while differential privacy strengthens client data protection, it can also compromise model performance. Despite the presence of noise, our detection methods preserve high global model accuracy. In particular, the TACC values achieved by our methods remain close to those of the noise-free FedAvg baseline, suggesting that our methods effectively balances privacy protection and model utility.


\section{Conclusion and Future Work} 
\label{sec:conclusion}

FL is vulnerable to poisoning attacks due to its decentralized nature. Existing detection methods often perform poorly. To address this, we propose \alg, a detection method where the server uses a synthetic dataset, generated from global model trajectories, to distinguish between benign and malicious clients. Experiments confirm its effectiveness.
A limitation of \alg is potential privacy concerns from the server's actions. Future work will focus on privacy-preserving detection. We also plan to extend \alg to decentralized FL~\cite{beltran2022decentralized,kalra2023decentralized}, where no trusted central server exists.

\begin{acks}
We thank the anonymous reviewers for their comments.
\end{acks}

\bibliographystyle{ACM-Reference-Format}
\bibliography{refs}


\appendix

\begin{table}[http]
\footnotesize
\centering
\caption{The CNN architecture.}
\begin{tabular}{|c|c|}
\hline
Layer&Size \\
\hline
Input&28 $\times$ 28 $\times$ 1 \\
\hline
Convolution $+$ ReLU& 3 $\times$ 3 $\times$ 30  \\
\hline
Max Pooling& 2 $\times$ 2 \\
\hline
Convolution $+$ ReLU& 3 $\times$ 3 $\times$ 5 \\
\hline
Max Pooling& 2 $\times$ 2 \\
\hline
Fully Connected $+$ ReLU& 100 \\
\hline
Soft& 10 (62 for FEMNIST) \\
\hline
\end{tabular}
\label{cnn}
\end{table}

\section{Details of Datasets, Poisoning Attacks, Compared Defenses} 
\label{sec:setting_app}

\subsection{Details of Datasets} 
\label{sec:setting_app_dataset}

\myparatight{a)~CIFAR-10~\cite{cifar10data}}
The CIFAR-10 dataset is a color image classification dataset containing 50,000 training examples and 10,000 testing examples, each categorized into one of ten classes.

\myparatight{b)~MNIST~\cite{mnist}} 
MNIST dataset contains 10 different classes and includes 60,000 examples for training and 10,000 for testing.

\myparatight{c)~FEMNIST~\cite{femnist}}
The FEMNIST dataset, derived from the extended MNIST dataset, is specifically designed for FL purposes. It contains a meticulously selected collection of handwritten character images, encompassing a diverse range of characters and digits, totaling 62 distinct classes. 
This dataset is inherently heterogeneous.

\myparatight{d)~STL-10~\cite{coates2011analysis}} 
The STL-10 dataset comprises 13,000 labeled images distributed among 10 object classes such as birds, cats, and trucks. Among these, 5,000 images are allocated for training, while the remaining 8,000 are reserved for testing.

\myparatight{e)~Tiny-ImageNet~\cite{deng2009imagenet}} Tiny-ImageNet is a subset of the ImageNet dataset, comprising 100,000 images distributed across 200 classes, with 500 images per class.

\subsection{Details of Poisoning Attacks} 
\label{sec:setting_app_attack}

\myparatight{a)~Trim attack~\cite{2019Local}} 
The Trim attack is an untargeted local model poisoning attack specifically designed to manipulate the Trimmed-mean and Median aggregation rules. We adopt the default parameter settings outlined in~\cite{2019Local} to execute the Trim attack.

\myparatight{b)~Scaling attack~\cite{2018How}} 
For this targeted attack, the attacker duplicates local training instances on malicious clients, adds a trigger, and assigns a chosen label. Local models are then computed using the augmented training data. Malicious clients amplify these local models before sending them to the server.

\myparatight{c)~Distributed backdoor (DBA) attack~\cite{xie2019dba}} 
DBA attack involves dividing the trigger pattern into four segments. These segments are then incorporated into the training data of four separate groups of malicious clients. Each client calculates its own local model and adjusts it using a scaling factor.

\myparatight{d)~Trim+DBA attack} 
This strategy involves a hybrid attack strategy where various malicious clients employ different methods to craft their local models. In our experiments, half of the malicious clients utilized the Trim attack to shape their local models, whereas the other half implemented the DBA strategy.

\myparatight{e)~Scaling+DBA attack} 
In this hybrid attack, half of the malicious clients used the Scaling attack to craft their local models, while the other half employed the DBA attack.

\myparatight{e)~Adaptive attack~\cite{shejwalkar2021manipulating}}
In the worst-case scenario, the attacker has complete information about the FL system, including the local models of all clients and the server's aggregation method, such as the \alg described in our work. Leveraging this knowledge, the attacker designs an adaptive attack to disrupt and deceive the FL process. In our experiments, we implement the Adaptive attack following the methodology outlined in ~\cite{shejwalkar2021manipulating}.

\subsection{Details of Compared Defenses} 
\label{sec:setting_app_defense}

\myparatight{a)~FLAME~\cite{2021FLAME}}FLAME is a defense strategy against targeted attacks, including backdoor attacks. It uses clustering to remove suspected malicious local models, truncates the remaining models to limit their influence, and adds random noise to the aggregated model to eliminate backdoors.

\myparatight{b)~FLDetector~\cite{zhang2022fldetector}}FLDetector identifies malicious clients by analyzing the consistency of their local models. It leverages the observation that benign clients follow the FL algorithm and their local data, while malicious clients deviate by crafting inconsistent models across training rounds.

\myparatight{c)~FLTrust~\cite{2021FLTrust}}FLTrust assumes the server has a clean validation dataset from the same distribution as the clients' training data. The server trains a server model on this dataset, and a client's local model is deemed benign if it aligns positively with the server model.

\myparatight{d)~DeepSight~\cite{rieger2022deepsight}}DeepSight analyzes model updates, examining output parameters and data homogeneity to detect backdoor attacks. Using classifiers and clustering, it distinguishes malicious updates from benign ones, even with diverse data distributions.

\myparatight{e)~BackdoorIndicator~\cite{li2024backdoorindicator}}This proactive FL backdoor detection method uses out-of-distribution (OOD) data to identify malicious updates. The server injects an OOD indicator task into the global model, and after client training, evaluates its accuracy, adjusting for batch normalization shifts. Updates exceeding a set accuracy threshold are flagged as suspicious and excluded from aggregation.

\myparatight{f)~FreqFed~\cite{fereidooni2024freqfed}}FreqFed mitigates targeted and untargeted poisoning attacks by analyzing model weights in the frequency domain, detecting and removing malicious updates while maintaining global model performance.

\myparatight{g)~FedREDefense~\cite{xie2024fedredefense}}FedREDefense detects malicious models by measuring the reconstruction error of each client's model. Using distilled local knowledge, it reconstructs models and compares the error to a threshold. Updates with errors exceeding the threshold are flagged as malicious.

\myparatight{h)~Median~\cite{yin2018byzantine}}For every dimension, the server calculates the median value for each coordinate from all clients' local models.

\myparatight{i)~Trimmed mean (TrMean)~\cite{yin2018byzantine}}Like the Median method, the server removes the largest and smallest $k$ values for each dimension, then averages the remaining $n$ values, where $k$ is the number of malicious clients, and $n$ is the total number of clients.

\myparatight{j)~Krum~\cite{blanchard2017machine}}Upon receiving local models from clients, the server outputs a single local model that minimizes the sum of distances to its neighboring subset.

\begin{table*}[H]
\caption{Flip attack on CIFAR-10}
\begin{tabular}{c|ccc|ccc|ccc|ccc}
\hline
\multirow{2}{*}{Defense} & \multicolumn{3}{c|}{FedAvg} & \multicolumn{3}{c|}{Krum} & \multicolumn{3}{c|}{Trimmed-mean} & \multicolumn{3}{c}{Median} \\ \cline{2-13} 
                          & DACC    & FPR     & FNR     & DACC    & FPR    & FNR    & DACC     & FPR      & FNR      & DACC    & FPR     & FNR    \\ \hline
FLAME                     & 0.84    & 0.02    & 0.22    & 0.87    & 0.03   & 0.17   & 0.84     & 0.06     & 0.20     & 0.87    & 0.03    & 0.17   \\
FLDector                  & 0.91    & 0.04    & 0.11    & 0.94    & 0.00   & 0.09   & 0.87     & 0.02     & 0.18     & 0.93    & 0.04    & 0.08   \\
\algFirst                 & 0.89    & 0.02    & 0.15    & 0.91    & 0.03   & 0.12   & 0.90     & 0.06     & 0.12     & 0.91    & 0.00    & 0.13   \\
\algSecond                      & 0.95    & 0.00    & 0.07    & 0.97    & 0.00   & 0.04   & 0.94     & 0.00     & 0.09     & 1.00    & 0.00    & 0.00   \\ \hline
\end{tabular}
\label{flips}
\end{table*}

\begin{table*}[htbp]
        \footnotesize
  \addtolength{\tabcolsep}{-1.625pt} 
\caption{The impact of the malicious client ratio, total client number, and Non-IID degree is analyzed using the CIFAR-10 dataset. DACC values are reported for the Trim+DBA attack, Scaling+DBA attack, and Adaptive attack. ``BDIndicator'' refers to the ``BackdoorIndicator'' method.}
\resizebox{\linewidth}{!}{
\subfloat[Impact of fraction of malicious clients.]
{\tabcolsep 0.1cm
\begin{tabular}{|c|c|ccccc|}
\hline
\multirow{2}{*}{Attack}          & \multirow{2}{*}{Defense} & \multicolumn{5}{c|}{Malicious client ratio} \\ \cline{3-7} 
                                 &                          & 0\%      & 10\%     & 20\%     & 30\%    & 40\%    \\ \hline
\multirow{9}{*}{{\makecell {Trim+DBA\\ attack}}}    & FLAME        & 0.57                    & 0.74                     & 0.77                     & 0.81                     & 0.87                      \\
                             & FLDetector   & 0.65                    & 0.82                     & 0.89                     & 0.87                     & 0.84                      \\
                             & FLTrust      & 0.80                    & 0.83                     & 0.91                     & 0.88                     & 0.85                      \\
                             & DeepSight    & 1.00                    & 0.72                     & 0.79                     & 0.75                     & 0.83                      \\
                             & BDIndicator    & 0.95                    & 0.75                     & 0.72                     & 0.78                     & 0.80                      \\
                             & FreqFed      & 0.81                    & 0.83                     & 0.72                     & 0.80                     & 0.76                      \\
\hline
\end{tabular}
\label{fraction_mali_app}
}
\subfloat[Impact of total number of clients.]
{
\begin{tabular}{|c|c|ccccc|}
\hline
\multirow{2}{*}{Attack}          & \multirow{2}{*}{Defense} & \multicolumn{5}{c|}{Total client number} \\ \cline{3-7} 
                                  &                          & 60     & 80     & 100    & 120   & 150   \\ \hline
\multirow{9}{*}{{\makecell {Trim+DBA\\ attack}}}    & FLAME        & 0.82                   & 0.85                   & 0.81                    & 0.83                    & 0.84                     \\
                             & FLDetector   & 0.84                   & 0.86                   & 0.87                    & 0.85                    & 0.82                     \\
                             & FLTrust      & 0.84                   & 0.85                   & 0.88                    & 0.81                    & 0.84                     \\
                             & DeepSight    & 0.77                   & 0.74                   & 0.75                    & 0.79                    & 0.83                     \\
                             & BDIndicator    & 0.72                   & 0.76                   & 0.78                    & 0.76                    & 0.75                     \\
                             & FreqFed      & 0.81                   & 0.84                   & 0.80                    & 0.80                    & 0.77                     \\
\hline
\end{tabular}
\label{total_number_clients_app}
}
\subfloat[Impact of degree of Non-IID.]
{
\begin{tabular}{|c|c|ccccc|}
\hline
\multirow{2}{*}{Attack}          & \multirow{2}{*}{Defense} & \multicolumn{5}{c|}{Non-IID degree} \\ \cline{3-7} 
                                 &                          & 0.1      & 0.3      & 0.5      & 0.7     & 0.9     \\ \hline
\multirow{9}{*}{{\makecell {Trim+DBA\\ attack}}}    & FLAME        & 0.74                    & 0.77                    & 0.81                    & 0.83                    & 0.83                     \\
                             & FLDetector   & 0.82                    & 0.77                    & 0.87                    & 0.91                    & 0.93                     \\
                             & FLTrust      & 0.84                    & 0.86                    & 0.88                    & 0.85                    & 0.85                     \\
                             & DeepSight    & 0.70                    & 0.72                    & 0.75                    & 0.75                    & 0.77                     \\
                             & BDIndicator    & 0.79                    & 0.80                    & 0.78                    & 0.82                    & 0.83                     \\
                             & FreqFed      & 0.67                    & 0.74                    & 0.80                    & 0.81                    & 0.82                     \\
\hline
\end{tabular}
\label{non_iid_impact_app}
}}
\label{para_impact_app}
\end{table*}

\begin{table*}[]
\footnotesize
\caption{\textcolor{black}{The impact of the selection rate is analyzed using the CIFAR-10 dataset, where DACC values are reported.}}
\subfloat[Results under Trim, Scaling, and DBA attacks.]
{
\begin{tabular}{|c|c|llllc|}
\hline
\multirow{2}{*}{Attack}          & \multirow{2}{*}{Defense} & \multicolumn{5}{c|}{Selection rate} \\ \cline{3-7} 
                                 &                          & 20\%     & 30\%    & 50\%      & 80\%   & 100\%  \\ \hline
                                     & FLAME        &  0.75 & 0.76 & 0.77 & 0.76 & 0.78 \\
                                     & FLDetector   & 0.75 & 0.77 & 0.76 & 0.75 & 0.96 \\
                                     & FLTrust      & 0.93 & 0.96 & 0.95 & 0.94 & 0.85 \\
                                     & DeepSight    & 0.84 & 0.87 & 0.86 & 0.86 & 0.88 \\
                                     & BackdoorIndicator  & 0.85 & 0.87 & 0.89 & 0.87 & 0.73 \\
                                     & FreqFed      & 0.77 & 0.82 & 0.80 & 0.74 & 0.89 \\
\hline
\end{tabular}
}
\quad
\subfloat[Results under Trim+DBA, Scaling+DBA, and Adaptive attacks.]
{
\begin{tabular}{|c|c|llllc|}
\hline
\multirow{2}{*}{Attack}          & \multirow{2}{*}{Defense} & \multicolumn{5}{c|}{Selection rate} \\ \cline{3-7} 
                                 &                          & 20\%     & 30\%    & 50\%      & 80\%   & 100\%  \\ \hline
                                     & FLAME        & 0.84 & 0.87 & 0.83 & 0.85 & 0.81 \\
                                     & FLDetector   & 0.84 & 0.86 & 0.85 & 0.86 & 0.87 \\
                                     & FLTrust      & 0.90 & 0.87 & 0.91 & 0.85 & 0.88 \\
                                     & DeepSight    & 0.79 & 0.74 & 0.78 & 0.81 & 0.75 \\
                                     & BackdoorIndicator  & 0.82 & 0.78 & 0.79 & 0.81 & 0.78 \\
                                     & FreqFed      & 0.78 & 0.83 & 0.80 & 0.82 & 0.80 \\
\hline
\end{tabular}
}
\label{exp_selection_rate}
\end{table*}

\begin{table*}[t]
\footnotesize
\caption{Detection results of various methods under advanced untargeted attacks on CIFAR-10, with DACC values reported.}
\footnotesize
\centering
\begin{tabular}{|c|c|c|c|c|c|c|c|c|c|}
\hline
Attack                & FLAME & FLDetector & FLTrust & DeepSight     & BackdoorIndicator & FreqFed & FedREDefense & \cellcolor{greyL}\algFirst & \cellcolor{greyL}\algSecond     \\ \hline
Label flipping attack & 0.89  & 0.91       & 0.82    & 1.00 & 0.76      & 0.79    & 0.85         & \cellcolor{greyL}0.96      & \cellcolor{greyL}1.00 \\ \hline
LIE attack            & 0.87  & 0.73       & 0.79    & 0.94          & 0.91      & 0.85    & 0.78         & \cellcolor{greyL}0.95      & \cellcolor{greyL}1.00 \\ \hline
\end{tabular}
\label{three_attacks}
\end{table*}

\begin{table*}[]
\centering
  \footnotesize
\caption{\textcolor{black}{Detection performance of various detection-based methods under Neurotoxin, Irreversible backdoor, and Clean-label backdoor attacks, evaluated using DACC (\(\uparrow\)), FPR (\(\downarrow\)), and FNR (\(\downarrow\)) metrics. Here, \(\uparrow\) denotes better detection performance with higher values, and \(\downarrow\) denotes better performance with lower values.}}
\begin{tabular}{|c|c|ccc|ccc|ccc|ccc|ccc|}
\hline
                                  &                                      & \multicolumn{3}{c|}{\textbf{CIFAR-10}} & \multicolumn{3}{c|}{\textbf{MNIST}} & \multicolumn{3}{c|}{\textbf{FEMNIST}} & \multicolumn{3}{c|}{\textbf{STL-10}} & \multicolumn{3}{c|}{\textbf{Tiny-ImageNet}} \\ \cline{3-17} 
\multirow{-2}{*}{\textbf{Attack}} & \multirow{-2}{*}{\textbf{Defense}} & DACC & FPR & FNR & DACC & FPR & FNR & DACC & FPR & FNR & DACC & FPR & FNR & DACC & FPR & FNR \\ \hline
\multirow{9}{*}{{\makecell {Neurotoxin\\ attack}}}
                                  & FLAME                                & 0.84 & 0.27 & 0.11 & 0.88 & 0.25 & 0.06 & 0.86 & 0.17 & 0.13 & 0.88 & 0.20 & 0.09 & 0.86 & 0.15 & 0.14 \\
                                  & FLDetector                           & 0.83 & 0.30 & 0.11 & 0.78 & 0.45 & 0.12 & 0.88 & 0.25 & 0.06 & 0.91 & 0.18 & 0.05 & 0.79 & 0.22 & 0.21 \\
                                  & FLTrust                              & 0.72 & 0.41 & 0.22 & 0.74 & 0.35 & 0.22 & 0.77 & 0.29 & 0.20 & 0.82 & 0.23 & 0.16 & 0.86 & 0.19 & 0.12 \\
                                  & DeepSight                            & 0.88 & 0.29 & 0.05 & 0.86 & 0.19 & 0.12 & 0.84 & 0.25 & 0.12 & 0.90 & 0.07 & 0.11 & 0.85 & 0.22 & 0.12 \\
                                  & BackdoorIndicator                    & 0.76 & 0.39 & 0.18 & 0.89 & 0.16 & 0.09 & 0.93 & 0.09 & 0.06 & 0.77 & 0.15 & 0.26 & 0.86 & 0.13 & 0.14 \\
                                  & FreqFed                              & 0.84 & 0.22 & 0.13 & 0.88 & 0.15 & 0.11 & 0.85 & 0.14 & 0.15 & 0.84 & 0.23 & 0.13 & 0.90 & 0.23 & 0.04 \\
\hline
\end{tabular}
\label{exp:three_attack_detection}
\end{table*}

\begin{table*}[t]
\centering
\footnotesize
\caption{\textcolor{black}{Performance of final global models obtained through various detection-based methods under Neurotoxin, Irreversible backdoor, and Clean-label backdoor attacks, where TACC (\(\uparrow\)) and ASR (\(\downarrow\)) metrics are considered. \(\uparrow\) denotes better performance with higher values, and \(\downarrow\) denotes better performance with lower values.}}
\begin{tabular}{|c|c|cc|cc|cc|cc|cc|}
\hline
                                  &                                      & \multicolumn{2}{c|}{\textbf{CIFAR-10}} & \multicolumn{2}{c|}{\textbf{MNIST}} & \multicolumn{2}{c|}{\textbf{FEMNIST}} & \multicolumn{2}{c|}{\textbf{STL-10}} & \multicolumn{2}{c|}{\textbf{Tiny-ImageNet}} \\ \cline{3-12} 
\multirow{-2}{*}{\textbf{Attack}} & \multirow{-2}{*}{\textbf{Defense}} & TACC & ASR & TACC & ASR & TACC & ASR & TACC & ASR & TACC & ASR \\ \hline
                                  & FLAME                                & 0.77 & 0.35 & 0.93 & 0.27 & 0.63 & 0.32 & 0.49 & 0.34 & 0.48 & 0.27 \\
                                  & FLDetector                           & 0.74 & 0.34 & 0.95 & 0.48 & 0.65 & 0.44 & 0.50 & 0.20 & 0.50 & 0.35 \\
                                  & FLTrust                              & 0.78 & 0.57 & 0.94 & 0.50 & 0.67 & 0.50 & 0.52 & 0.23 & 0.47 & 0.19 \\
                                  & DeepSight                            & 0.73 & 0.35 & 0.94 & 0.29 & 0.64 & 0.54 & 0.52 & 0.17 & 0.51 & 0.26 \\
                                  & BackdoorIndicator                            & 0.80 & 0.34 & 0.98 & 0.23 & 0.64 & 0.12 & 0.47 & 0.15 & 0.50 & 0.17 \\
                                  & FreqFed                              & 0.79 & 0.60 & 0.92 & 0.39 & 0.66 & 0.17 & 0.51 & 0.35 & 0.46 & 0.07 \\
\hline
\end{tabular}
\label{exp:three_attack_tacc}
\end{table*}

\begin{table*}[]
\footnotesize
  \addtolength{\tabcolsep}{-3.525pt}  
\caption{\textcolor{black}{Detection performance of various detection-based methods is assessed using Precision ($\uparrow$), Recall ($\uparrow$), and F1-score ($\uparrow$).}}
\begin{tabular}{|c|c|ccc|ccc|ccc|ccc|ccc|}
\hline
\multirow{2}{*}{\textbf{Attack}} & \multirow{2}{*}{\textbf{Detection}} & \multicolumn{3}{c|}{\textbf{CIFAR-10}} & \multicolumn{3}{c|}{\textbf{MNIST}} & \multicolumn{3}{c|}{\textbf{FEMNIST}} & \multicolumn{3}{c|}{\textbf{STL-10}} & \multicolumn{3}{c|}{\textbf{Tiny-ImageNet}} \\ \cline{3-17} 
                                 &                                     & Precision     & Recall     & F1-score       & Precision     & Recall    & F1-score      & Precision     & Recall     & F1-score       & Precision     & Recall     & F1-score      & Precision       & Recall       & F1-score         \\ \hline
\multirow{9}{*}{No attack}              & FLAME                               & NA            & NA         & NA       & NA            & NA        & NA      & NA            & NA         & NA       & NA            & NA         & NA      & NA              & NA           & NA         \\
                                 & FLDetector                          & NA            & NA         & NA       & NA            & NA        & NA      & NA            & NA         & NA       & NA            & NA         & NA      & NA              & NA           & NA         \\
                                 & FLTrust                             & NA            & NA         & NA       & NA            & NA        & NA      & NA            & NA         & NA       & NA            & NA         & NA      & NA              & NA           & NA         \\
                                 & DeepSight                           & NA            & NA         & NA       & NA            & NA        & NA      & NA            & NA         & NA       & NA            & NA         & NA      & NA              & NA           & NA         \\
\hline
\end{tabular}
\label{three_n_index}
\end{table*}

\begin{table*}[]
\centering
  \addtolength{\tabcolsep}{-2.575pt}  
  \footnotesize
\caption{\textcolor{black}{Detection performance of various detection-based methods is assessed using Precision ($\uparrow$), Recall ($\uparrow$), and F1-score ($\uparrow$), under three advanced backdoor attacks.}}
\begin{tabular}{|c|c|ccc|ccc|ccc|ccc|ccc|}
\hline
                                  &                                      & \multicolumn{3}{c|}{\textbf{CIFAR-10}} & \multicolumn{3}{c|}{\textbf{MNIST}} & \multicolumn{3}{c|}{\textbf{FEMNIST}} & \multicolumn{3}{c|}{\textbf{STL-10}} & \multicolumn{3}{c|}{\textbf{Tiny-ImageNet}} \\ \cline{3-17} 
\multirow{-2}{*}{\textbf{Attack}} & \multirow{-2}{*}{\textbf{Defense}} & Precision & Recall & F1-score & Precision & Recall & F1-score & Precision & Recall & F1-score & Precision & Recall & F1-score & Precision & Recall & F1-score \\ \hline
\multirow{9}{*}{{\makecell {Neurotoxin\\ attack}}}
                                  & FLAME                                & 0.86 & 0.89 & 0.87 & 0.89 & 0.94 & 0.91 & 0.90 & 0.87 & 0.88 & 0.91 & 0.91 & 0.91 & 0.92 & 0.86 & 0.89 \\
                                  & FLDetector                           & 0.84 & 0.89 & 0.86 & 0.79 & 0.88 & 0.83 & 0.86 & 0.94 & 0.90 & 0.93 & 0.95 & 0.94 & 0.85 & 0.79 & 0.82 \\
                                  & FLTrust                              & 0.76 & 0.78 & 0.77 & 0.78 & 0.78 & 0.78 & 0.81 & 0.80 & 0.80 & 0.85 & 0.84 & 0.85 & 0.92 & 0.88 & 0.90 \\
                                  & DeepSight                            & 0.88 & 0.95 & 0.91 & 0.90 & 0.88 & 0.89 & 0.87 & 0.88 & 0.88 & 0.94 & 0.89 & 0.91 & 0.89 & 0.88 & 0.89 \\
                                  & BackdoorIndicator                    & 0.81 & 0.82 & 0.82 & 0.92 & 0.91 & 0.91 & 0.96 & 0.94 & 0.95 & 0.83 & 0.74 & 0.78 & 0.92 & 0.86 & 0.89 \\
                                  & FreqFed                              & 0.88 & 0.87 & 0.87 & 0.91 & 0.89 & 0.90 & 0.89 & 0.85 & 0.87 & 0.86 & 0.87 & 0.87 & 0.93 & 0.90 & 0.91 \\
\hline
\end{tabular}
\label{exp:three_attack_detection_index}
\end{table*}

\begin{table*}[]
\footnotesize
\caption{\textcolor{black}{Storage costs of different methods.}}
\begin{tabular}{|c|c|c|c|c|c|}
\hline
\textbf{Defense} & \textbf{CIFAR-10} & \textbf{MNIST} & \textbf{FEMNIST} & \textbf{STL-10} & \textbf{Tiny-ImageNet} \\ \hline
FLDetector         & 21.73 GB & 15.76 GB & 22.95 GB & 38.46 GB & 31.28 GB \\ \hline
FedREDefense       &19.53 GB & 10.72 GB & 20.15 GB & 28.83 GB & 27.46 GB \\ \hline
\cellcolor{greyL}\algFirst          & \cellcolor{greyL}13.66 GB & \cellcolor{greyL}7.96 GB & \cellcolor{greyL}13.57 GB & \cellcolor{greyL}22.40 GB & \cellcolor{greyL}19.93 GB \\ \hline
\cellcolor{greyL}\algSecond          & \cellcolor{greyL}13.66 GB & \cellcolor{greyL}7.96 GB & \cellcolor{greyL}13.57 GB & \cellcolor{greyL}22.40 GB & \cellcolor{greyL}19.93 GB \\ \hline
\end{tabular}
\label{exp:memory}
\end{table*}

\begin{table*}[]
\footnotesize
\caption{\textcolor{black}{
Equalized odds and Demographic parity scores of our methods under different attacks, where the CIFAR-10 dataset is considered.}}
\begin{tabular}{|c|c|cc|cc|}
\hline
    \multicolumn{1}{|c|}{\multirow{2}{*}{Attack}} & \multicolumn{1}{c|}{\multirow{2}{*}{Defense}} & \multicolumn{2}{c|}{Setting I} & \multicolumn{2}{c|}{Setting II} \\
\cline{3-6}          &       & \multicolumn{1}{c}{Equalized odds} & \multicolumn{1}{c|}{Demographic parity} & \multicolumn{1}{c}{Equalized odds} & \multicolumn{1}{c|}{Demographic parity} \\
    \hline
                                       & FedAvg                      & 0.79 & 0.57 & 0.69 & 0.48 \\
                                       & \cellcolor{greyL}\algFirst                   & \cellcolor{greyL}0.77 & \cellcolor{greyL}0.59 & \cellcolor{greyL}0.66 & \cellcolor{greyL}0.47 \\
\multirow{-3}{*}{No attack}          & \cellcolor{greyL}\algSecond                   & \cellcolor{greyL}0.78 & \cellcolor{greyL}0.56 & \cellcolor{greyL}0.71 & \cellcolor{greyL}0.52 \\ \hline

& \cellcolor{greyL}\algFirst                   & \cellcolor{greyL}0.81 & \cellcolor{greyL}0.58 & \cellcolor{greyL}0.71 & \cellcolor{greyL}0.49 \\
\multirow{-2}{*}{Trim attack}          & \cellcolor{greyL}\algSecond                   & \cellcolor{greyL}0.77 & \cellcolor{greyL}0.57 & \cellcolor{greyL}0.70 & \cellcolor{greyL}0.49 \\

\hline
                                       & \cellcolor{greyL}\algFirst                   & \cellcolor{greyL}0.77 & \cellcolor{greyL}0.56 & \cellcolor{greyL}0.70 & \cellcolor{greyL}0.50 \\
\hline
\end{tabular}
\label{exp:bias}
\end{table*}

\begin{table*}[]
\footnotesize
\caption{\textcolor{black}{The impact of the noise level is analyzed using the CIFAR-10 dataset, where DACC and TACC values are reported.}}
\centering
\begin{tabular}{|c|c|cc|cc|cc|cc|cc|}
   \hline
    \multirow{3}{*}{Attack} & \multirow{3}{*}{Defense} & \multicolumn{10}{c|}{Noise level} \\
\cline{3-12}          &       & \multicolumn{2}{c|}{0} & \multicolumn{2}{c|}{0.5} & \multicolumn{2}{c|}{1} & \multicolumn{2}{c|}{1.5} & \multicolumn{2}{c|}{2} \\
\cline{3-12}          &       & DACC  & TACC  & DACC  & TACC  & DACC  & TACC  & DACC  & TACC  & DACC  & TACC \\
   \hline
&FedAvg & NA & 0.85 & NA & 0.81 &NA & 0.80 & NA & 0.76 & NA & 0.70 \\
& FLAME                       & 0.57 & 0.72 & 0.55 & 0.71 & 0.56 & 0.70 & 0.52 & 0.66 & 0.58 & 0.62 \\
                                     & FLDetector                  & 0.65 & 0.77 & 0.66 & 0.74 & 0.64 & 0.73 & 0.62 & 0.66 & 0.59 & 0.63 \\
                                     & FLTrust                     & 0.82 & 0.81 & 0.77 & 0.74 & 0.75 & 0.72 & 0.70 & 0.67 & 0.70 & 0.65 \\
                                     & DeepSight                   & 1.00 & 0.75 & 0.94 & 0.74 & 0.86 & 0.70 & 0.83 & 0.67 & 0.81 & 0.66 \\
\hline
\end{tabular}
\label{exp:noise_level}
\end{table*}

\begin{figure*}[t]
  \centering
    \begin{subfigure}{0.163\textwidth}
    \includegraphics[width=\textwidth]{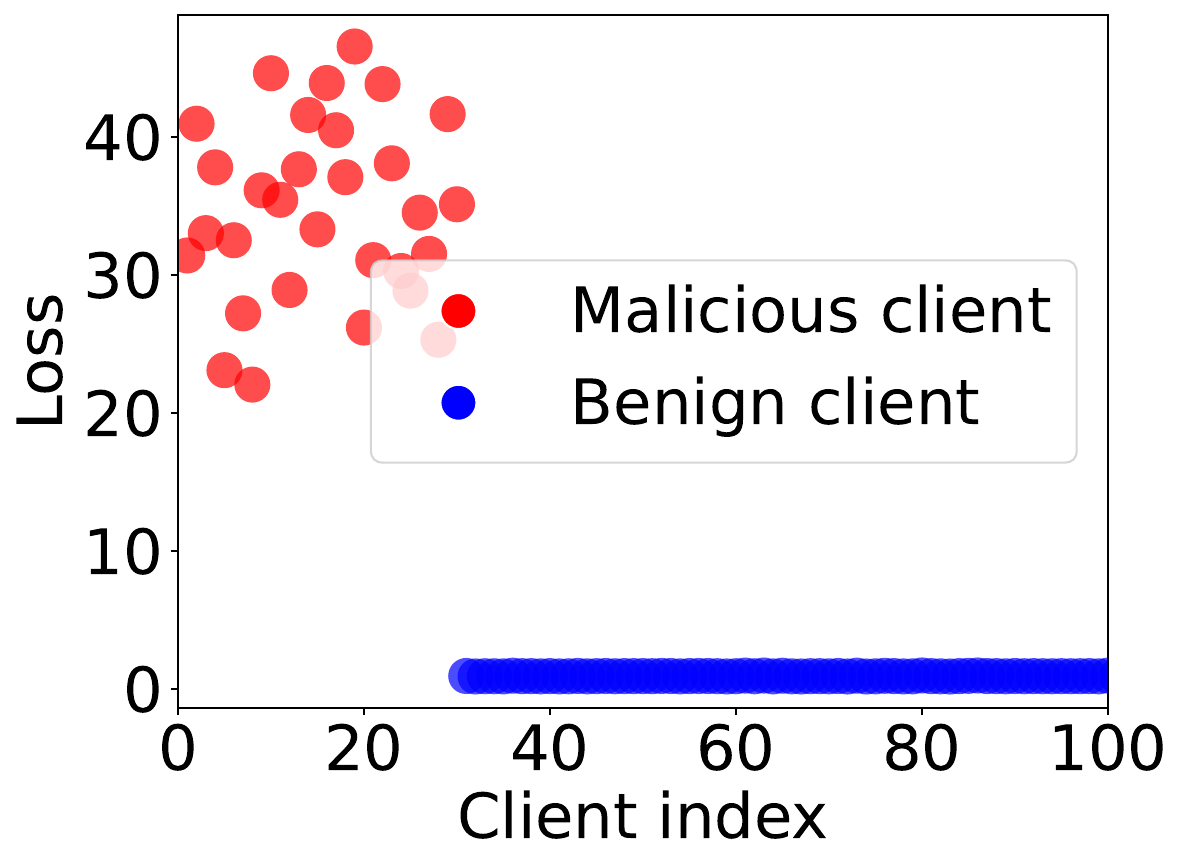}
    \caption{Trim attack}
  \end{subfigure}
    \begin{subfigure}{0.163\textwidth}
    \includegraphics[width=\textwidth]{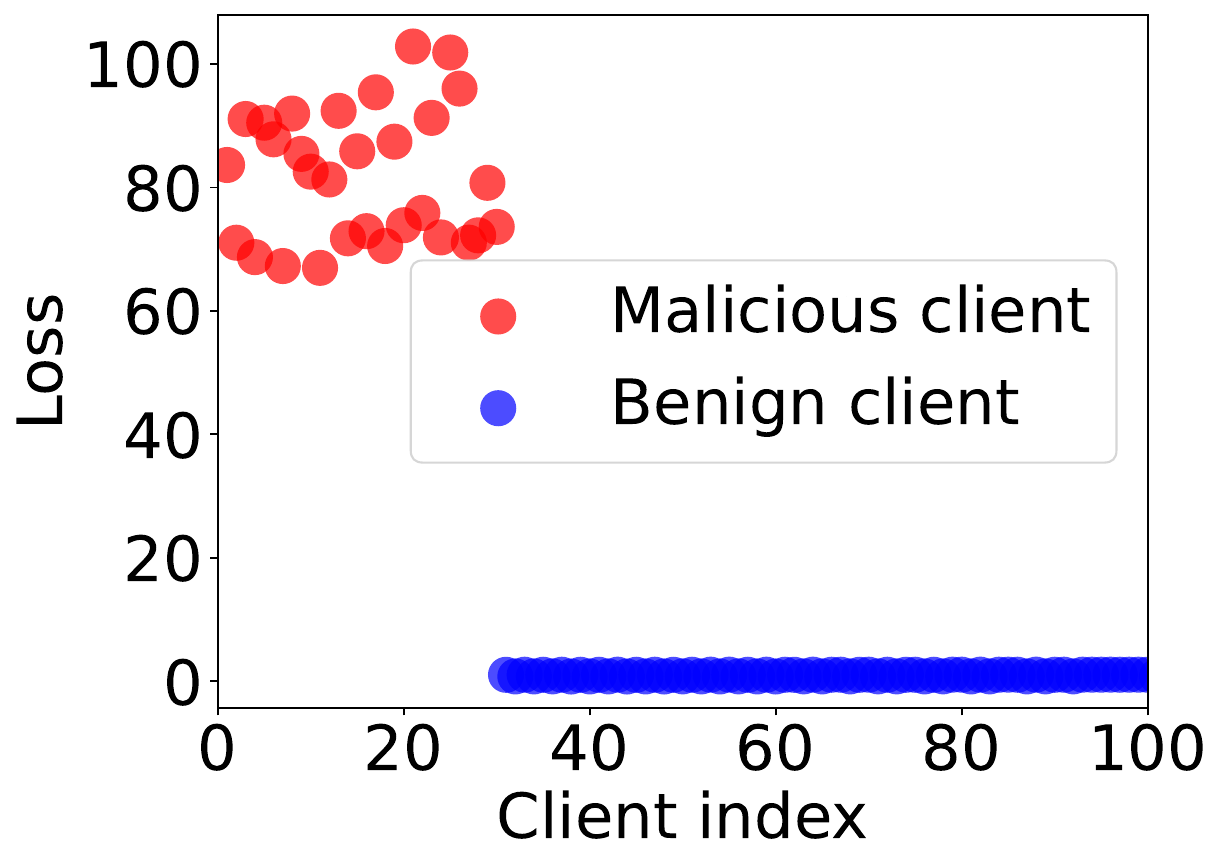}
    \caption{Scaling attack}
  \end{subfigure}
  \begin{subfigure}{0.163\textwidth}
    \includegraphics[width=\textwidth]{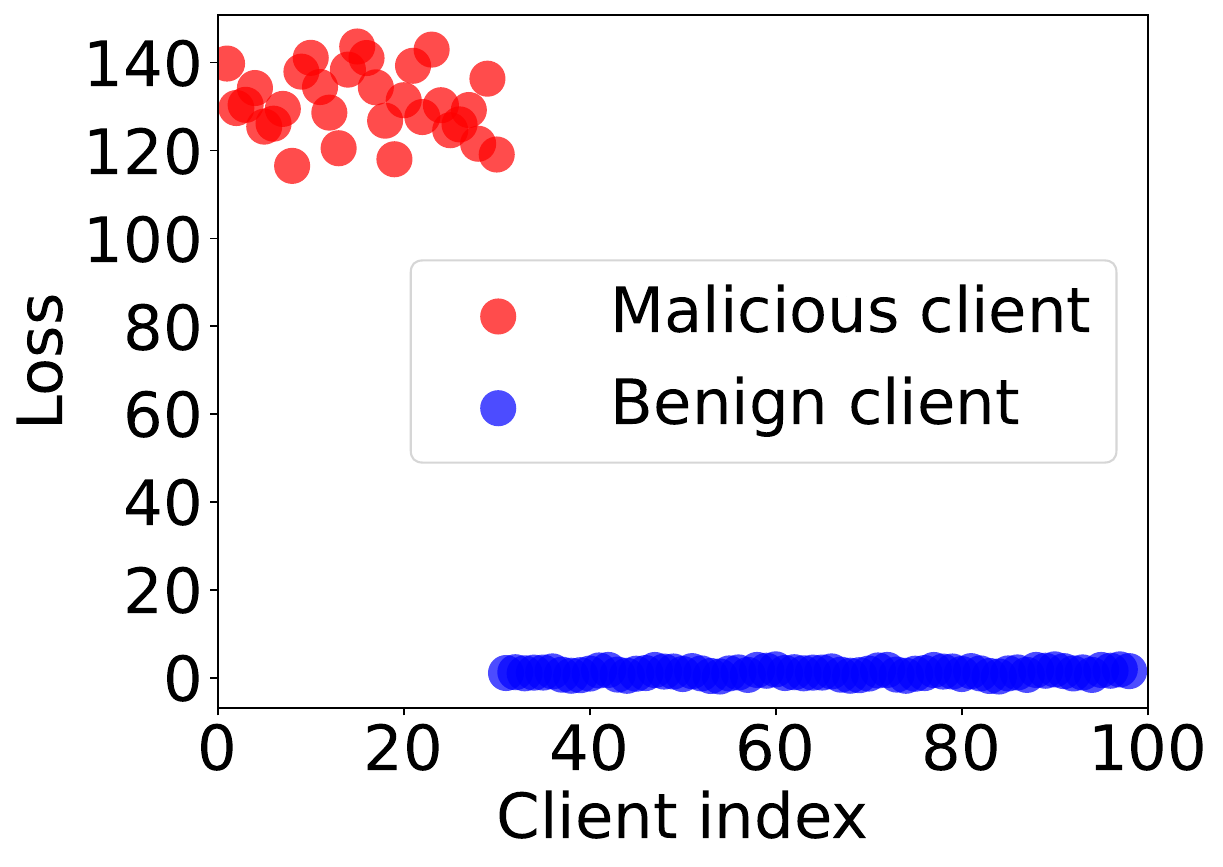}
    \caption{DBA attack}
  \end{subfigure}
  \begin{subfigure}{0.163\textwidth}
    \includegraphics[width=\textwidth]{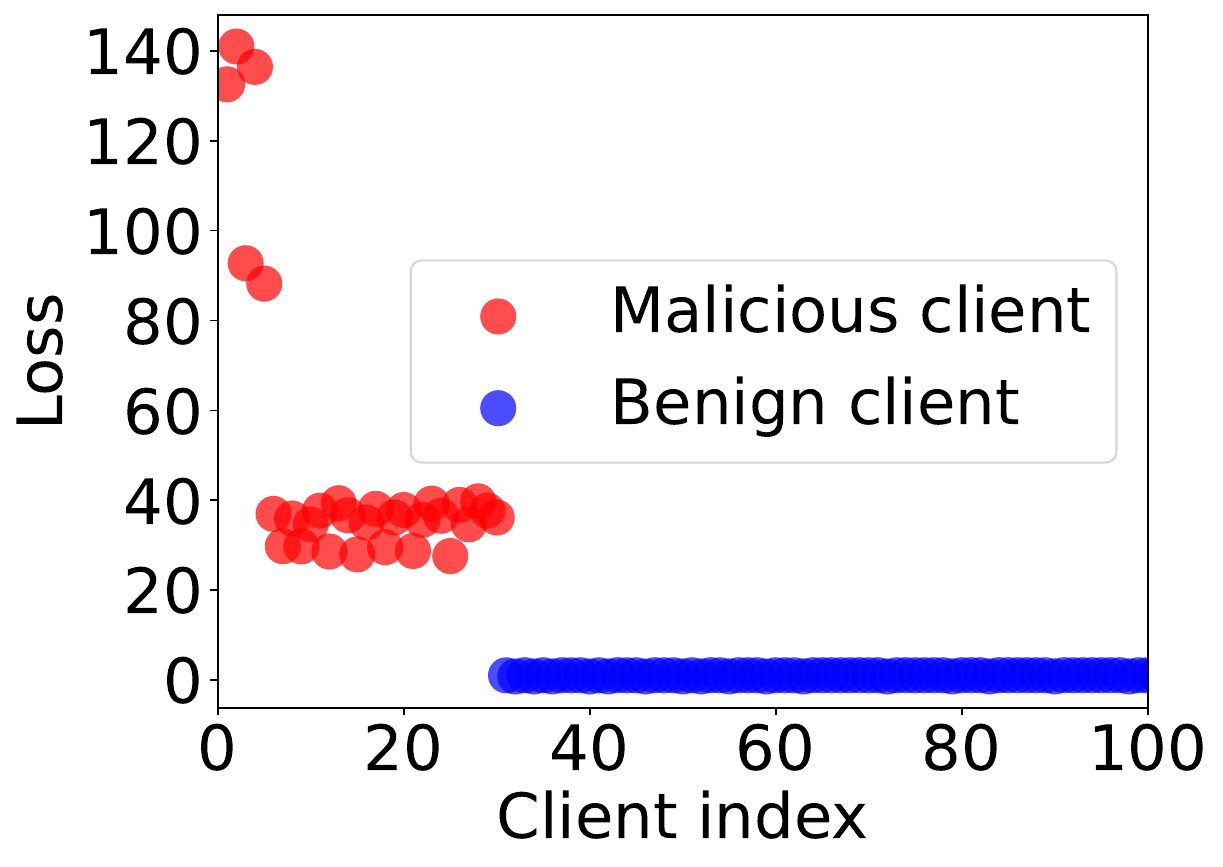}
    \caption{Trim+DBA attack}
  \end{subfigure}
    \begin{subfigure}{0.163\textwidth}
    \includegraphics[width=\textwidth]{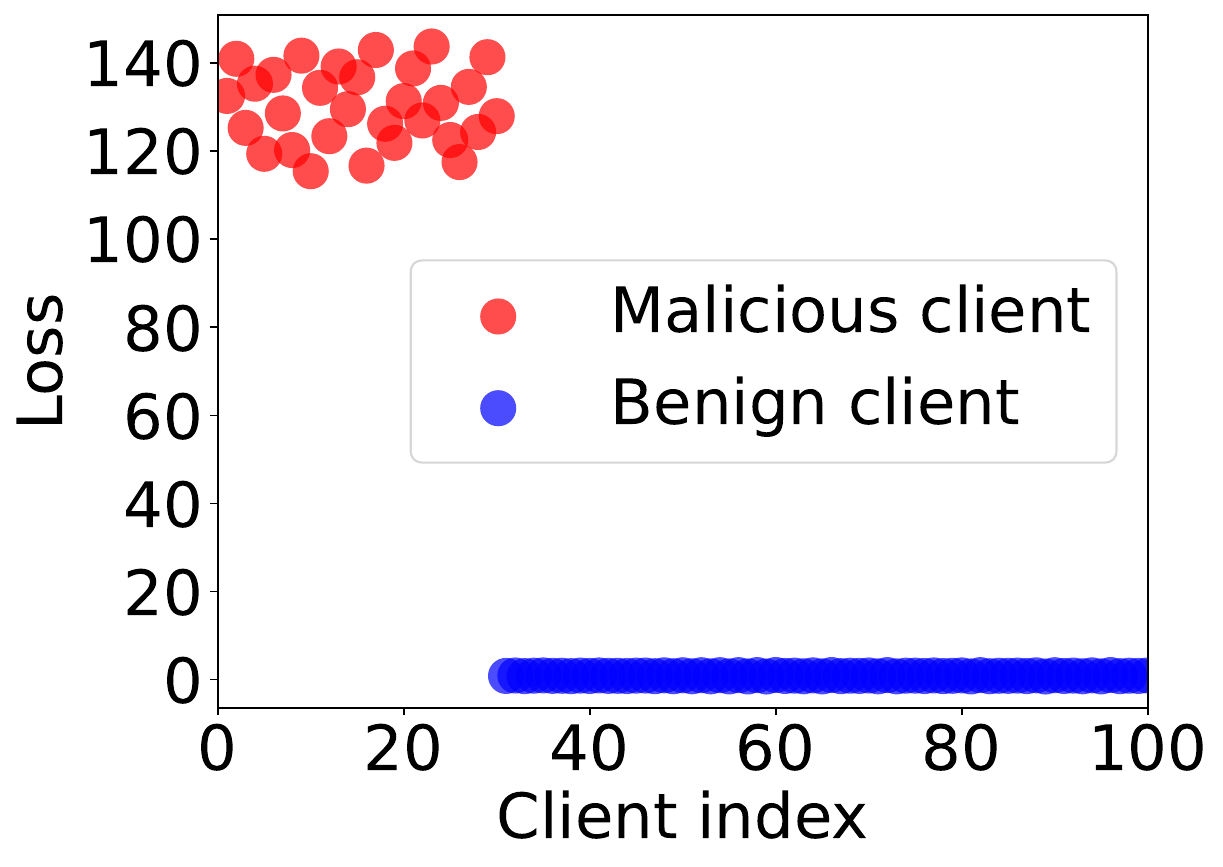}
    \caption{Scaling+DBA attack}
  \end{subfigure}
    \begin{subfigure}{0.163\textwidth}
    \includegraphics[width=\textwidth]{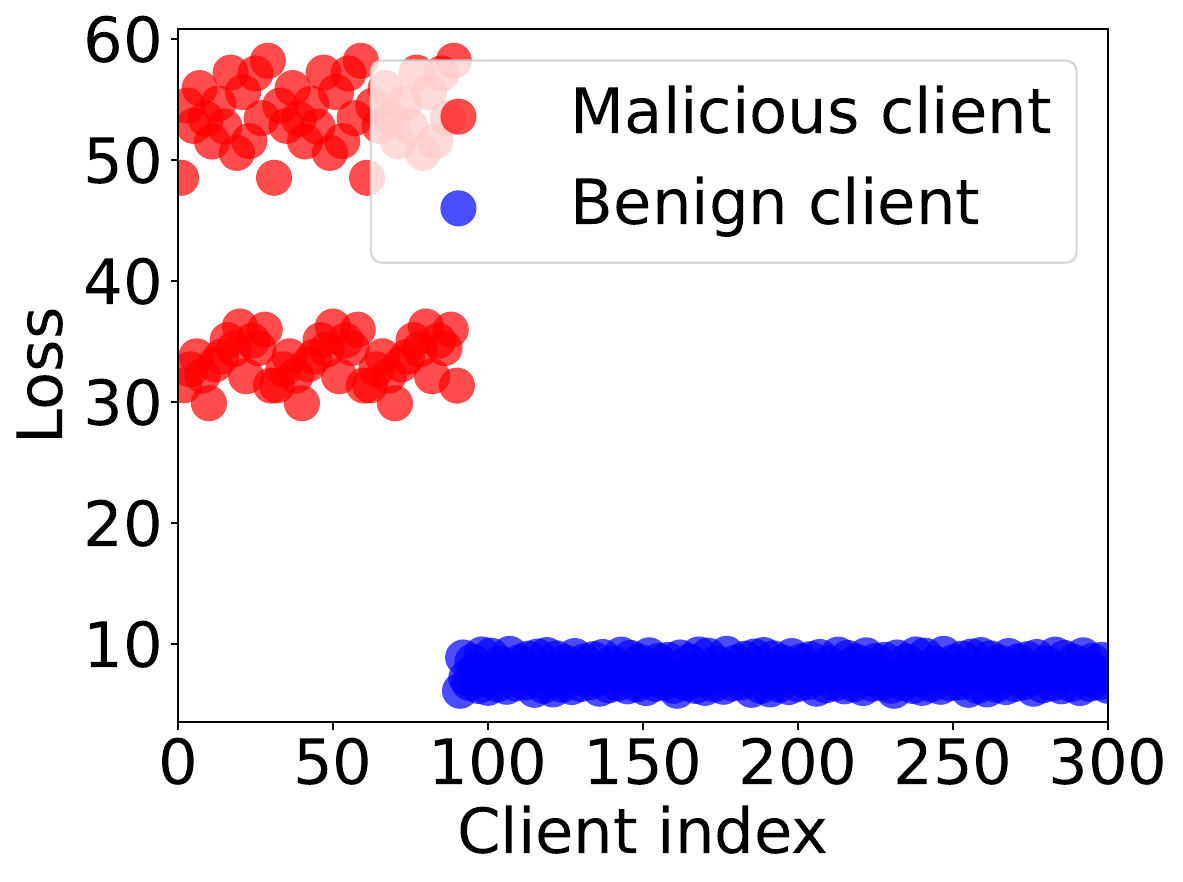}
    \caption{Adaptive attack}
  \end{subfigure}
  \caption{\textcolor{black}{The loss values of benign and malicious clients’ local models computed on the synthetic dataset, using \algFirst with the CIFAR-10 dataset.}
  \label{exp:safefl-ml-cifar}}
\end{figure*}

\begin{figure*}[t]
  \centering
    \begin{subfigure}{0.163\textwidth}
    \includegraphics[width=\textwidth]{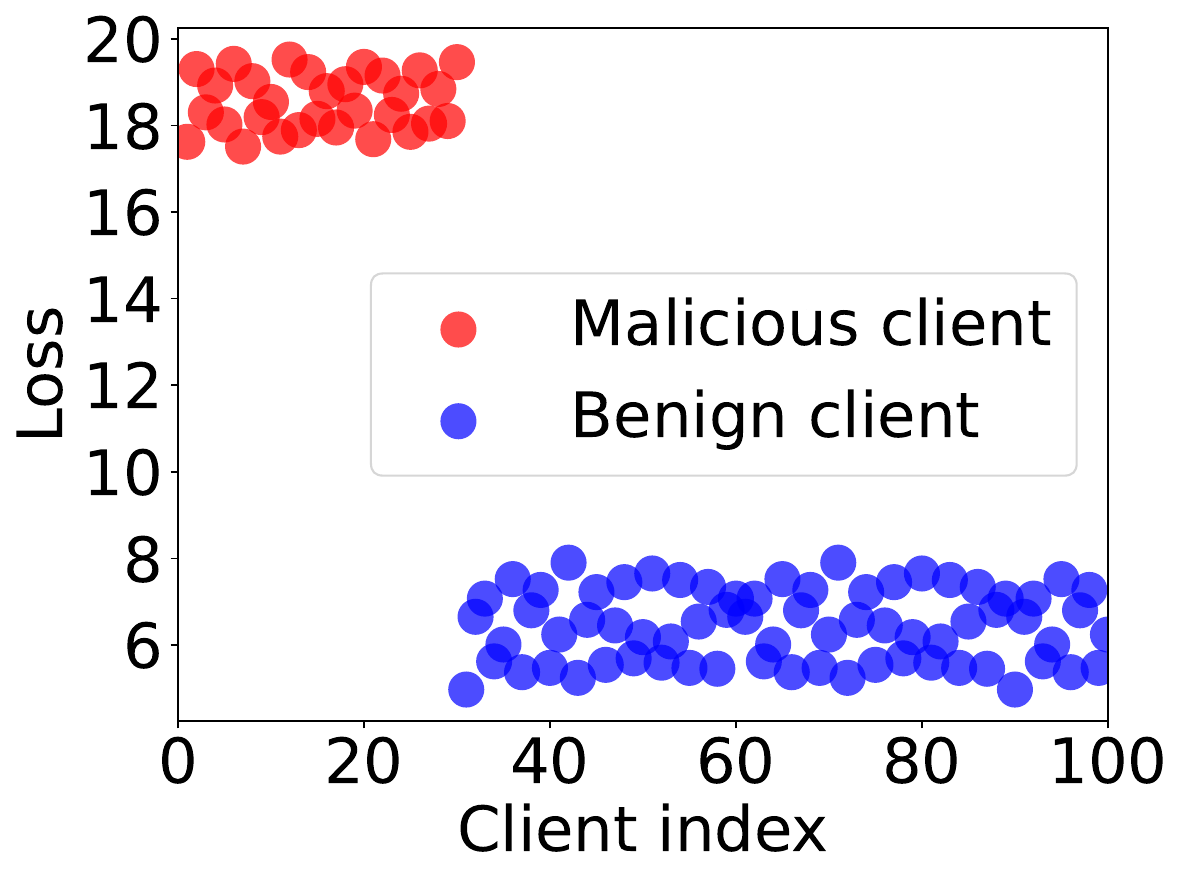}
    \caption{Trim attack}
  \end{subfigure}
    \begin{subfigure}{0.163\textwidth}
    \includegraphics[width=\textwidth]{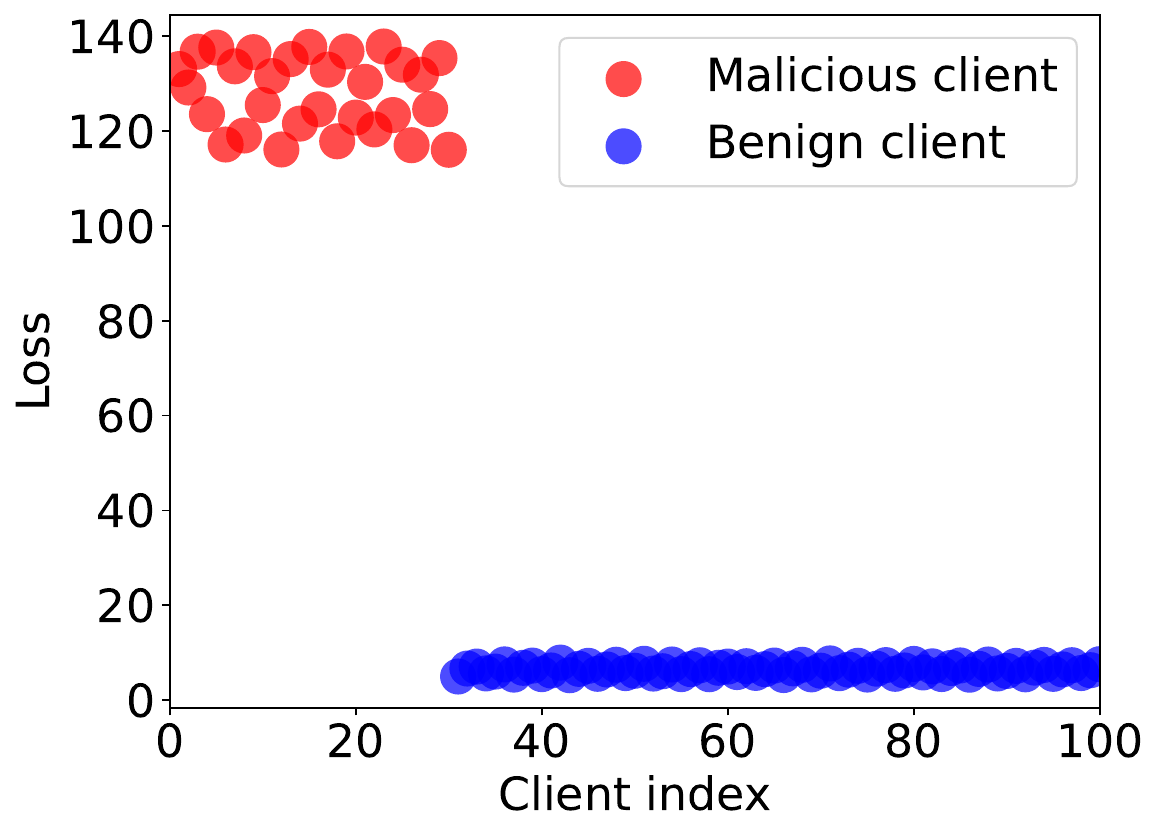}
    \caption{Scaling attack}
  \end{subfigure}
  \begin{subfigure}{0.163\textwidth}
    \includegraphics[width=\textwidth]{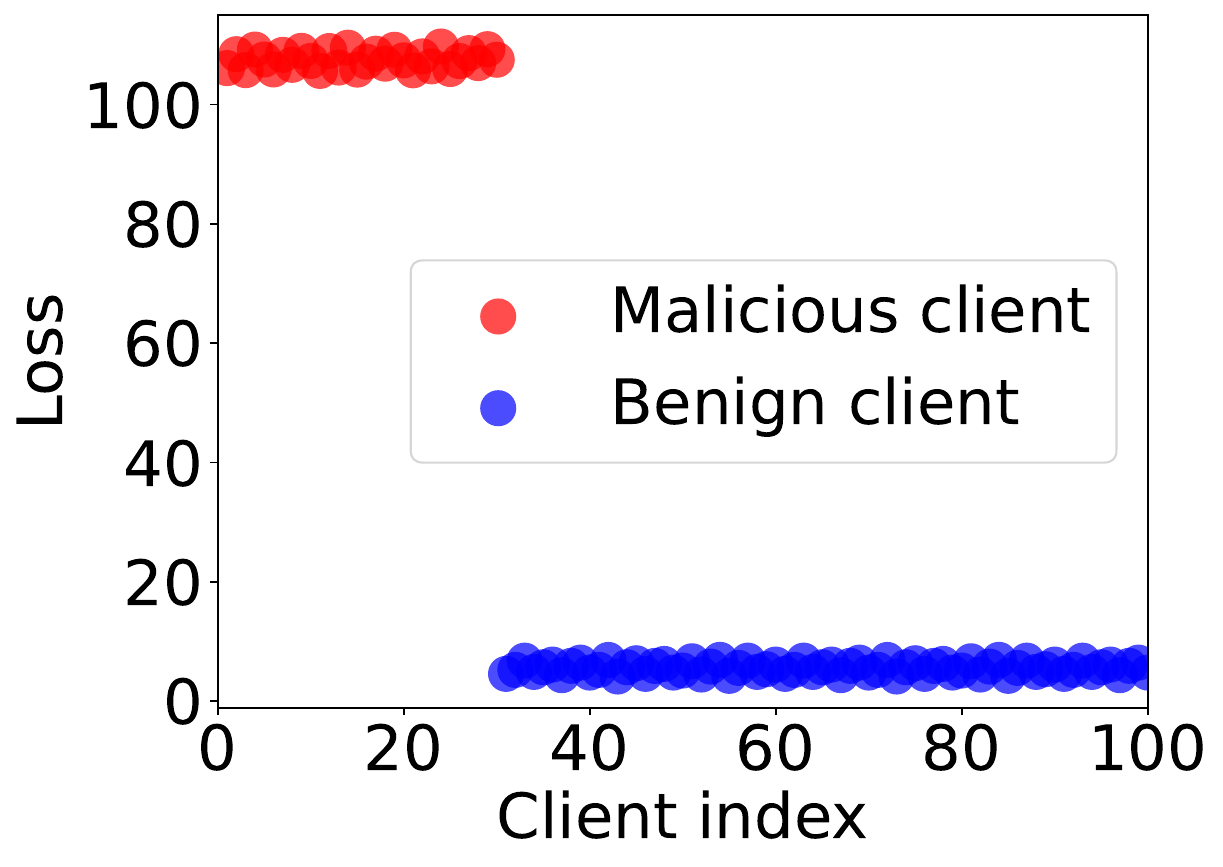}
    \caption{DBA attack}
  \end{subfigure}
  \begin{subfigure}{0.163\textwidth}
    \includegraphics[width=\textwidth]{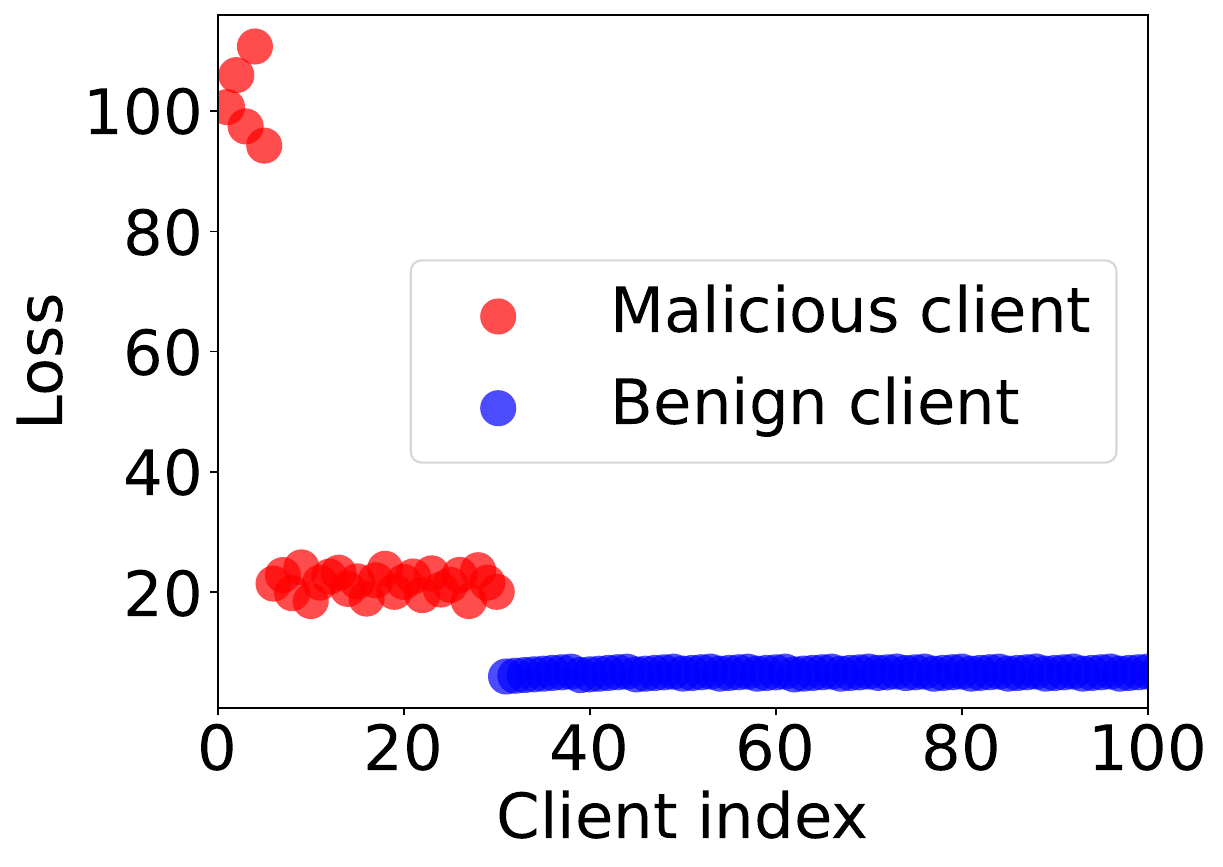}
    \caption{Trim+DBA attack}
  \end{subfigure}
    \begin{subfigure}{0.163\textwidth}
    \includegraphics[width=\textwidth]{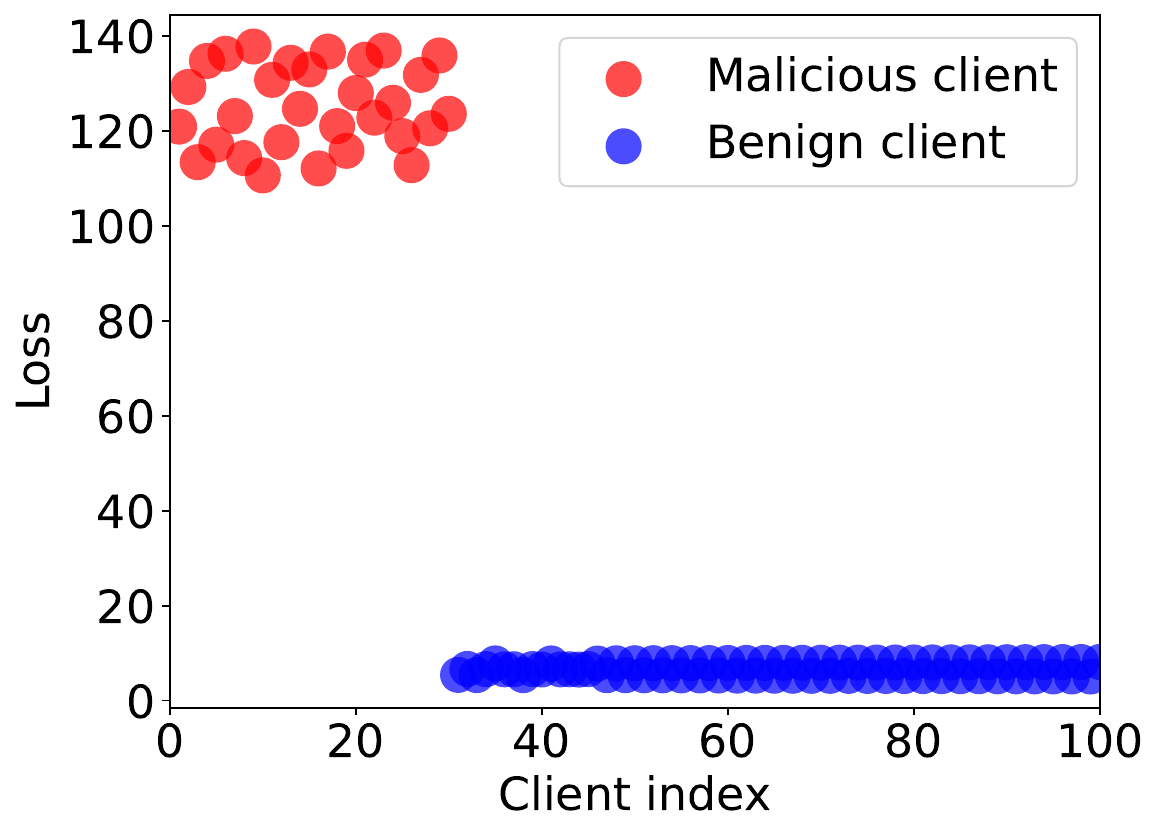}
    \caption{Scaling+DBA attack}
  \end{subfigure}
    \begin{subfigure}{0.163\textwidth}
    \includegraphics[width=\textwidth]{ML_imgs/MNIST_dba_ml.pdf}
    \caption{Adaptive attack}
  \end{subfigure}
  \caption{\textcolor{black}{The loss values of benign and malicious clients’ local models computed on the synthetic dataset, using \algFirst with the MNIST dataset.}}
  \label{exp:safefl-ml-mnist}
\end{figure*}

\begin{figure*}[t]
  \centering
    \begin{subfigure}{0.163\textwidth}
    \includegraphics[width=\textwidth]{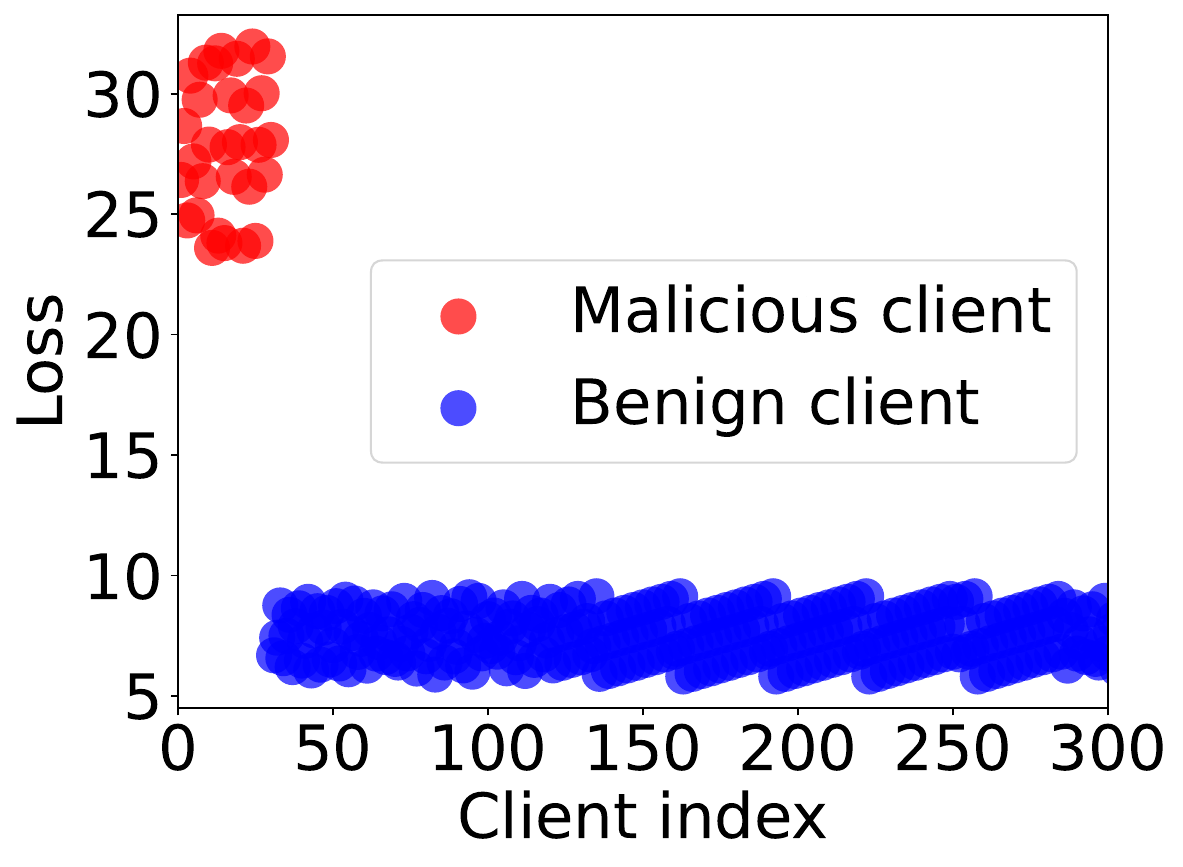}
    \caption{Trim attack}
  \end{subfigure}
    \begin{subfigure}{0.163\textwidth}
    \includegraphics[width=\textwidth]{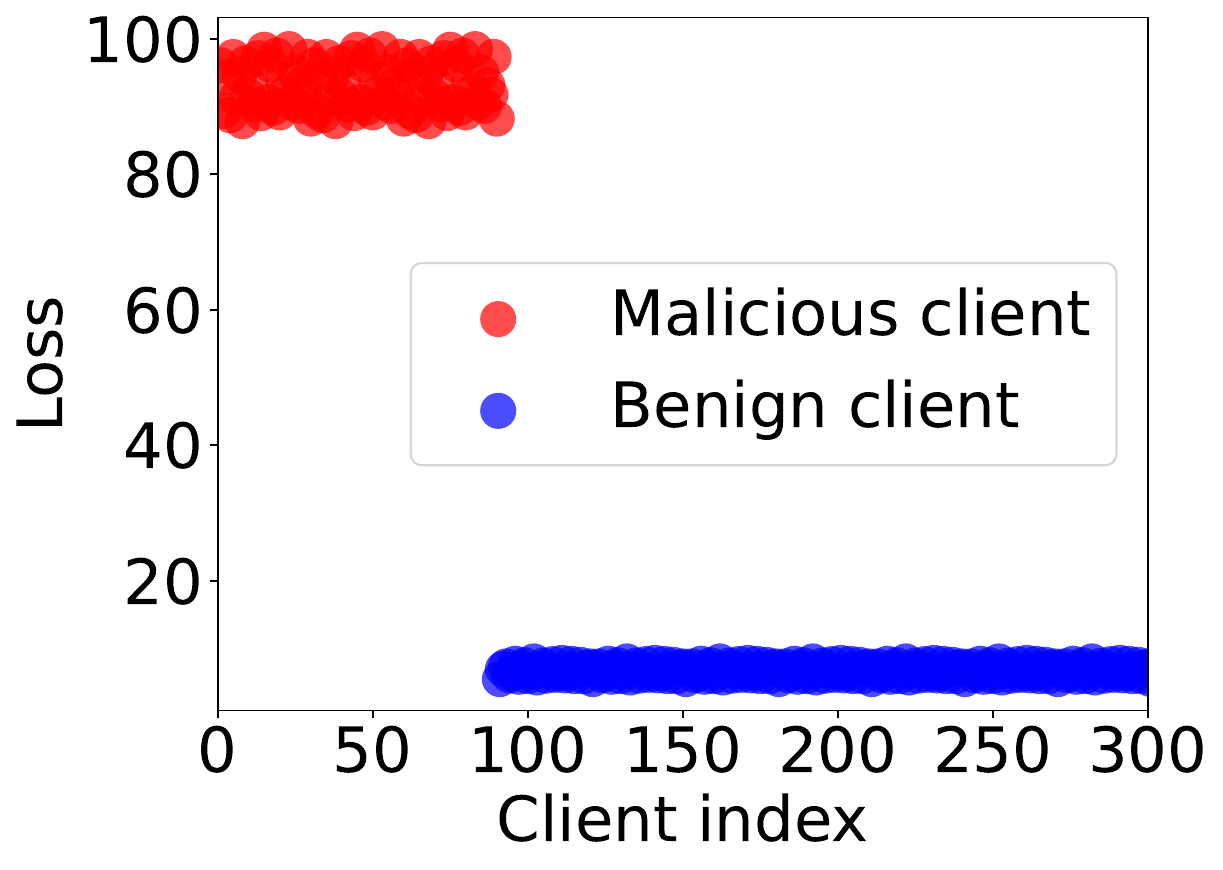}
    \caption{Scaling attack}
  \end{subfigure}
  \begin{subfigure}{0.163\textwidth}
    \includegraphics[width=\textwidth]{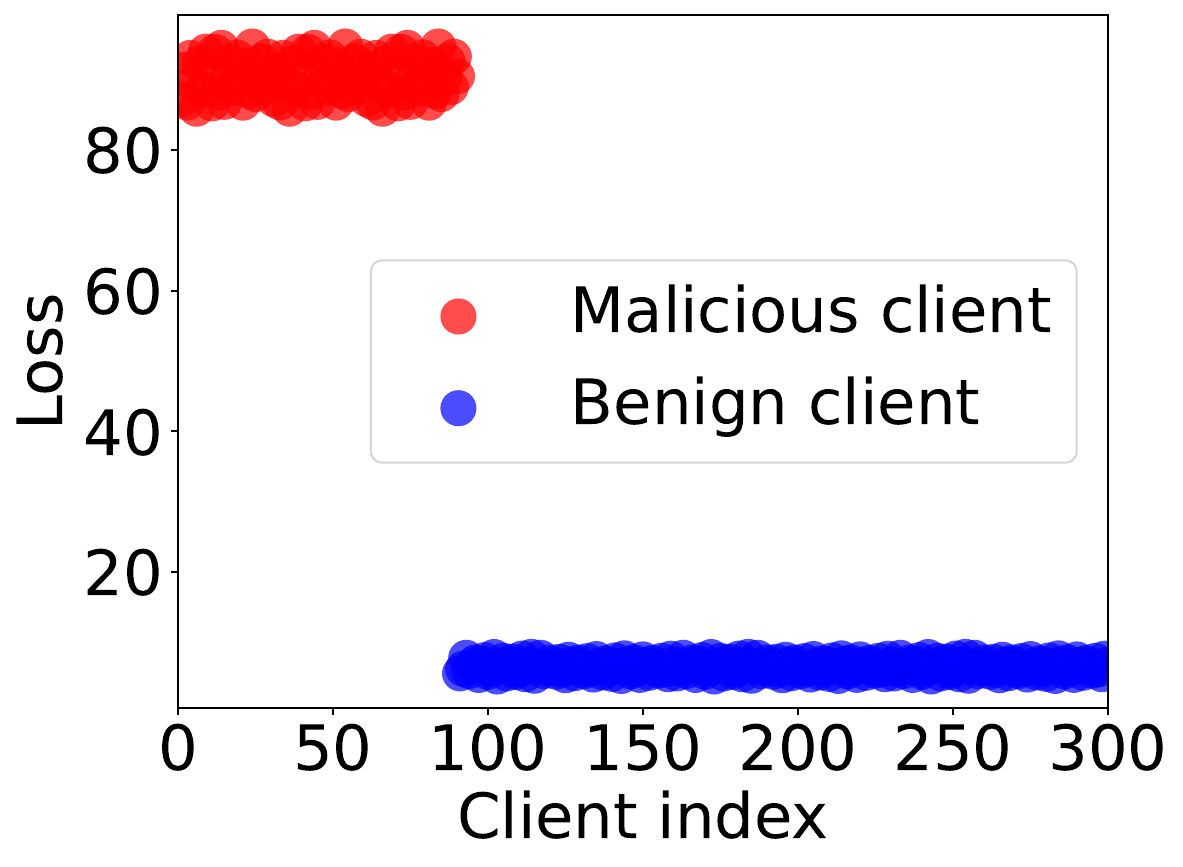}
    \caption{DBA attack}
  \end{subfigure}
  \begin{subfigure}{0.163\textwidth}
    \includegraphics[width=\textwidth]{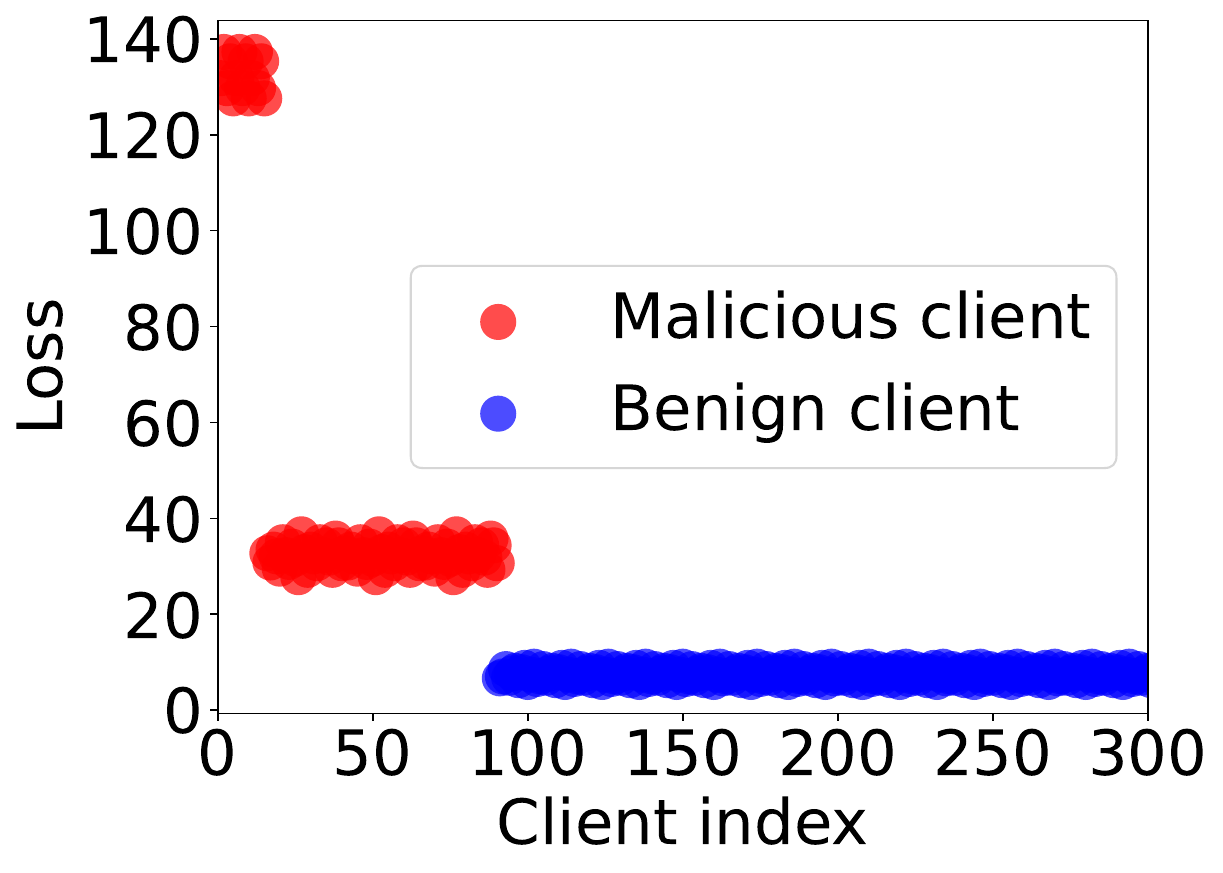}
    \caption{Trim+DBA attack}
  \end{subfigure}
    \begin{subfigure}{0.163\textwidth}
    \includegraphics[width=\textwidth]{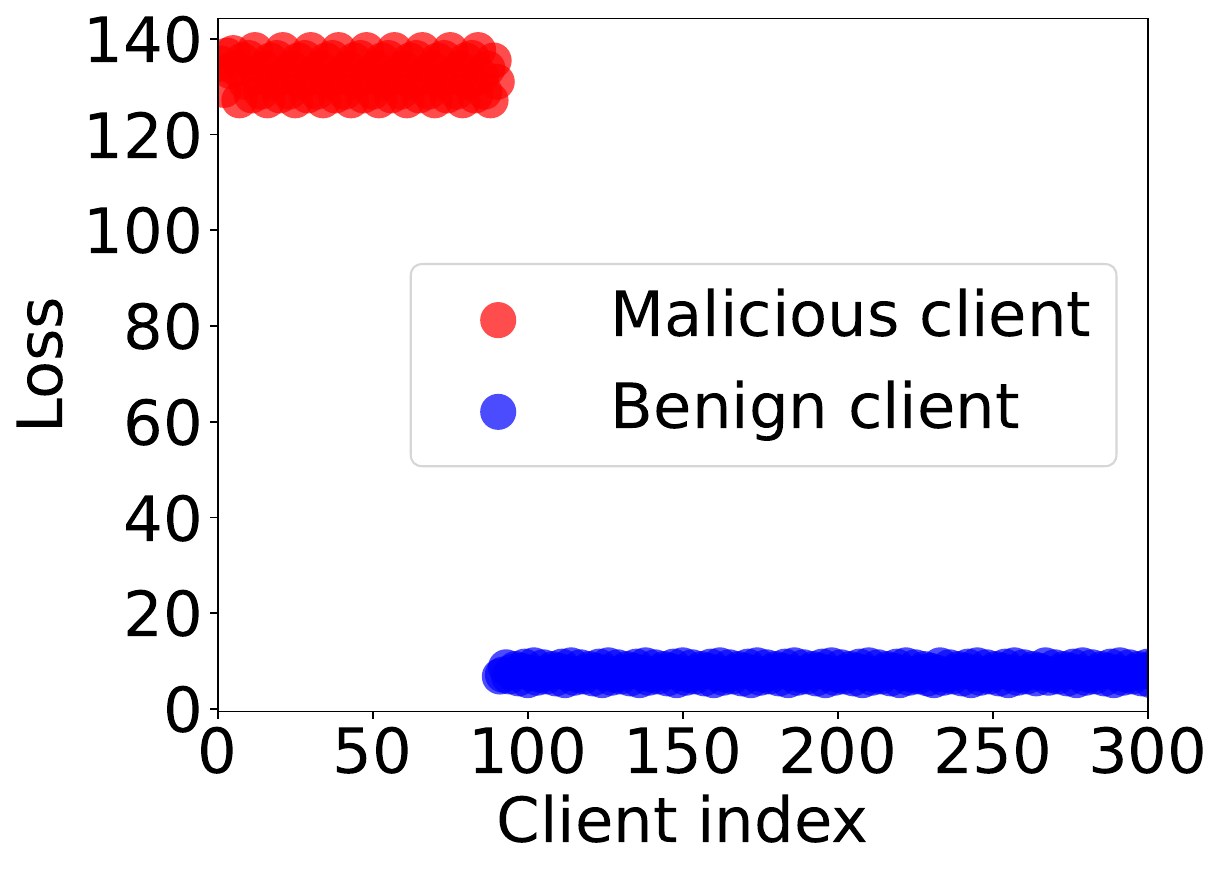}
    \caption{Scaling+DBA attack}
  \end{subfigure}
    \begin{subfigure}{0.163\textwidth}
    \includegraphics[width=\textwidth]{ML_imgs/FEMNIST_adap_ML.pdf}
    \caption{Adaptive attack}
  \end{subfigure}
  \caption{\textcolor{black}{The loss values of benign and malicious clients’ local models computed on the synthetic dataset, using \algFirst with the FEMNIST dataset.}}
  \label{exp:safefl-ml-femnist}
\end{figure*}

\begin{figure*}[t]
  \centering
    \begin{subfigure}{0.163\textwidth}
    \includegraphics[width=\textwidth]{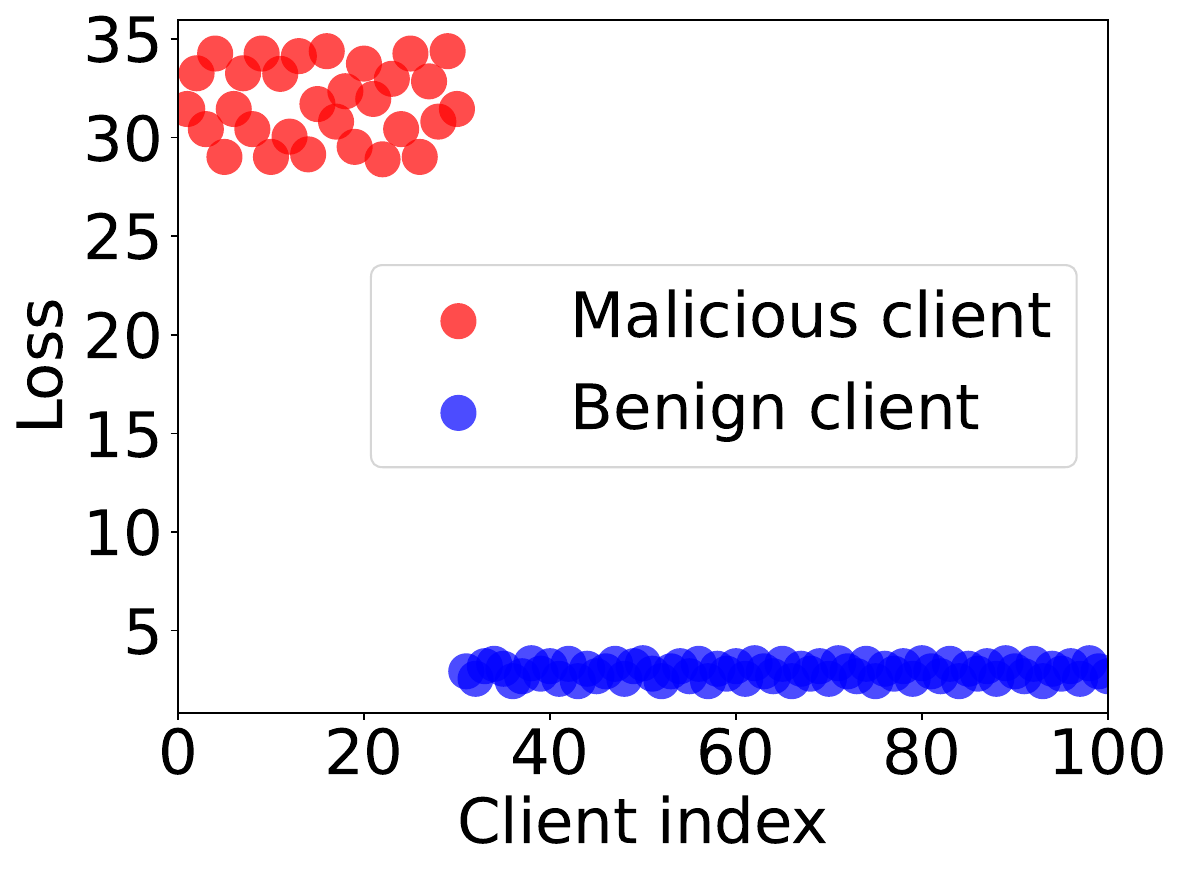}
    \caption{Trim attack}
  \end{subfigure}
    \begin{subfigure}{0.163\textwidth}
    \includegraphics[width=\textwidth]{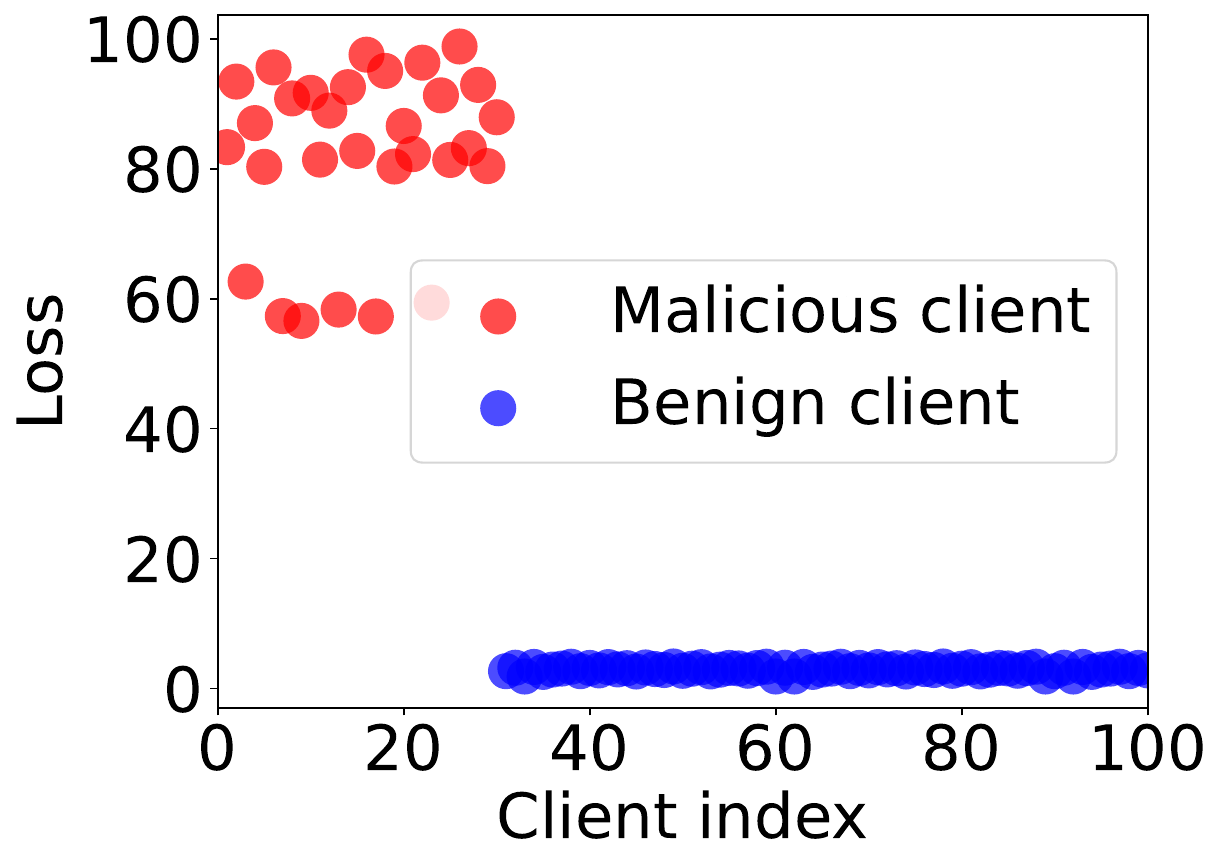}
    \caption{Scaling attack}
  \end{subfigure}
  \begin{subfigure}{0.163\textwidth}
    \includegraphics[width=\textwidth]{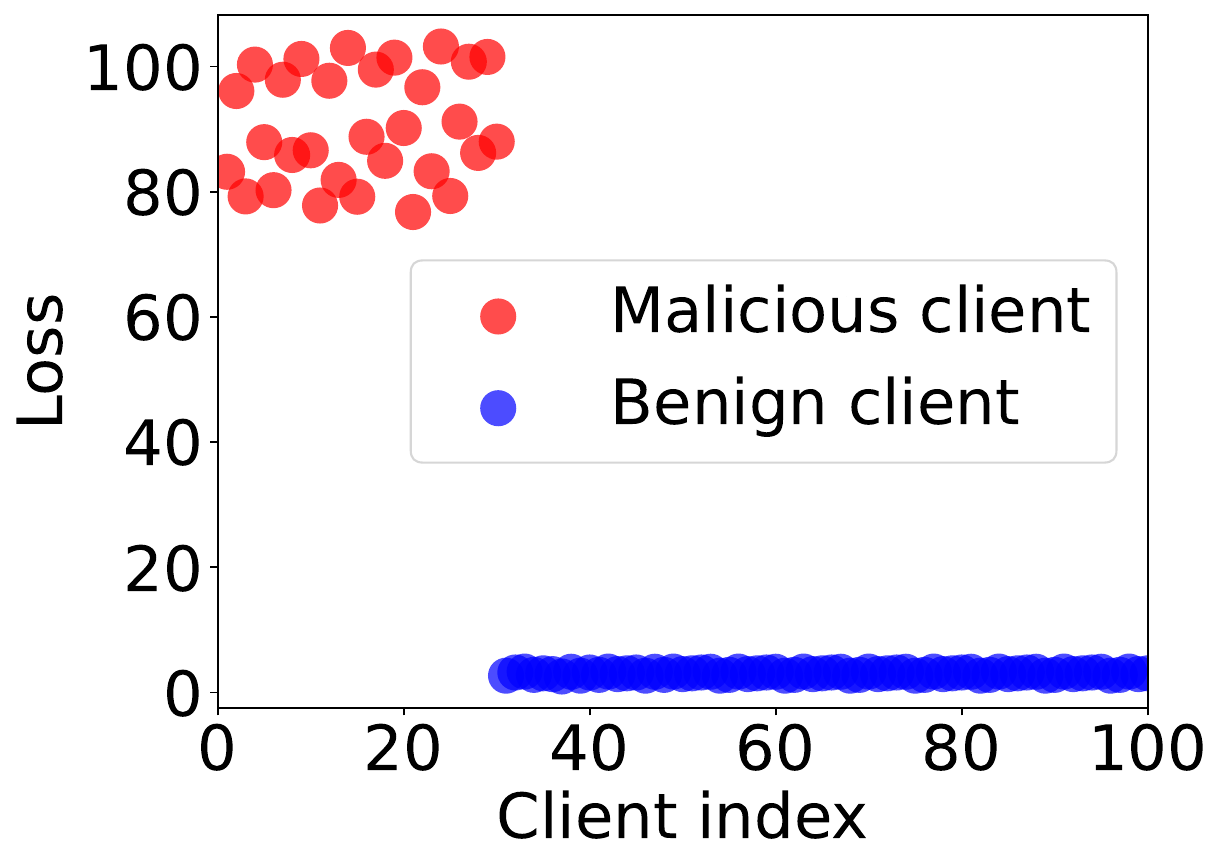}
    \caption{DBA attack}
  \end{subfigure}
  \begin{subfigure}{0.163\textwidth}
    \includegraphics[width=\textwidth]{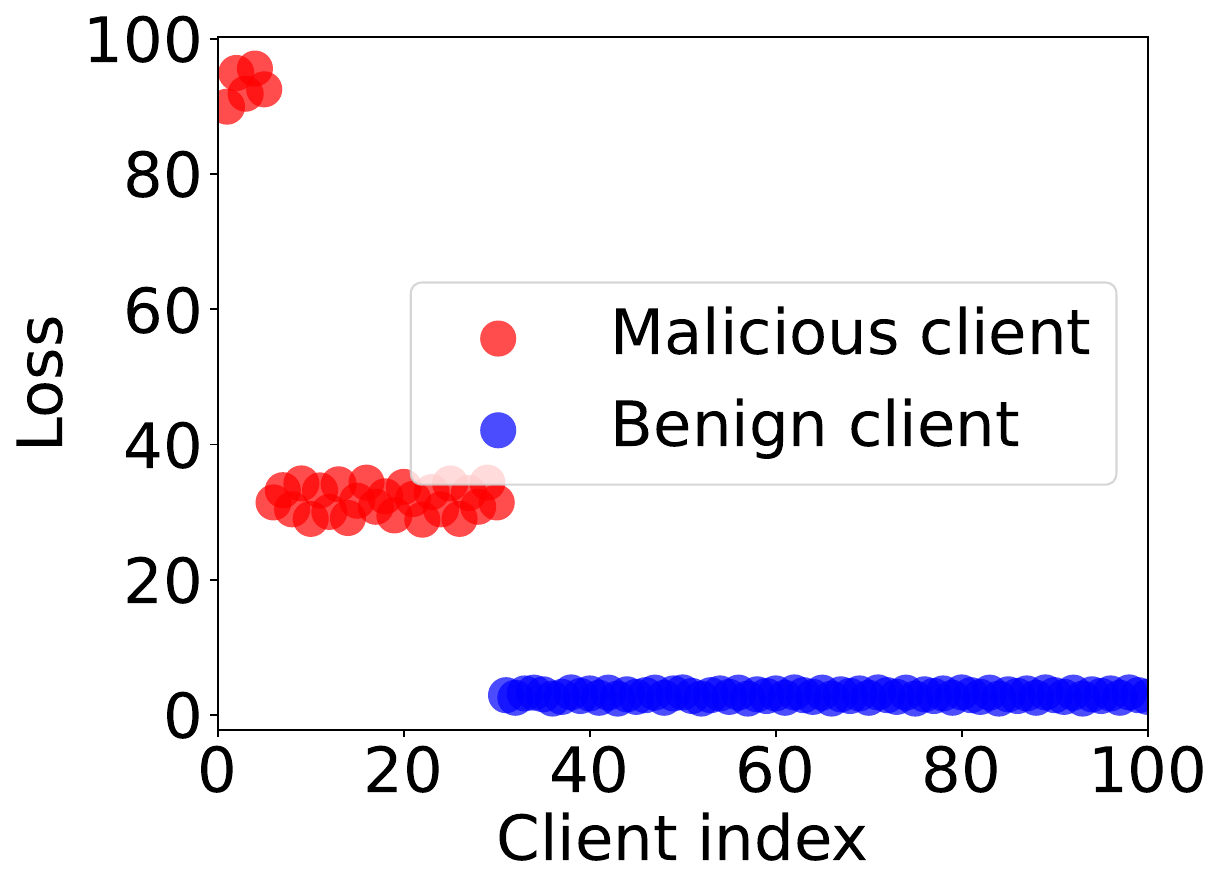}
    \caption{Trim+DBA attack}
  \end{subfigure}
    \begin{subfigure}{0.163\textwidth}
    \includegraphics[width=\textwidth]{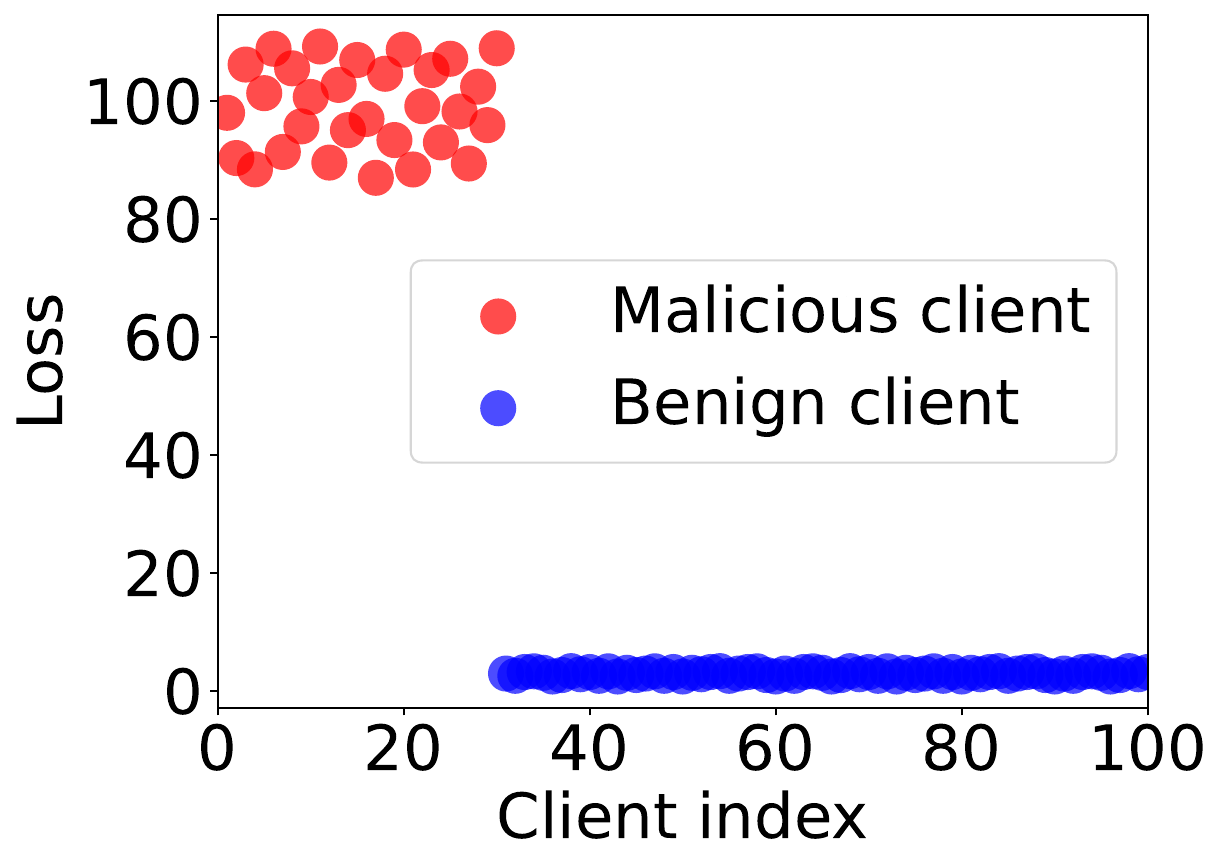}
    \caption{Scaling+DBA attack}
  \end{subfigure}
    \begin{subfigure}{0.163\textwidth}
    \includegraphics[width=\textwidth]{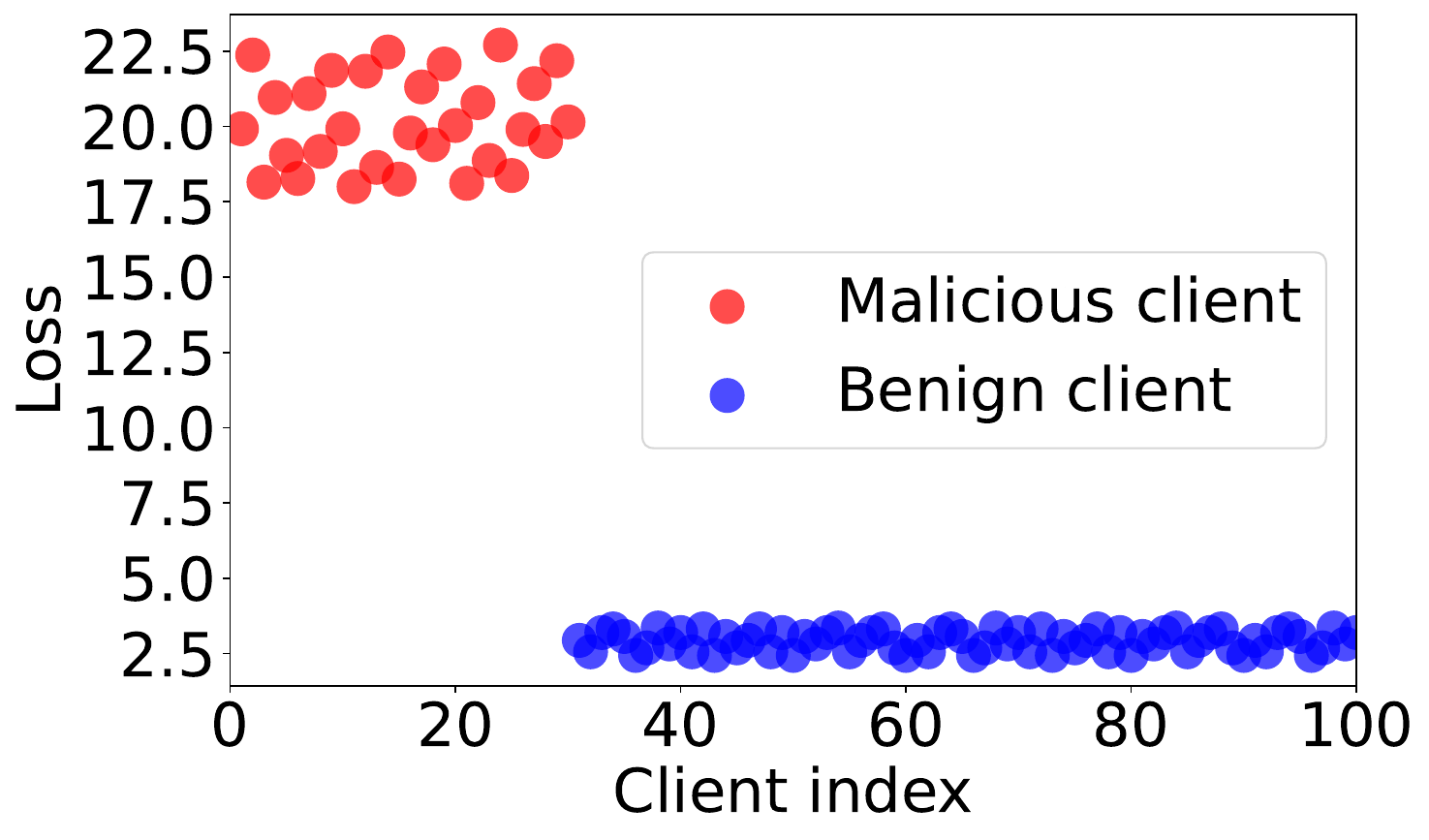}
    \caption{Adaptive attack}
  \end{subfigure}
  \caption{\textcolor{black}{The loss values of benign and malicious clients’ local models computed on the synthetic dataset, using \algFirst with the STL-10 dataset.}}
  \label{exp:safefl-ml-STL10}
\end{figure*}

\begin{figure*}[t]
  \centering
    \begin{subfigure}{0.163\textwidth}
    \includegraphics[width=\textwidth]{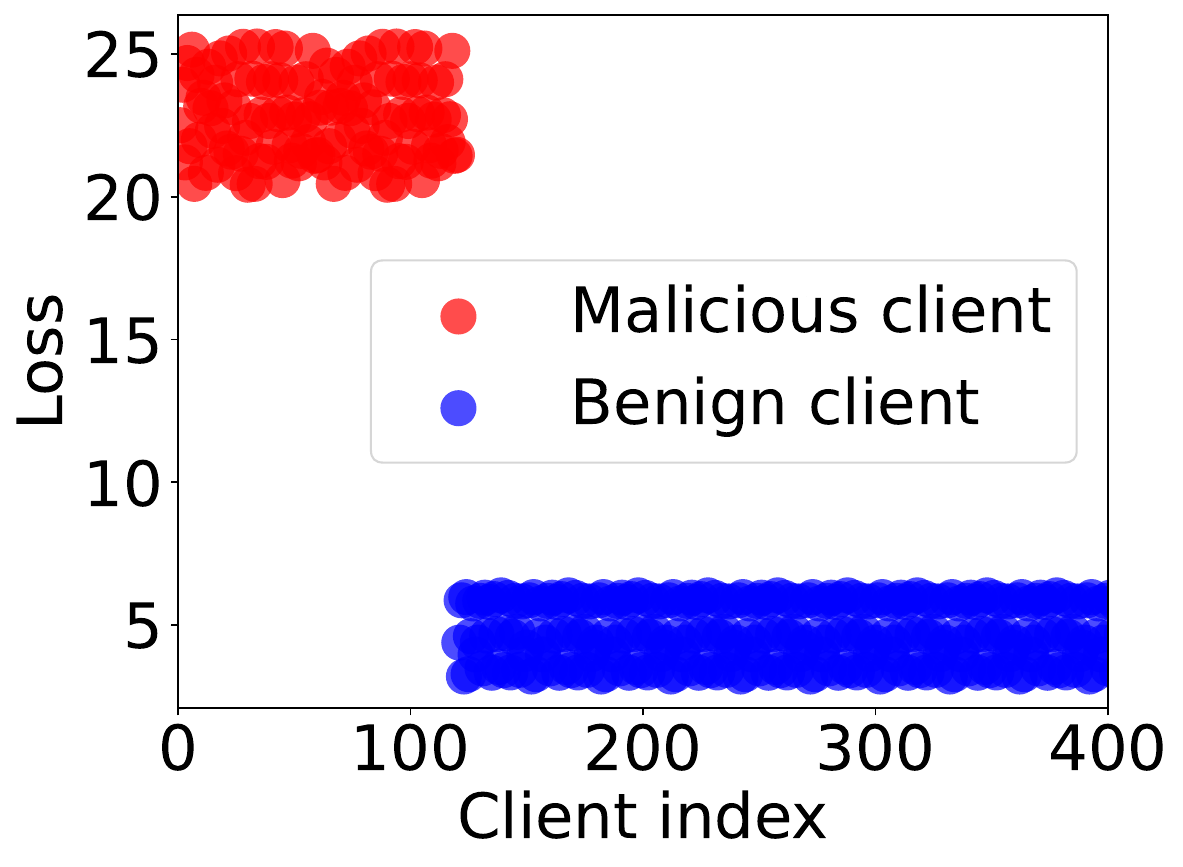}
    \caption{Trim attack}
  \end{subfigure}
    \begin{subfigure}{0.163\textwidth}
    \includegraphics[width=\textwidth]{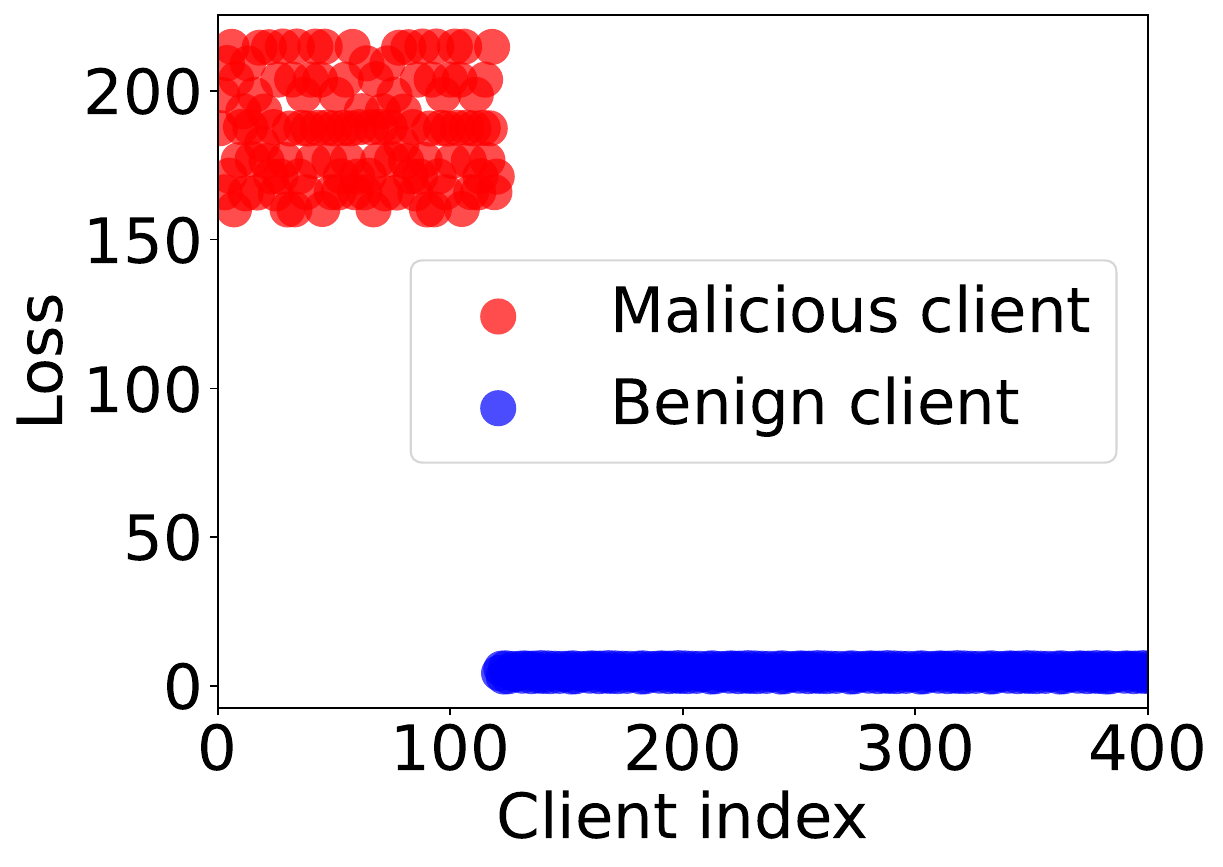}
    \caption{Scaling attack}
  \end{subfigure}
  \begin{subfigure}{0.163\textwidth}
    \includegraphics[width=\textwidth]{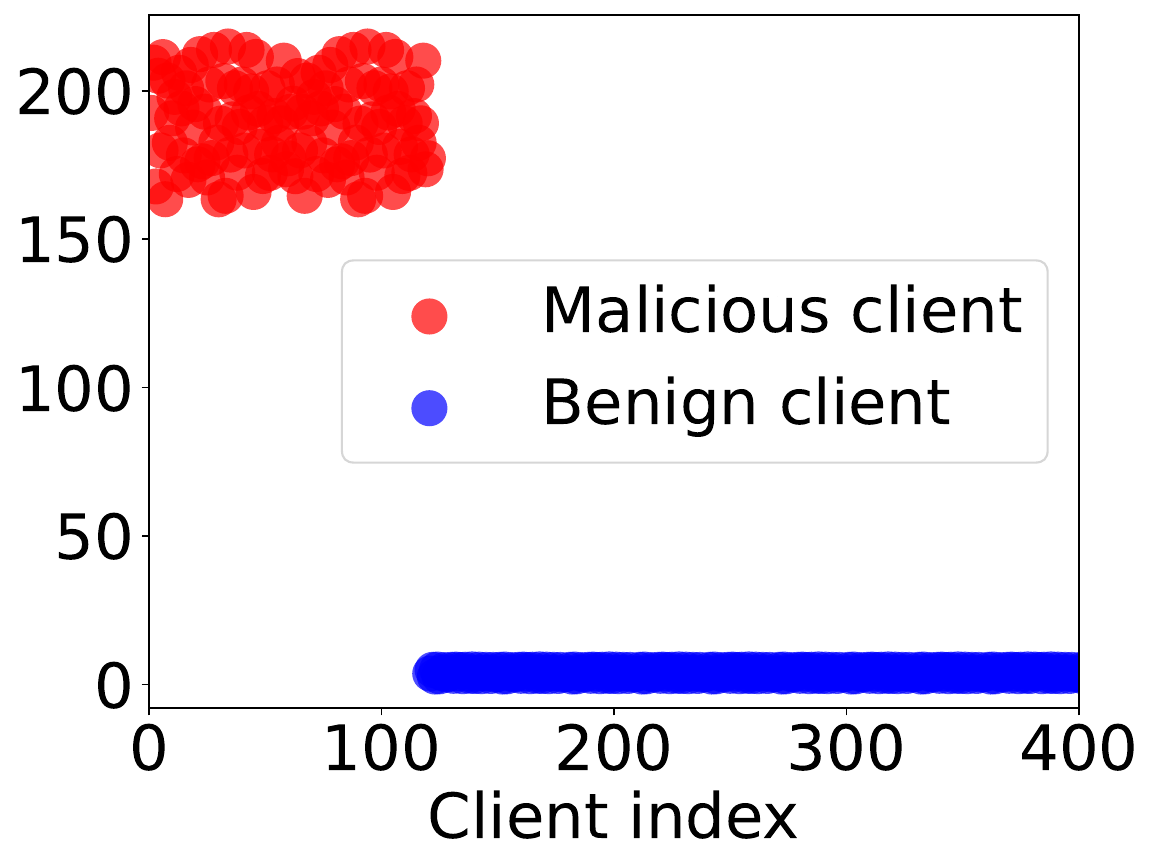}
    \caption{DBA attack}
  \end{subfigure}
  \begin{subfigure}{0.163\textwidth}
    \includegraphics[width=\textwidth]{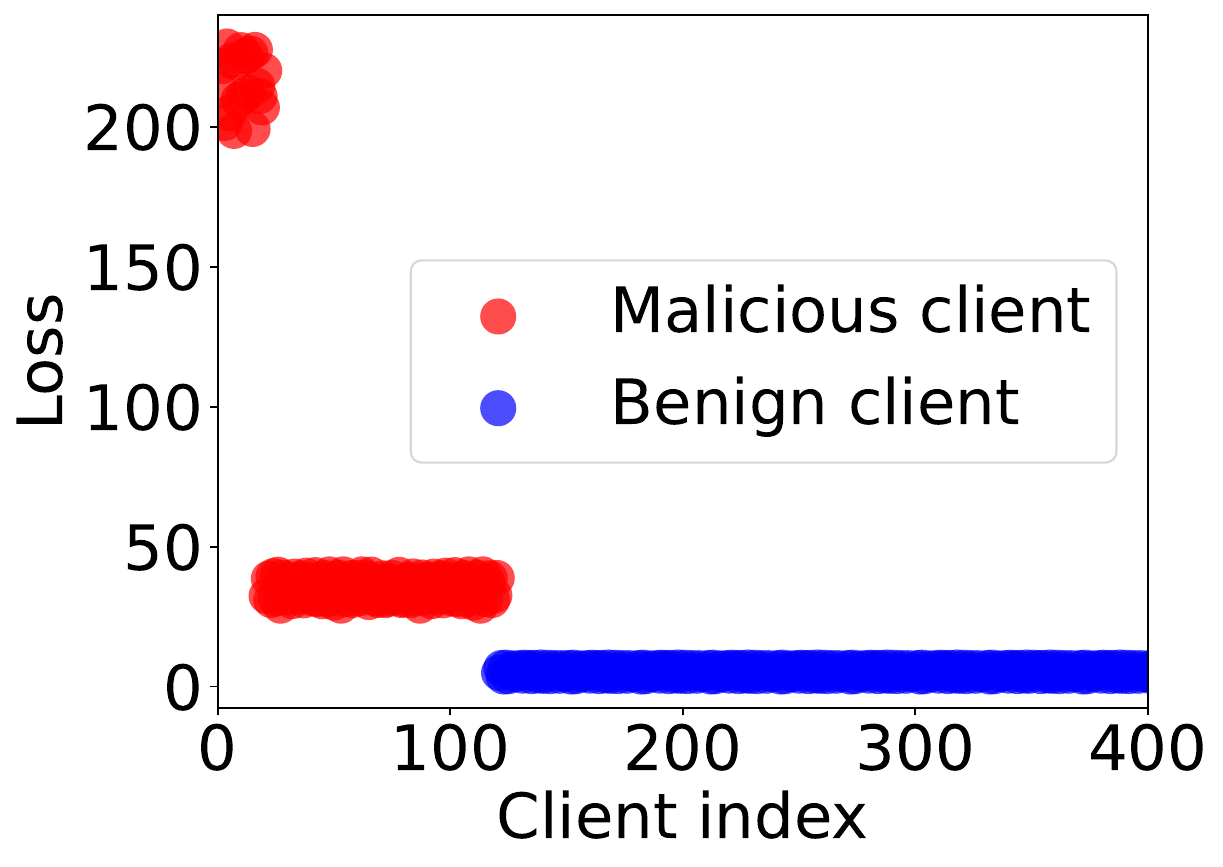}
    \caption{Trim+DBA attack}
  \end{subfigure}
    \begin{subfigure}{0.163\textwidth}
    \includegraphics[width=\textwidth]{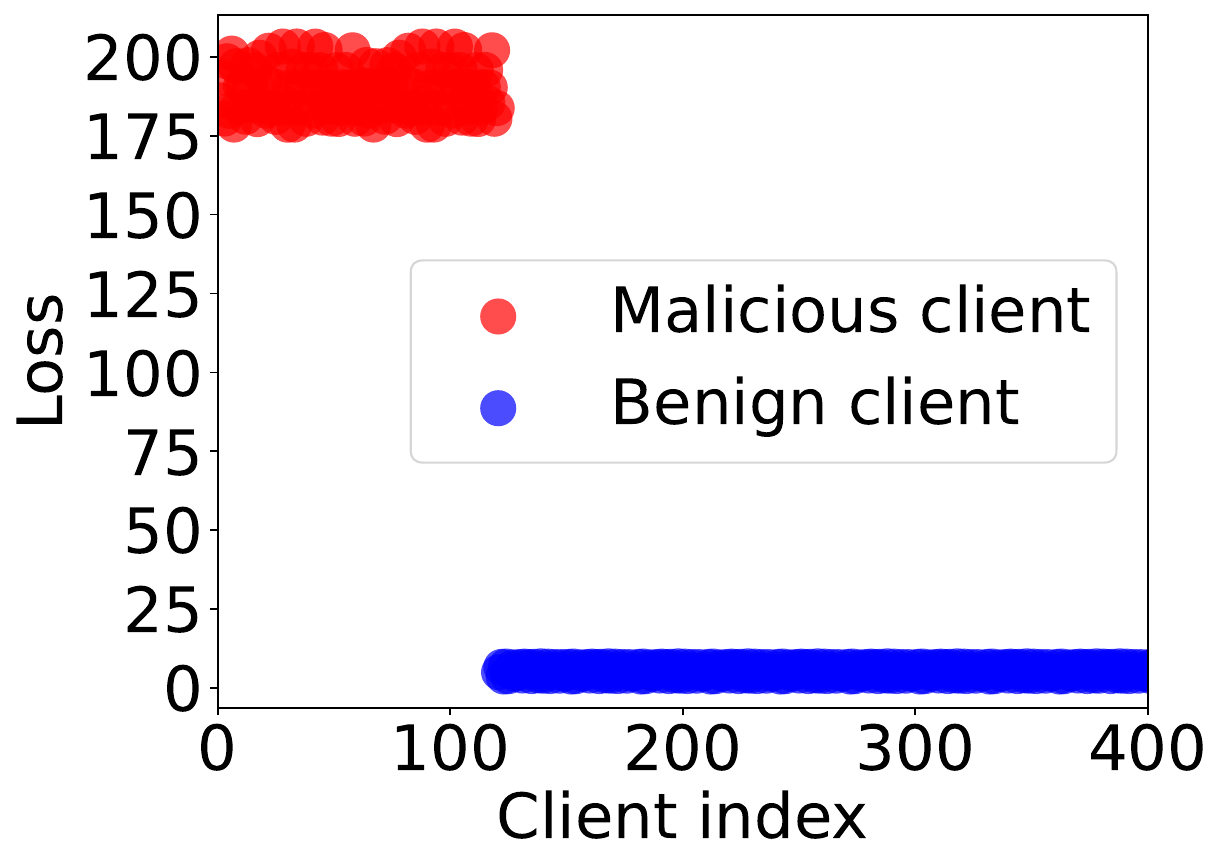}
    \caption{Scaling+DBA attack}
  \end{subfigure}
    \begin{subfigure}{0.163\textwidth}
    \includegraphics[width=\textwidth]{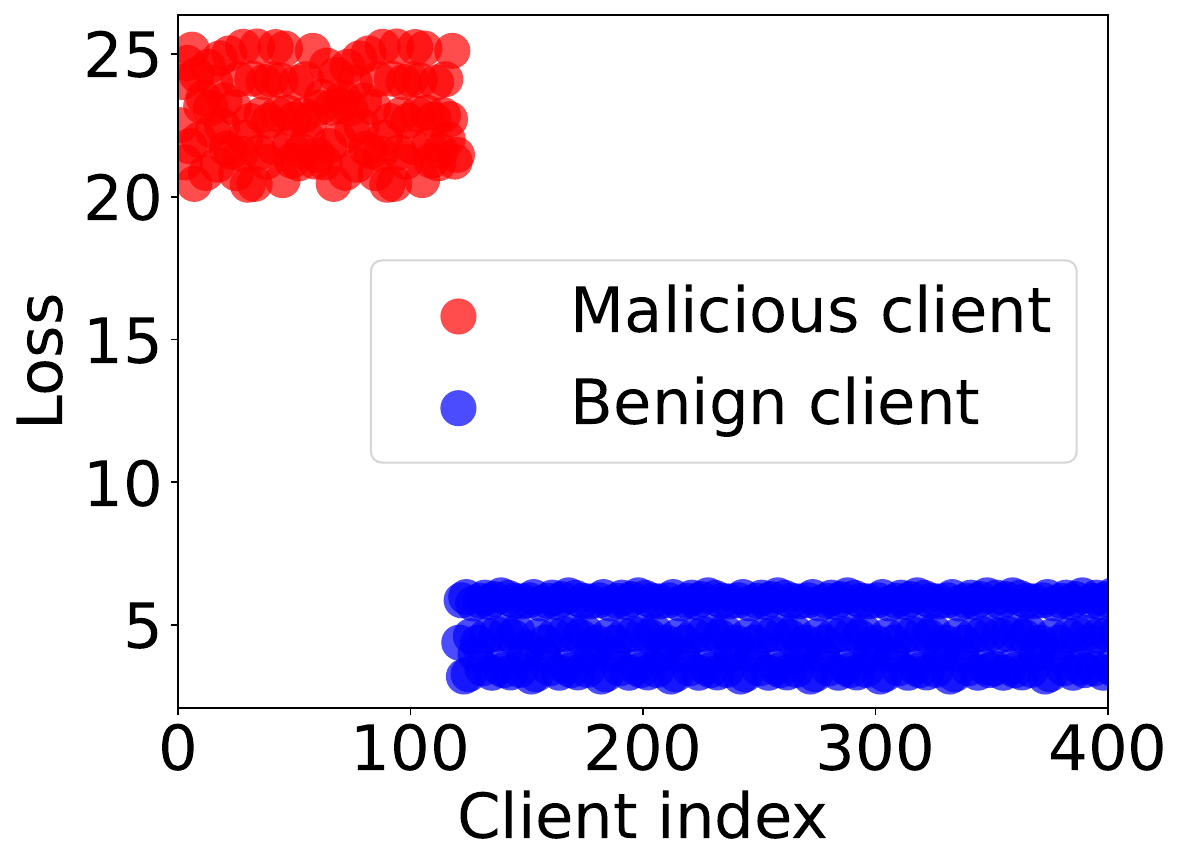}
    \caption{Adaptive attack}
  \end{subfigure}
  \caption{\textcolor{black}{The loss values of benign and malicious clients’ local models computed on the synthetic dataset, using \algFirst with the Tiny-ImageNet dataset.}}
  \label{exp:safefl-ml-Tiny}
\end{figure*}

\begin{figure*}[t]
  \centering
    \begin{subfigure}{0.163\textwidth}
    \includegraphics[width=\textwidth]{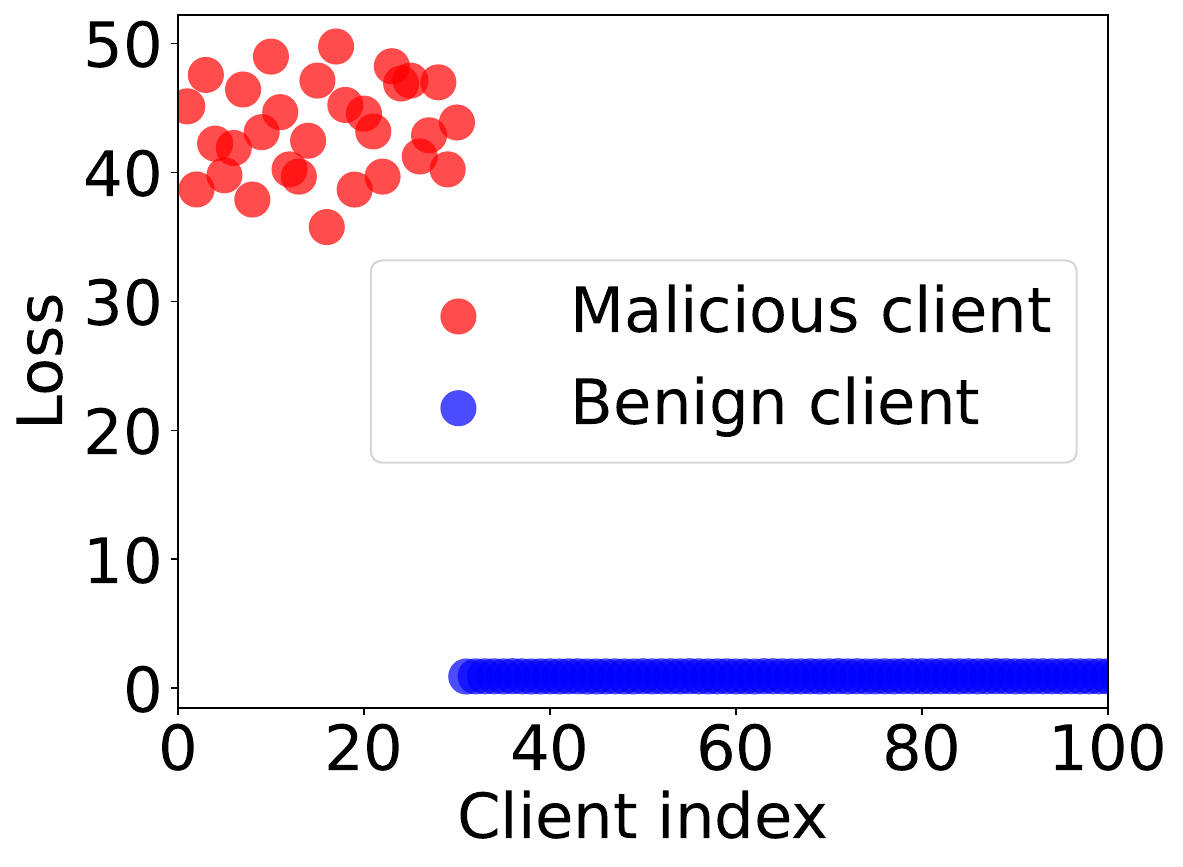}
    \caption{Trim attack}
  \end{subfigure}
    \begin{subfigure}{0.163\textwidth}
    \includegraphics[width=\textwidth]{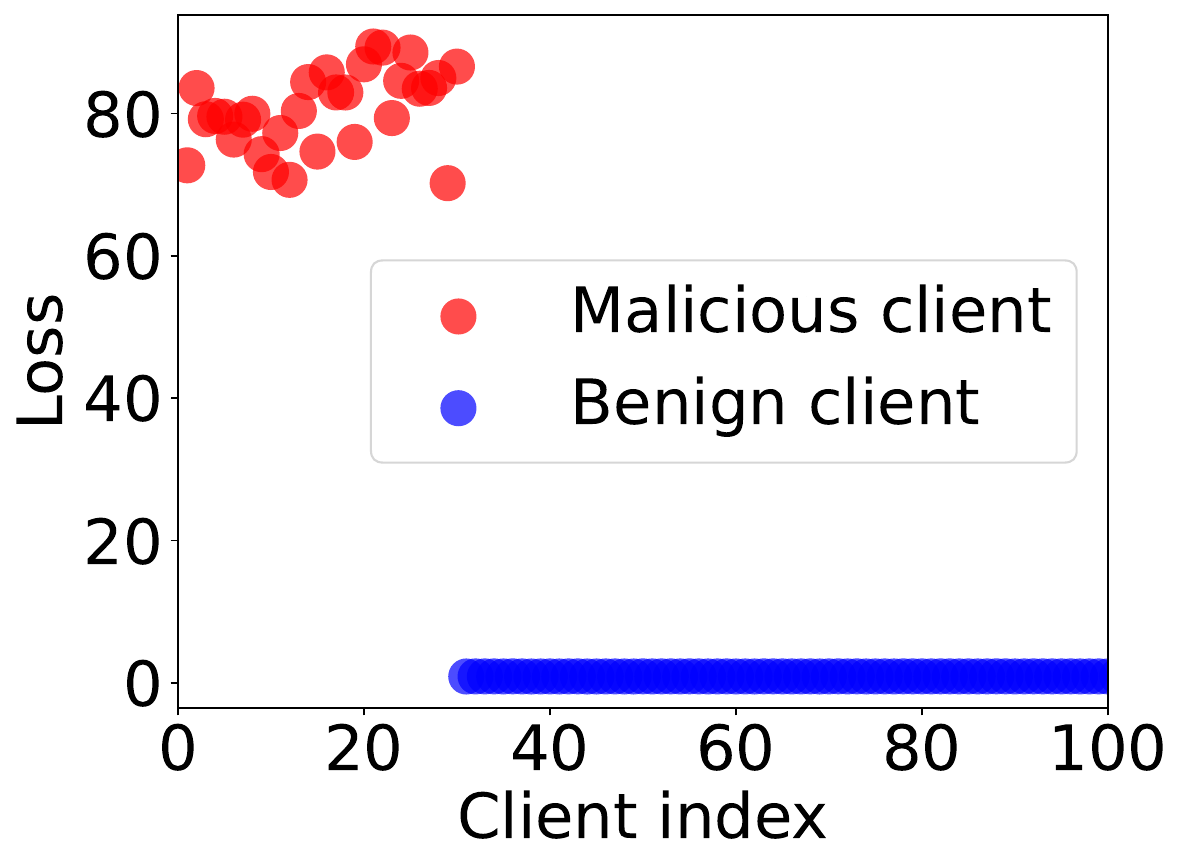}
    \caption{Scaling attack}
  \end{subfigure}
  \begin{subfigure}{0.163\textwidth}
    \includegraphics[width=\textwidth]{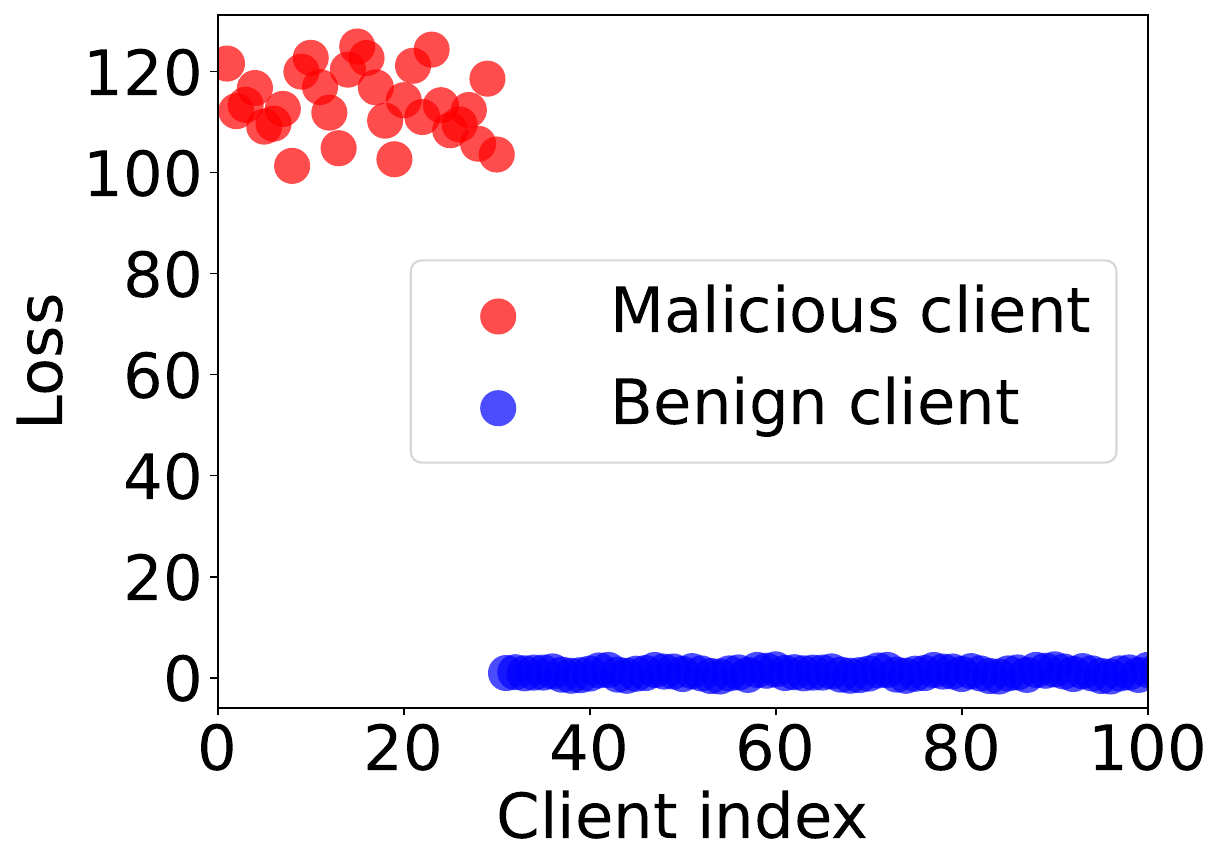}
    \caption{DBA attack}
  \end{subfigure}
  \begin{subfigure}{0.163\textwidth}
    \includegraphics[width=\textwidth]{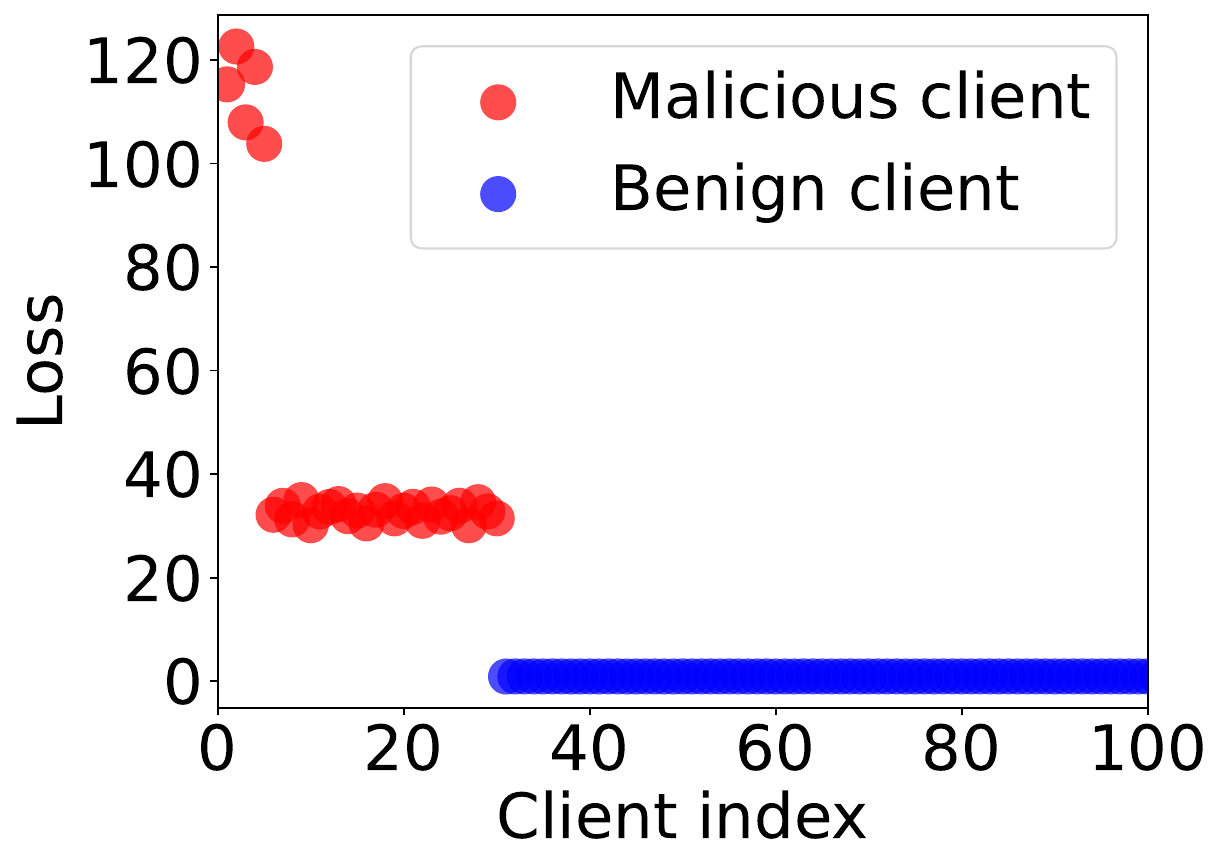}
    \caption{Trim+DBA attack}
  \end{subfigure}
    \begin{subfigure}{0.163\textwidth}
    \includegraphics[width=\textwidth]{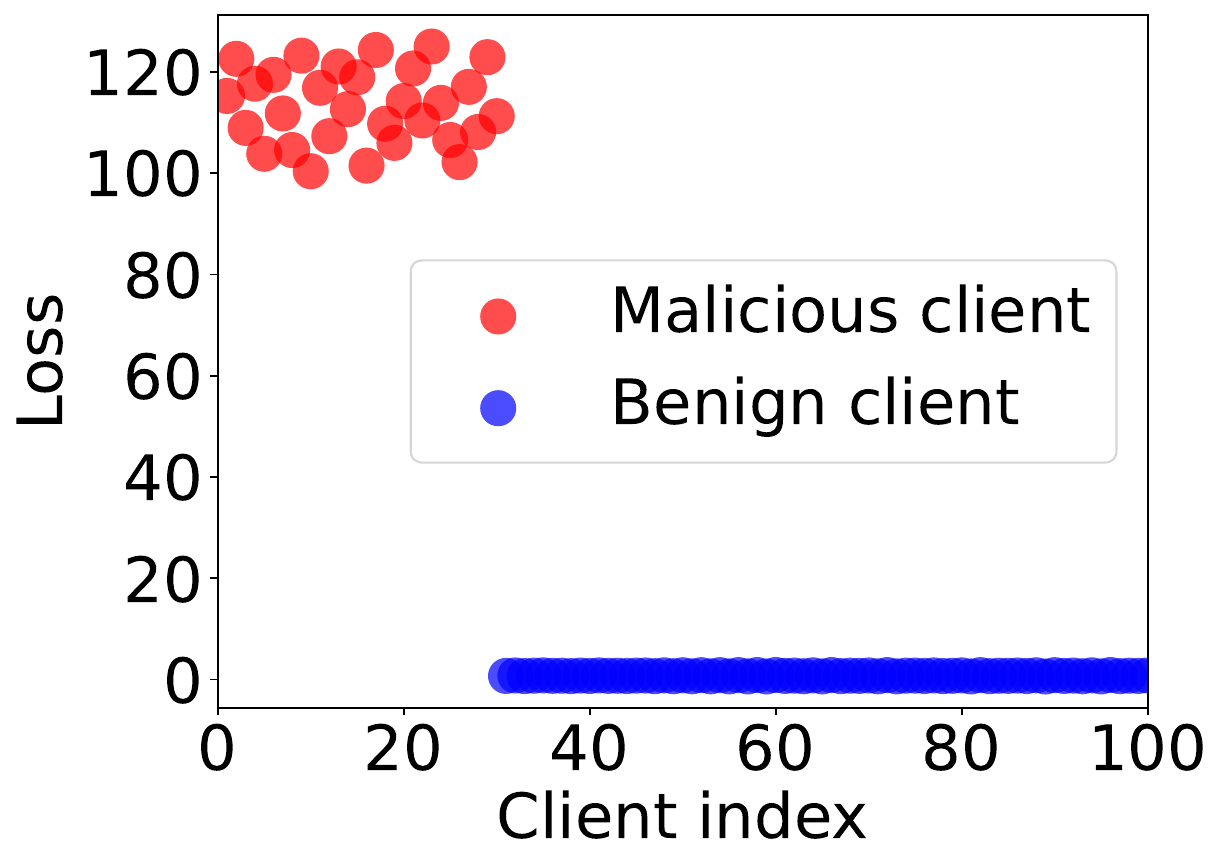}
    \caption{Scaling+DBA attack}
  \end{subfigure}
    \begin{subfigure}{0.163\textwidth}
    \includegraphics[width=\textwidth]{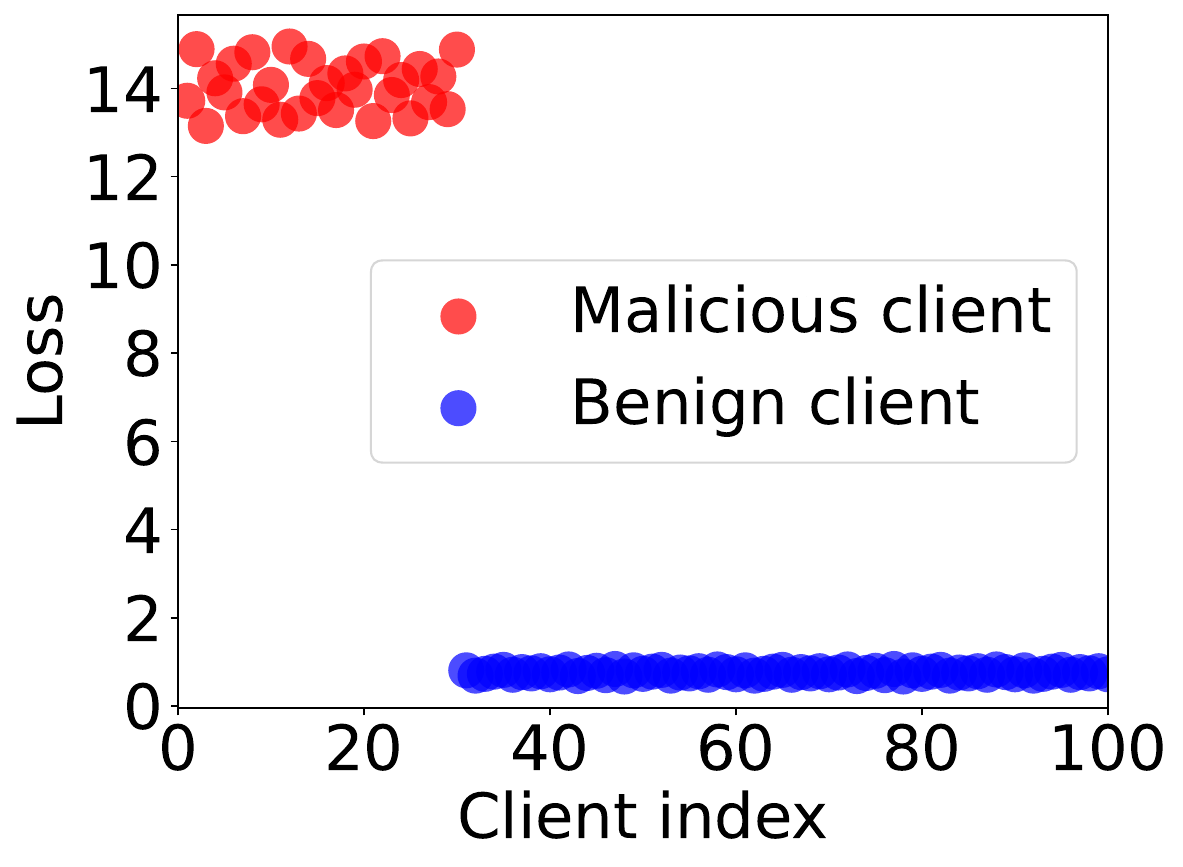}
    \caption{Adaptive attack}
  \end{subfigure}
  \caption{\textcolor{black}{The loss values of benign and malicious clients’ local models computed on the synthetic dataset, using \algSecond with the CIFAR-10 dataset.}}
  \label{exp:safefl-cl-cifar}
\end{figure*}

\begin{figure*}[t]
  \centering
    \begin{subfigure}{0.163\textwidth}
    \includegraphics[width=\textwidth]{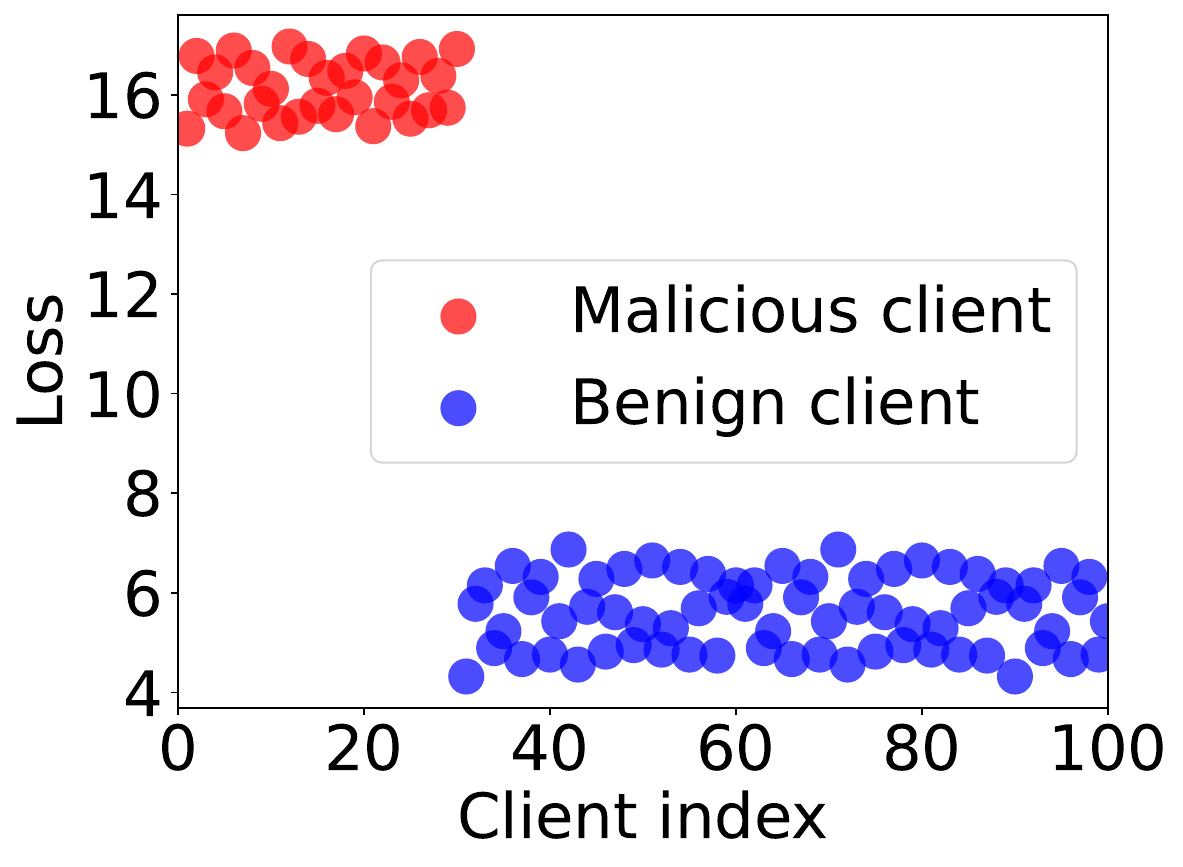}
    \caption{Trim attack}
  \end{subfigure}
    \begin{subfigure}{0.163\textwidth}
    \includegraphics[width=\textwidth]{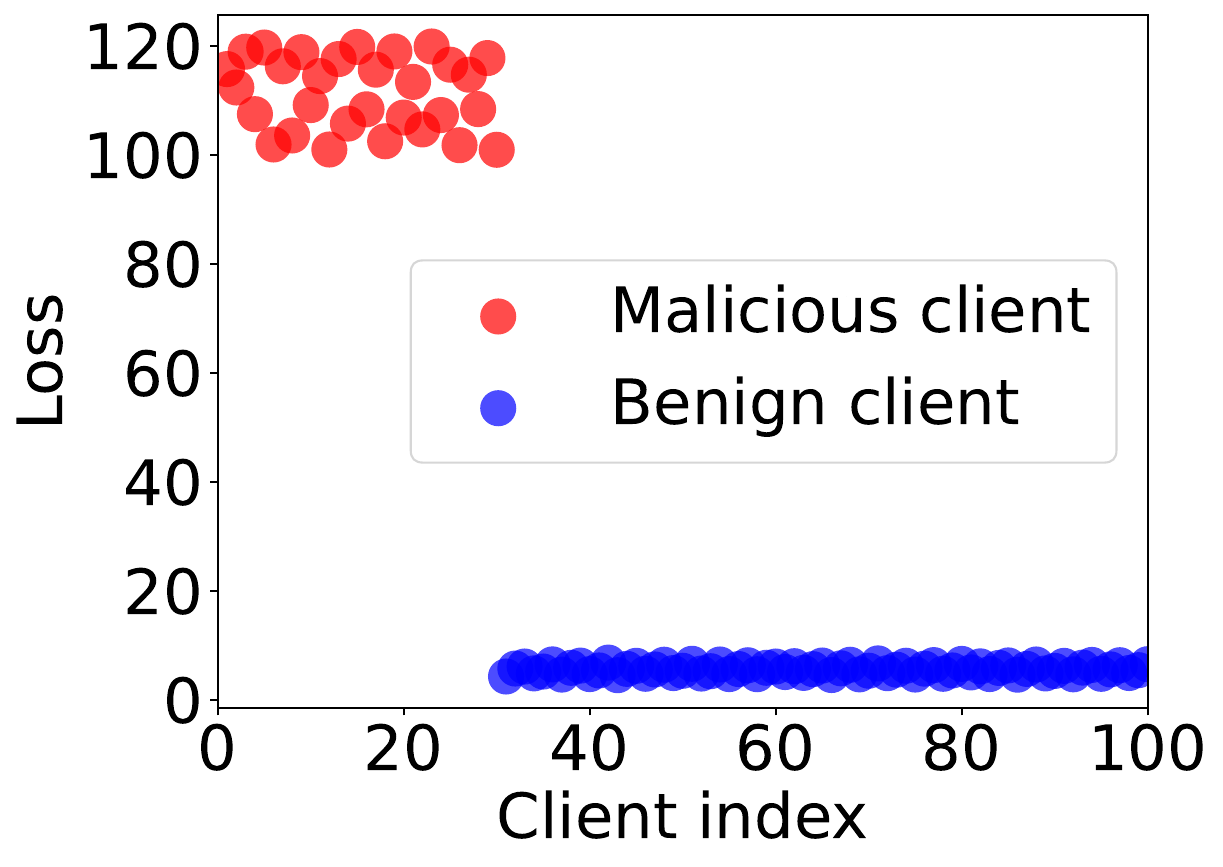}
    \caption{Scaling attack}
  \end{subfigure}
  \begin{subfigure}{0.163\textwidth}
    \includegraphics[width=\textwidth]{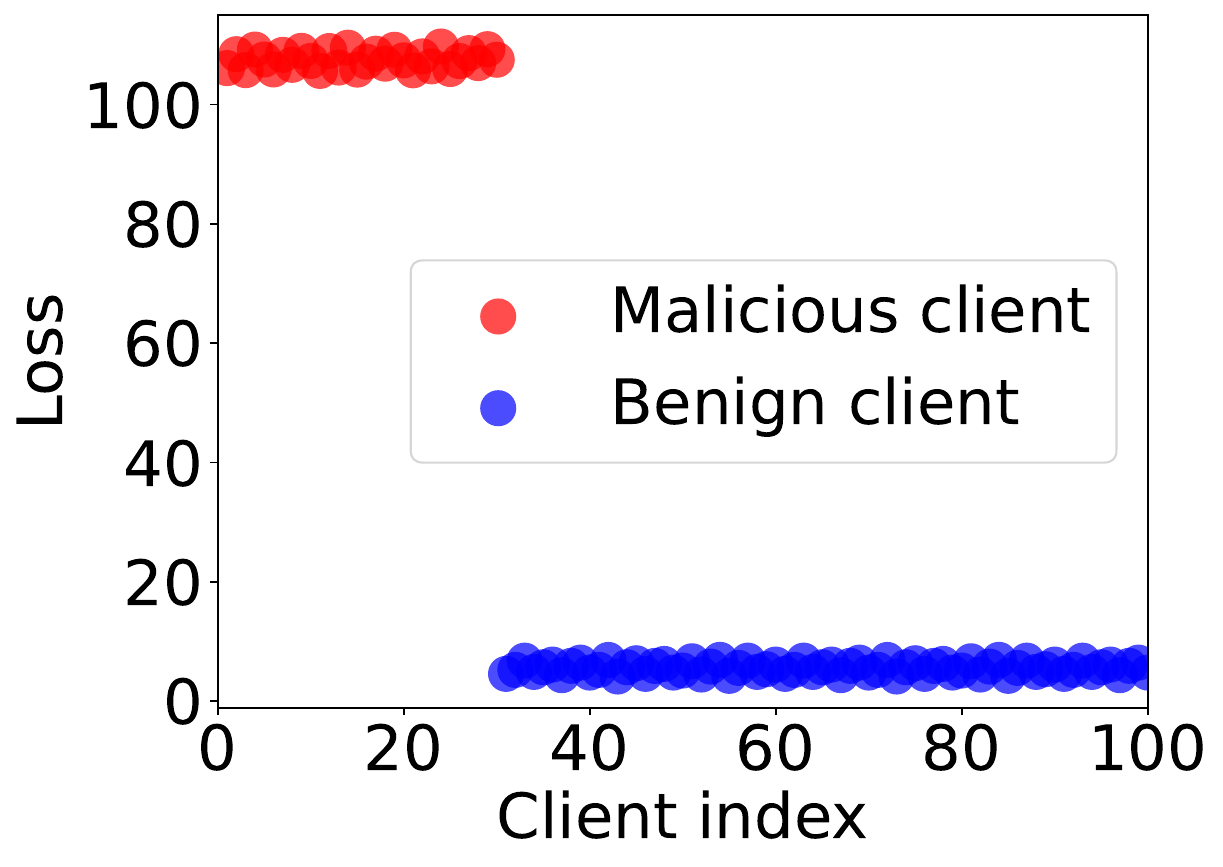}
    \caption{DBA attack}
  \end{subfigure}
  \begin{subfigure}{0.163\textwidth}
    \includegraphics[width=\textwidth]{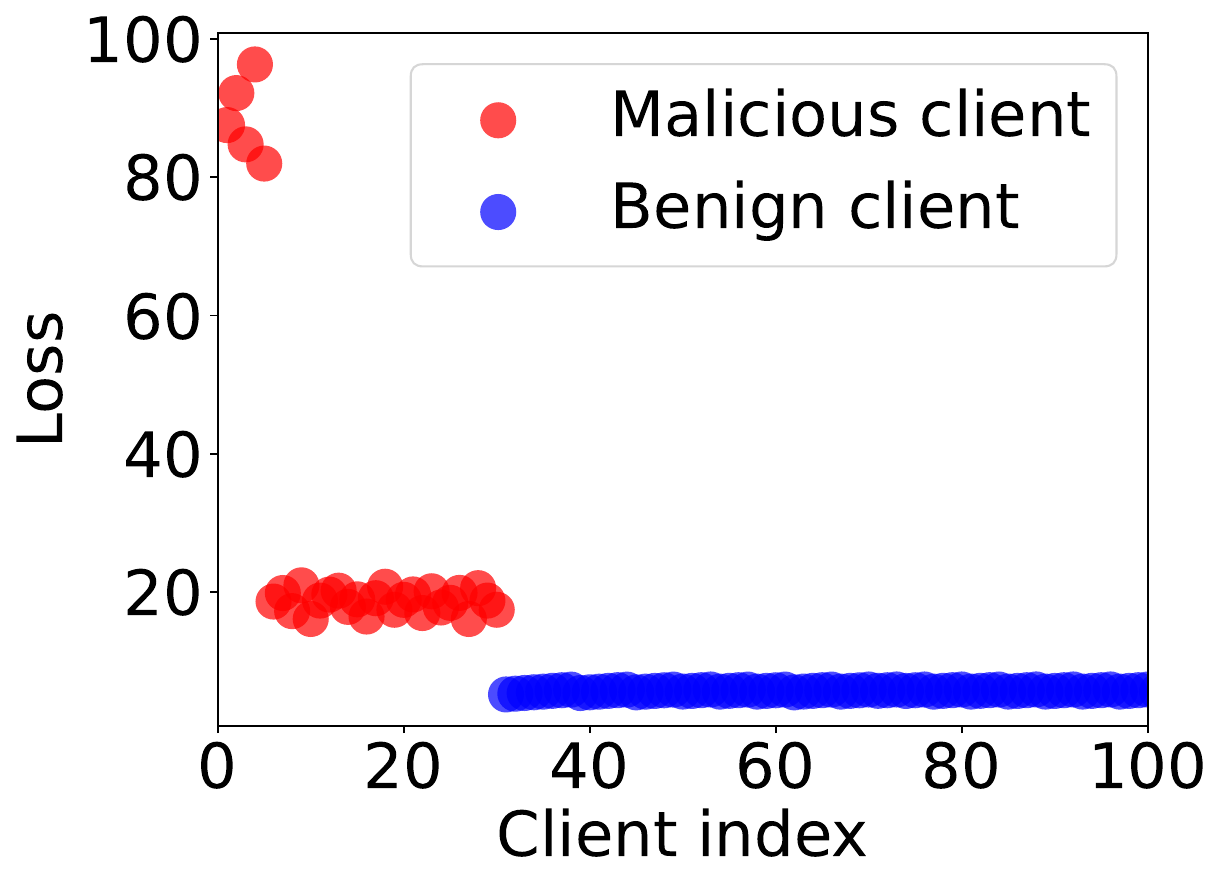}
    \caption{Trim+DBA attack}
  \end{subfigure}
    \begin{subfigure}{0.163\textwidth}
    \includegraphics[width=\textwidth]{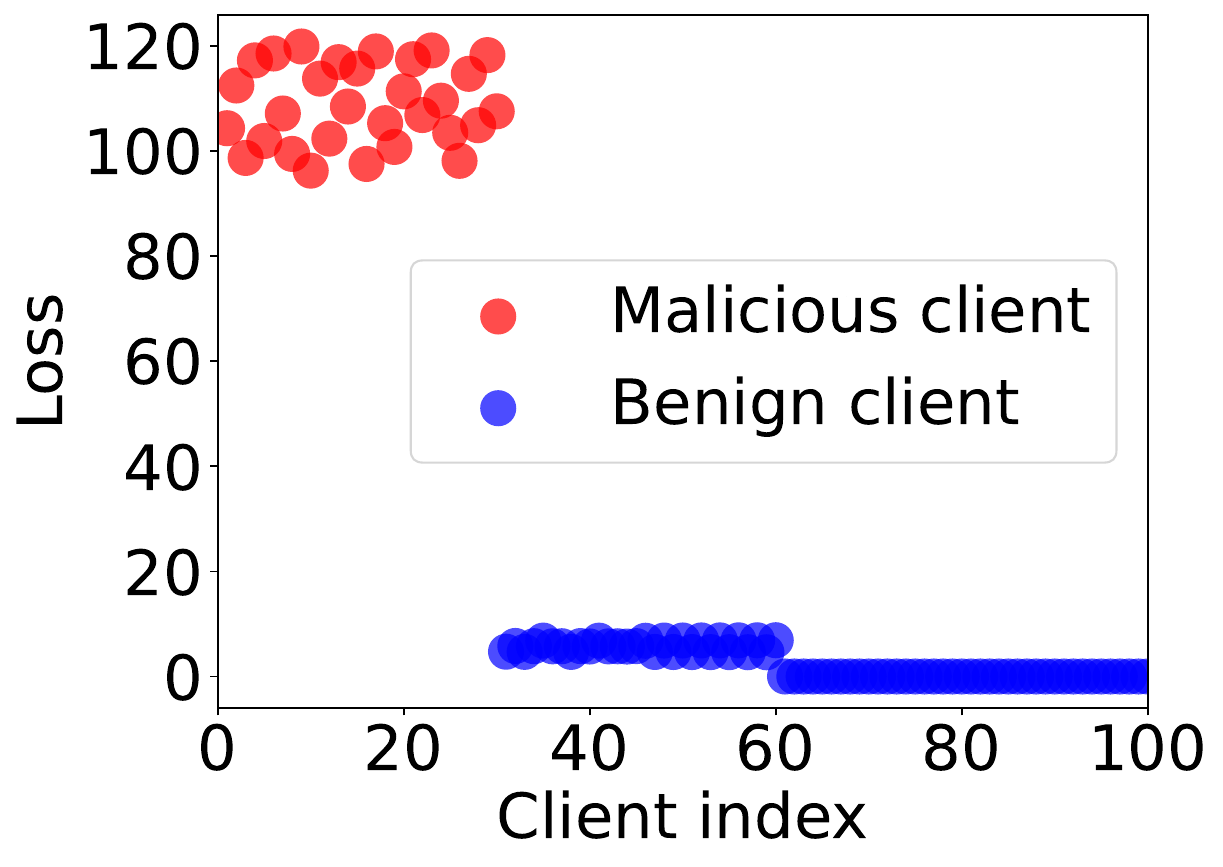}
    \caption{Scaling+DBA attack}
  \end{subfigure}
    \begin{subfigure}{0.163\textwidth}
    \includegraphics[width=\textwidth]{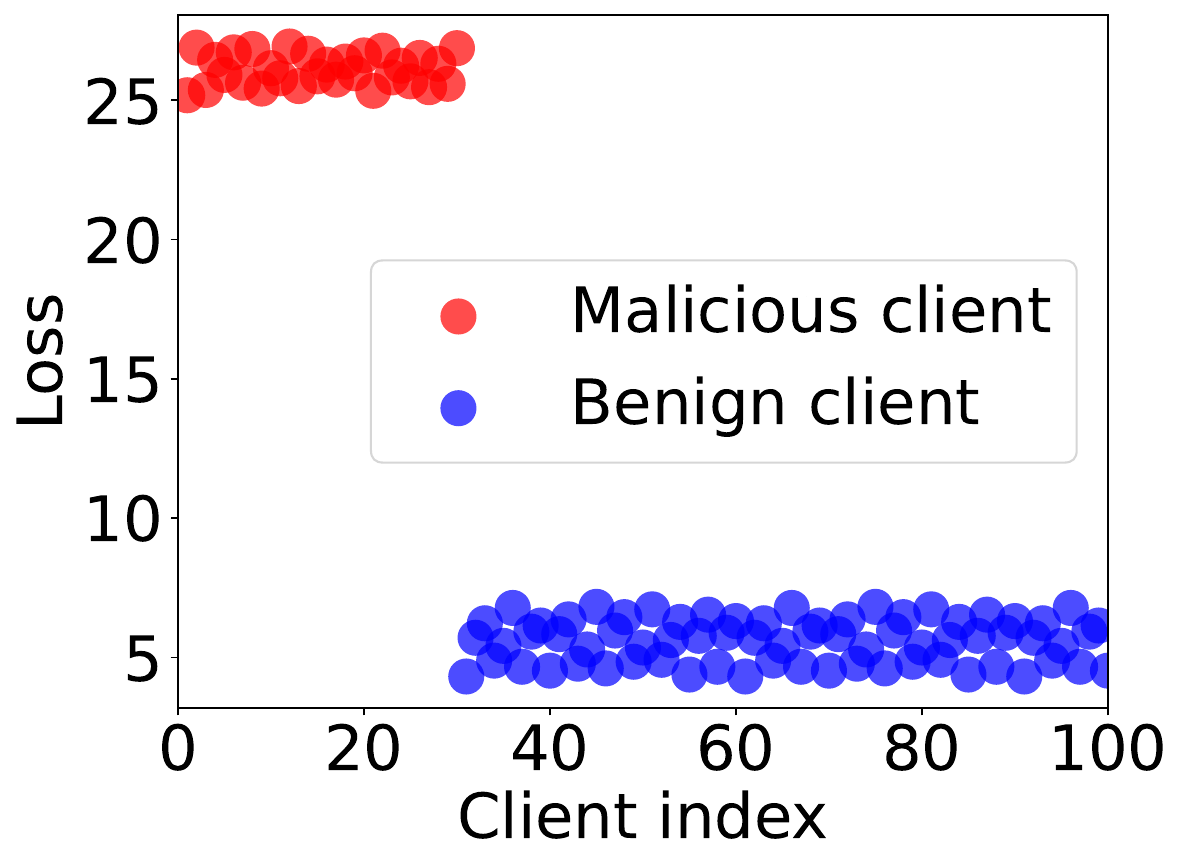}
    \caption{Adaptive attack}
  \end{subfigure}
  \caption{\textcolor{black}{The loss values of benign and malicious clients’ local models computed on the synthetic dataset, using \algSecond with the MNIST dataset.}}
  \label{exp:safefl-cl-mnist}
\end{figure*}

\begin{figure*}[!t]
  \centering
    \begin{subfigure}{0.163\textwidth}
    \includegraphics[width=\textwidth]{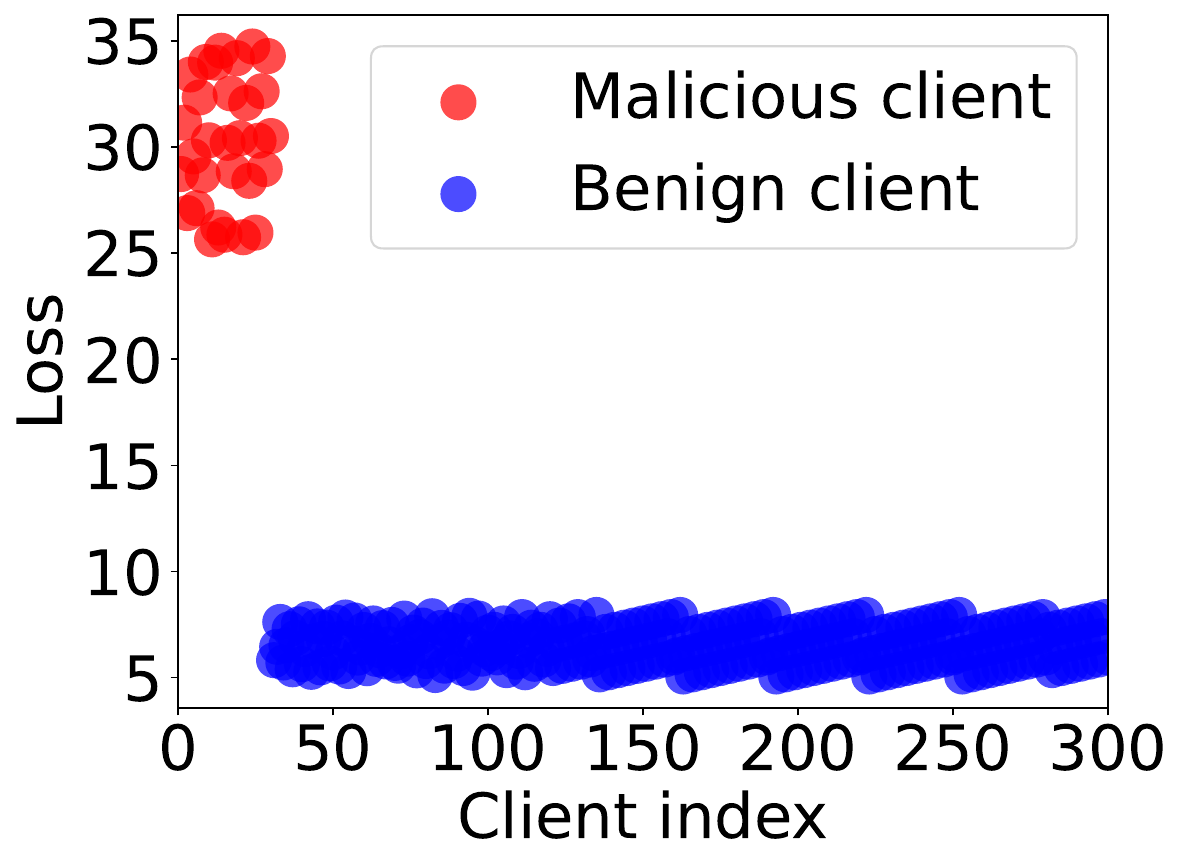}
    \caption{Trim attack}
  \end{subfigure}
    \begin{subfigure}{0.163\textwidth}
    \includegraphics[width=\textwidth]{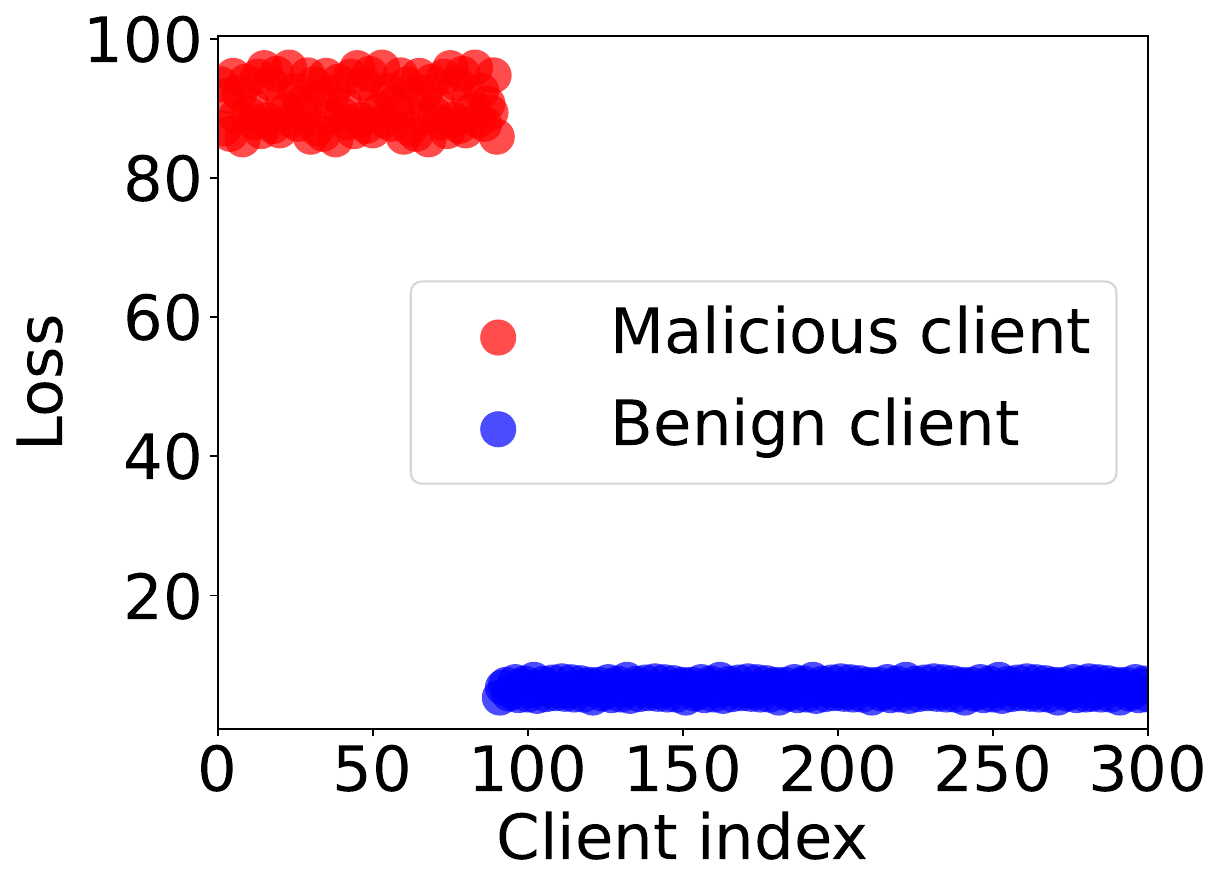}
    \caption{Scaling attack}
  \end{subfigure}
  \begin{subfigure}{0.163\textwidth}
    \includegraphics[width=\textwidth]{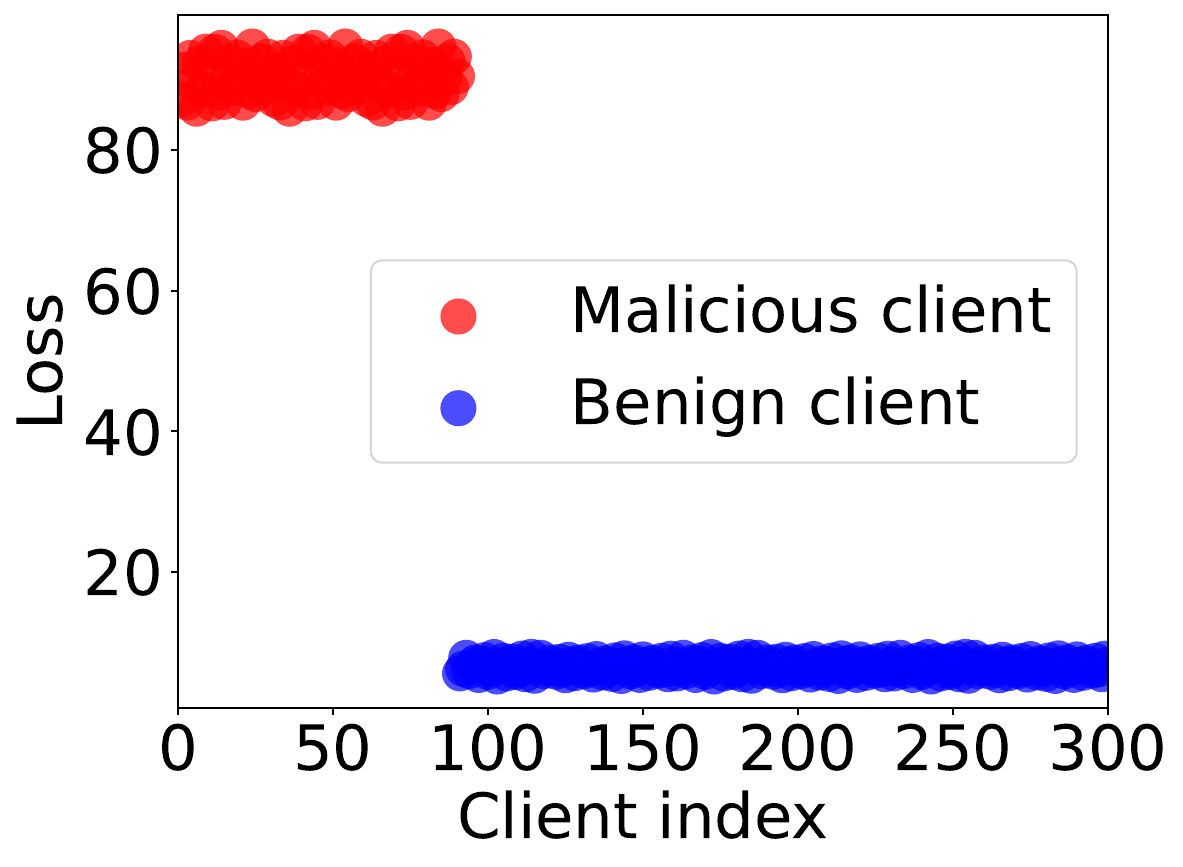}
    \caption{DBA attack}
  \end{subfigure}
  \begin{subfigure}{0.163\textwidth}
    \includegraphics[width=\textwidth]{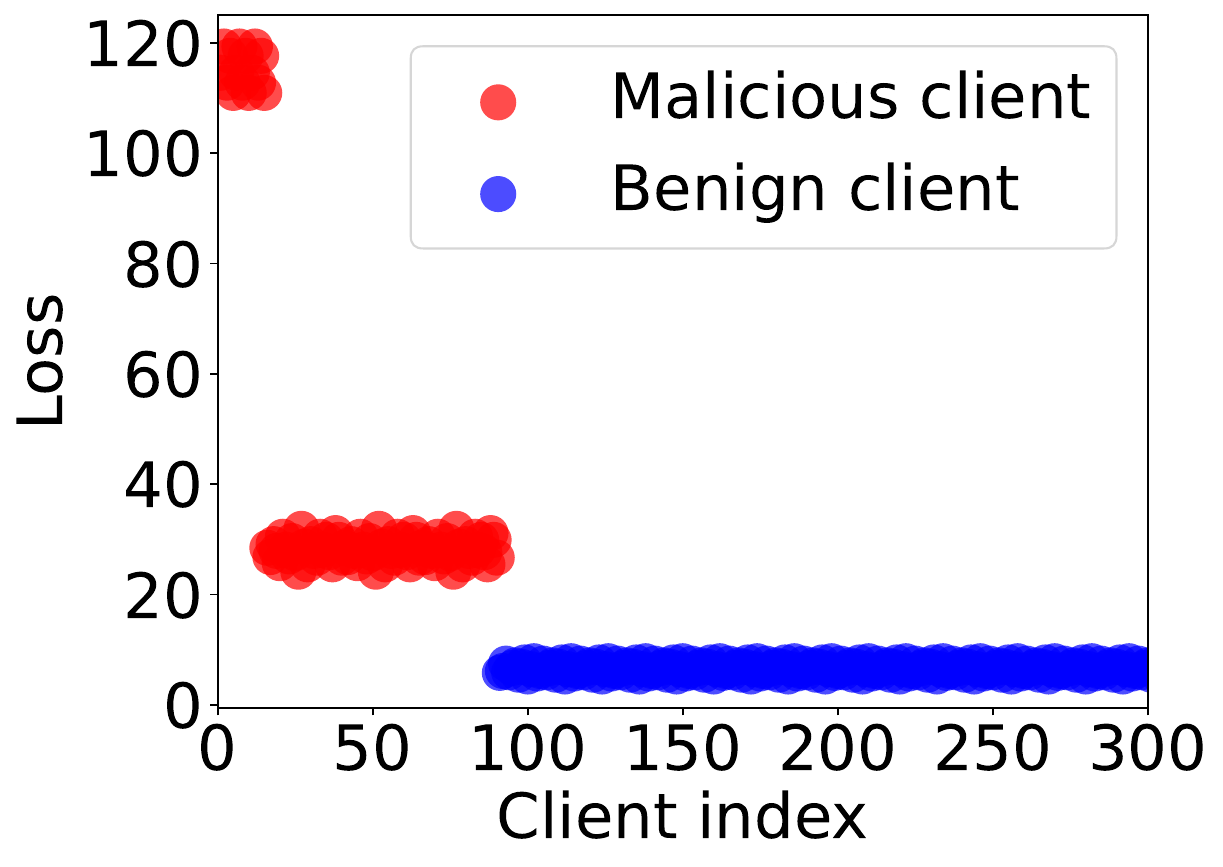}
    \caption{Trim+DBA attack}
  \end{subfigure}
    \begin{subfigure}{0.163\textwidth}
    \includegraphics[width=\textwidth]{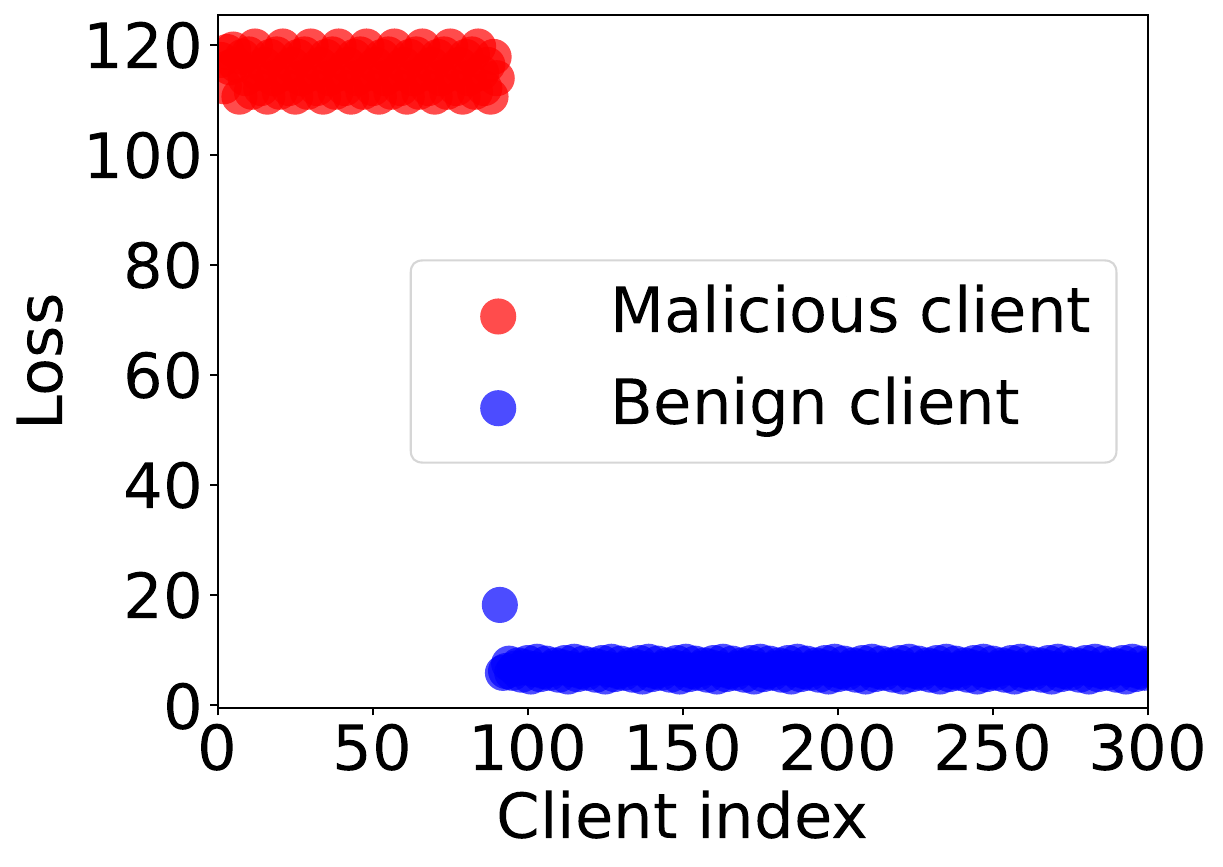}
    \caption{Scaling+DBA attack}
  \end{subfigure}
    \begin{subfigure}{0.163\textwidth}
    \includegraphics[width=\textwidth]{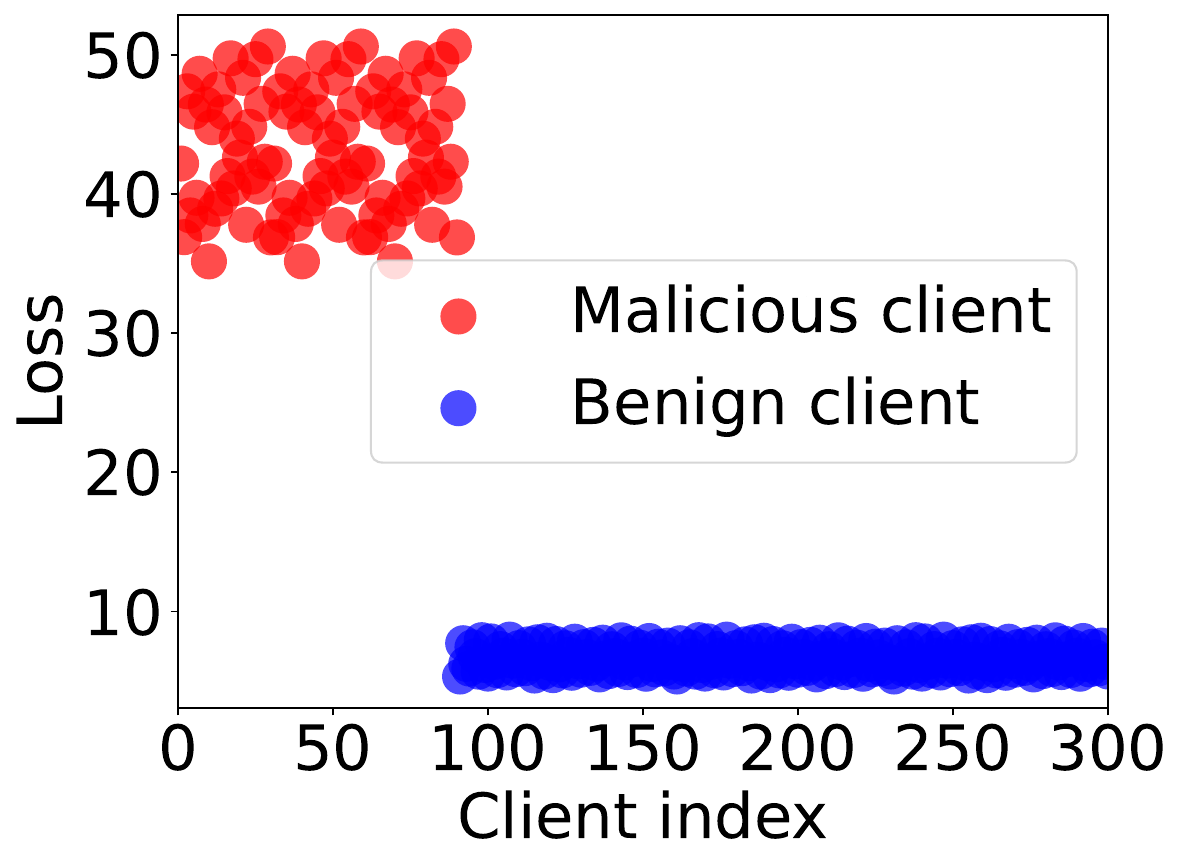}
    \caption{Adaptive attack}
  \end{subfigure}
  \caption{\textcolor{black}{The loss values of benign and malicious clients’ local models computed on the synthetic dataset, using \algSecond with the FEMNIST dataset.}}
  \label{exp:safefl-cl-femnist}
\end{figure*}

\begin{figure*}[!t]
  \centering
    \begin{subfigure}{0.163\textwidth}
    \includegraphics[width=\textwidth]{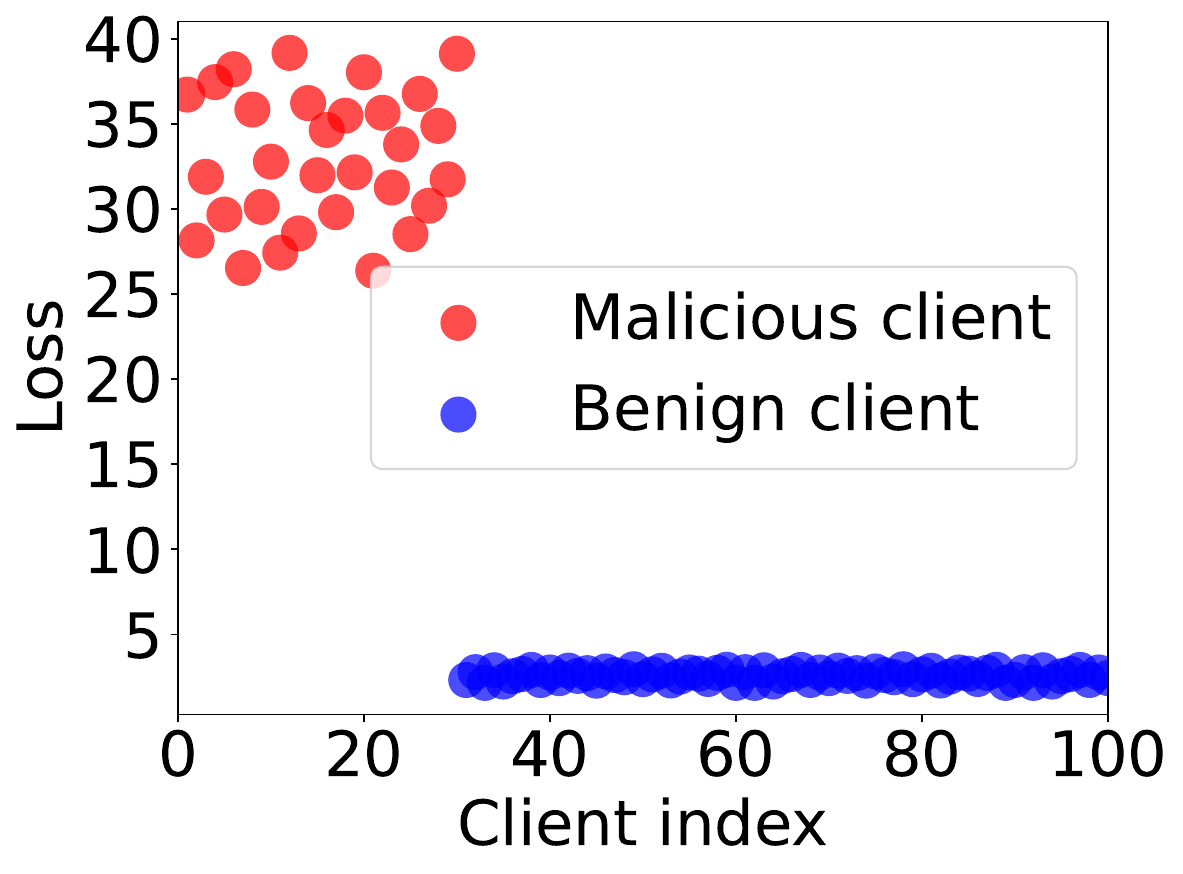}
    \caption{Trim attack}
  \end{subfigure}
    \begin{subfigure}{0.163\textwidth}
    \includegraphics[width=\textwidth]{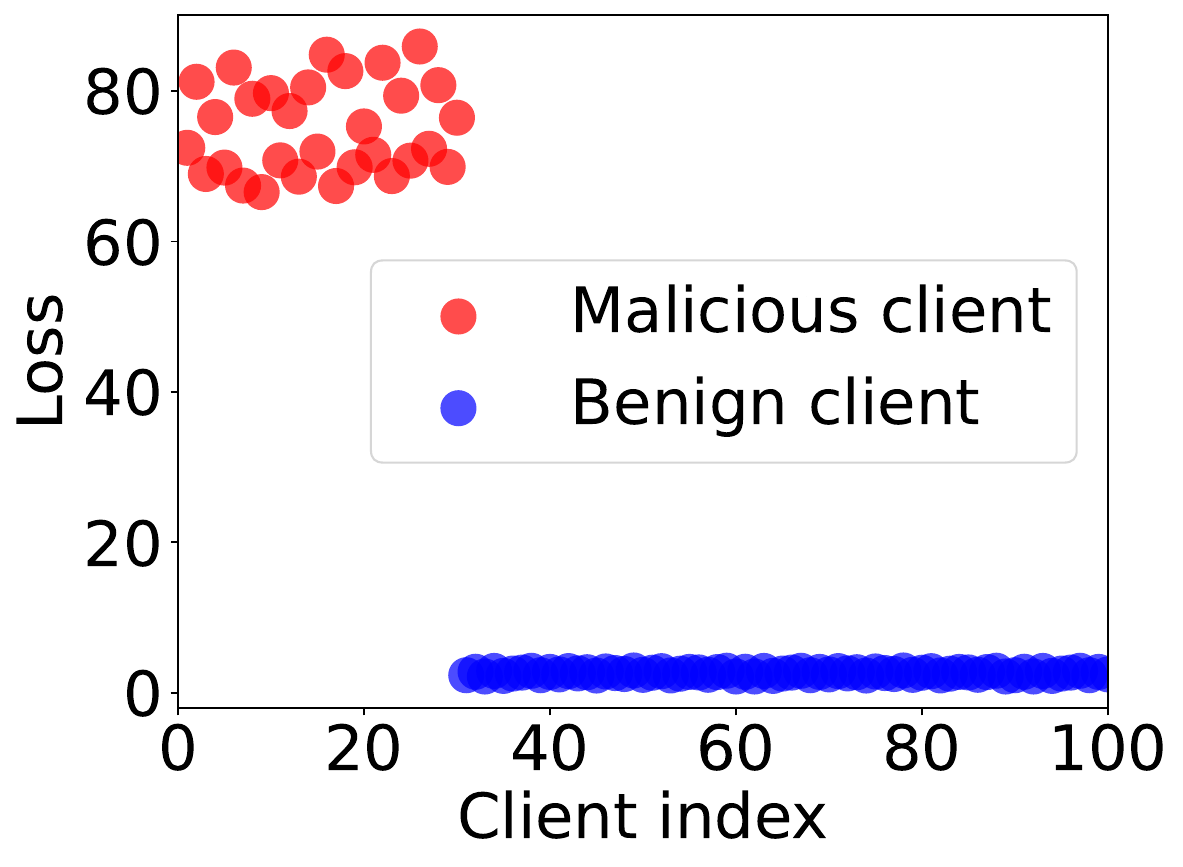}
    \caption{Scaling attack}
  \end{subfigure}
  \begin{subfigure}{0.163\textwidth}
    \includegraphics[width=\textwidth]{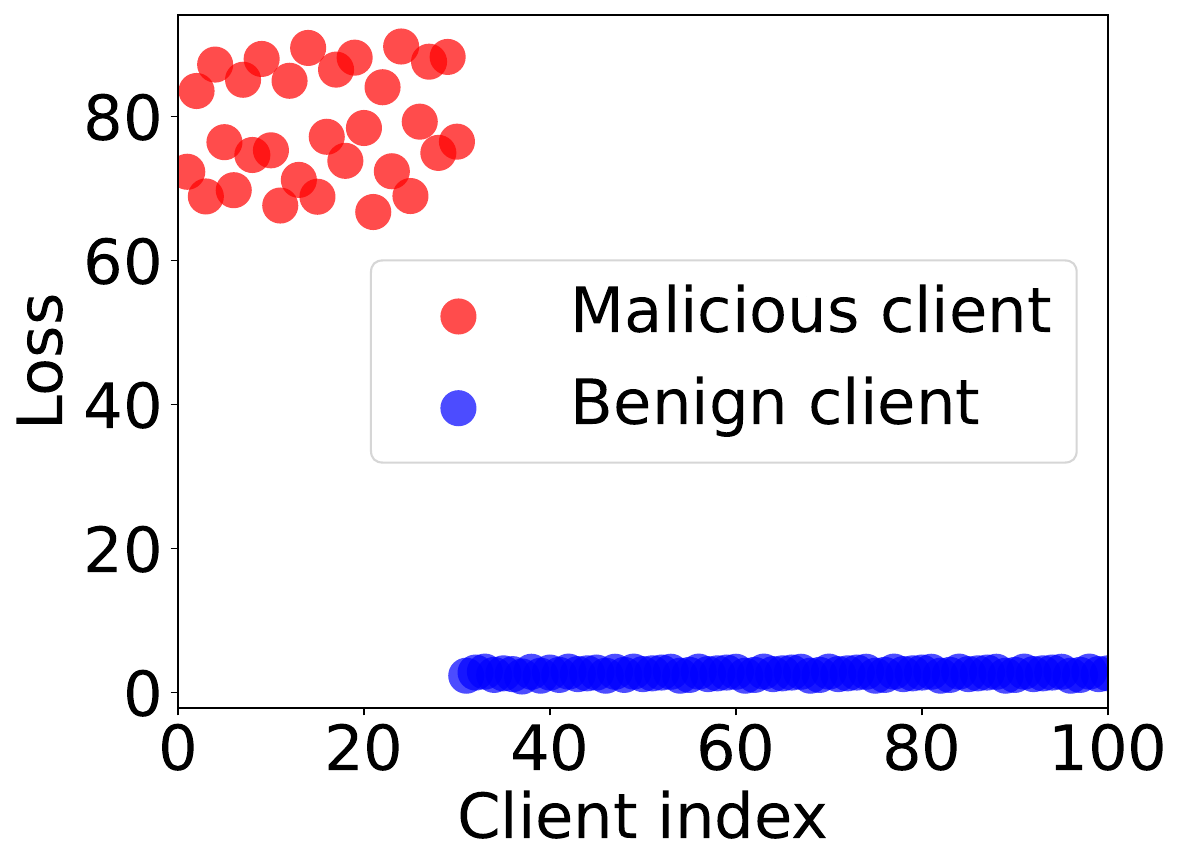}
    \caption{DBA attack}
  \end{subfigure}
  \begin{subfigure}{0.163\textwidth}
    \includegraphics[width=\textwidth]{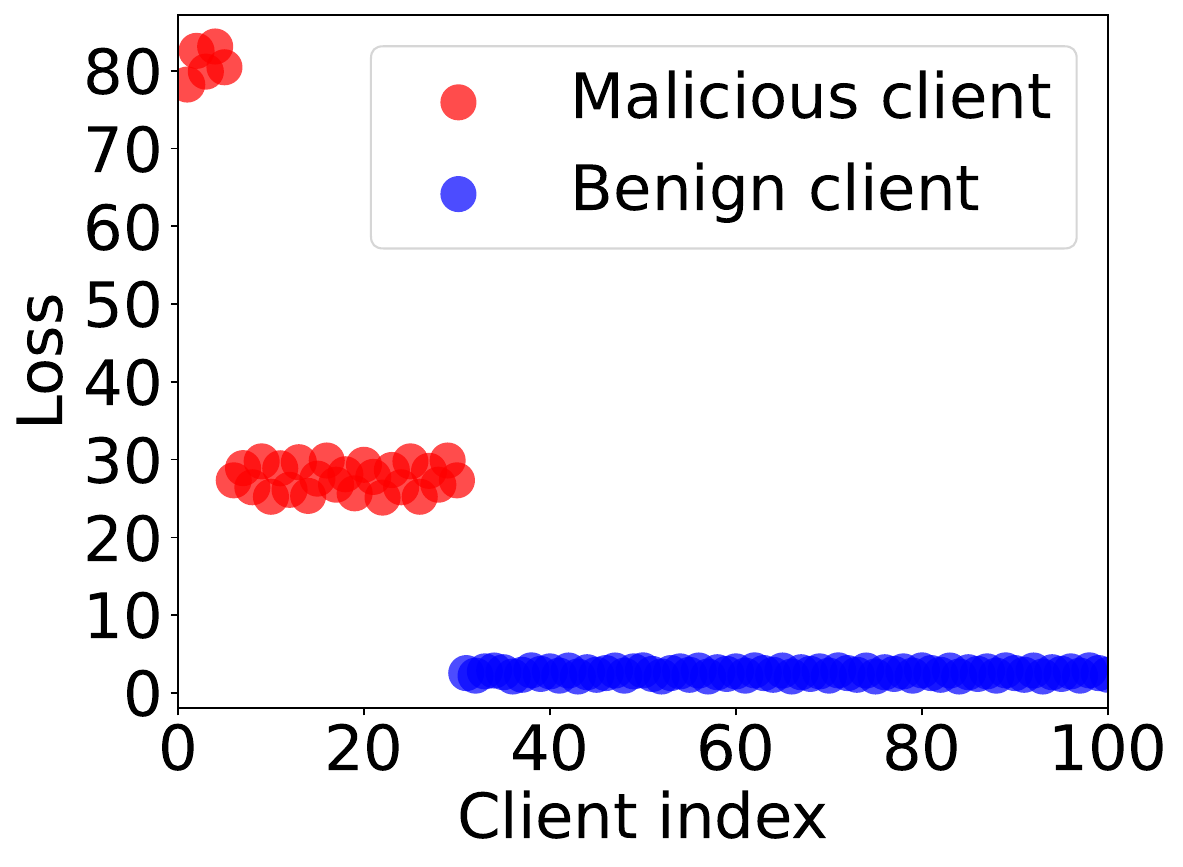}
    \caption{Trim+DBA attack}
  \end{subfigure}
    \begin{subfigure}{0.163\textwidth}
    \includegraphics[width=\textwidth]{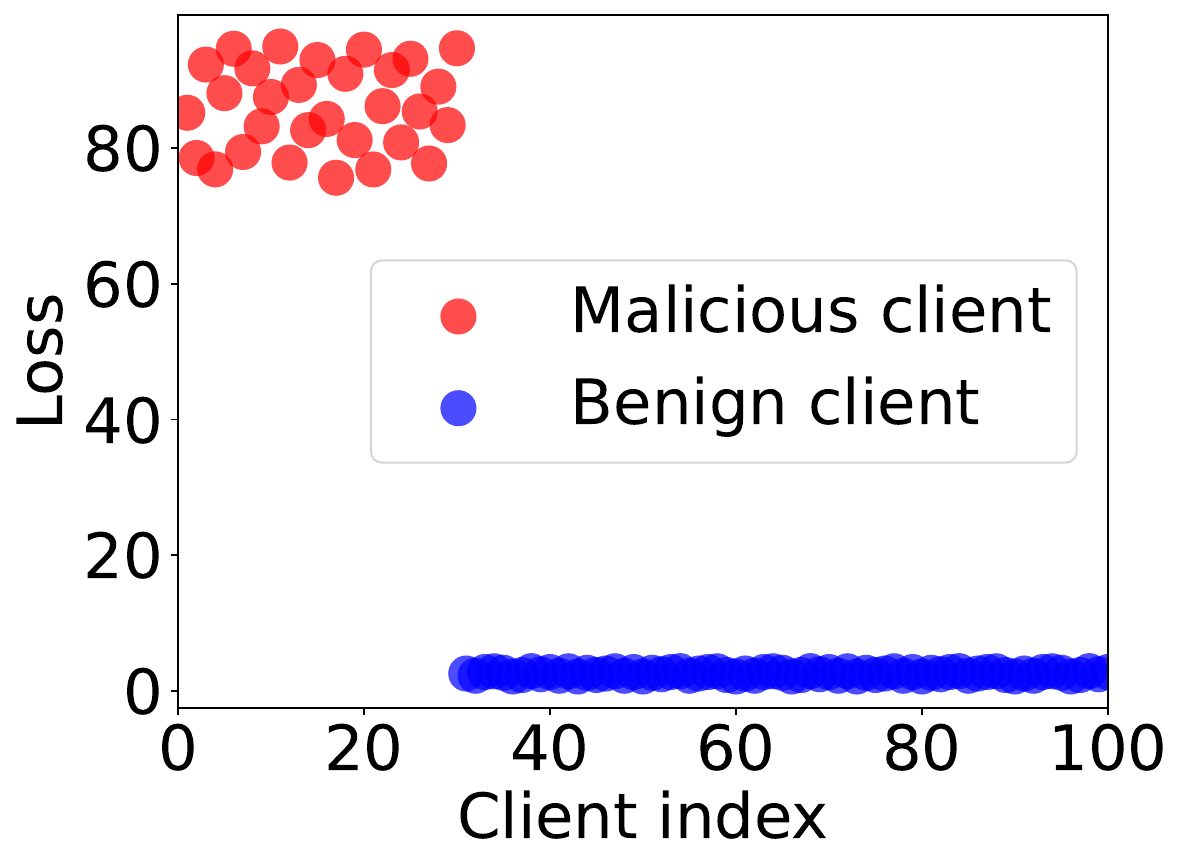}
    \caption{Scaling+DBA attack}
  \end{subfigure}
    \begin{subfigure}{0.163\textwidth}
    \includegraphics[width=\textwidth]{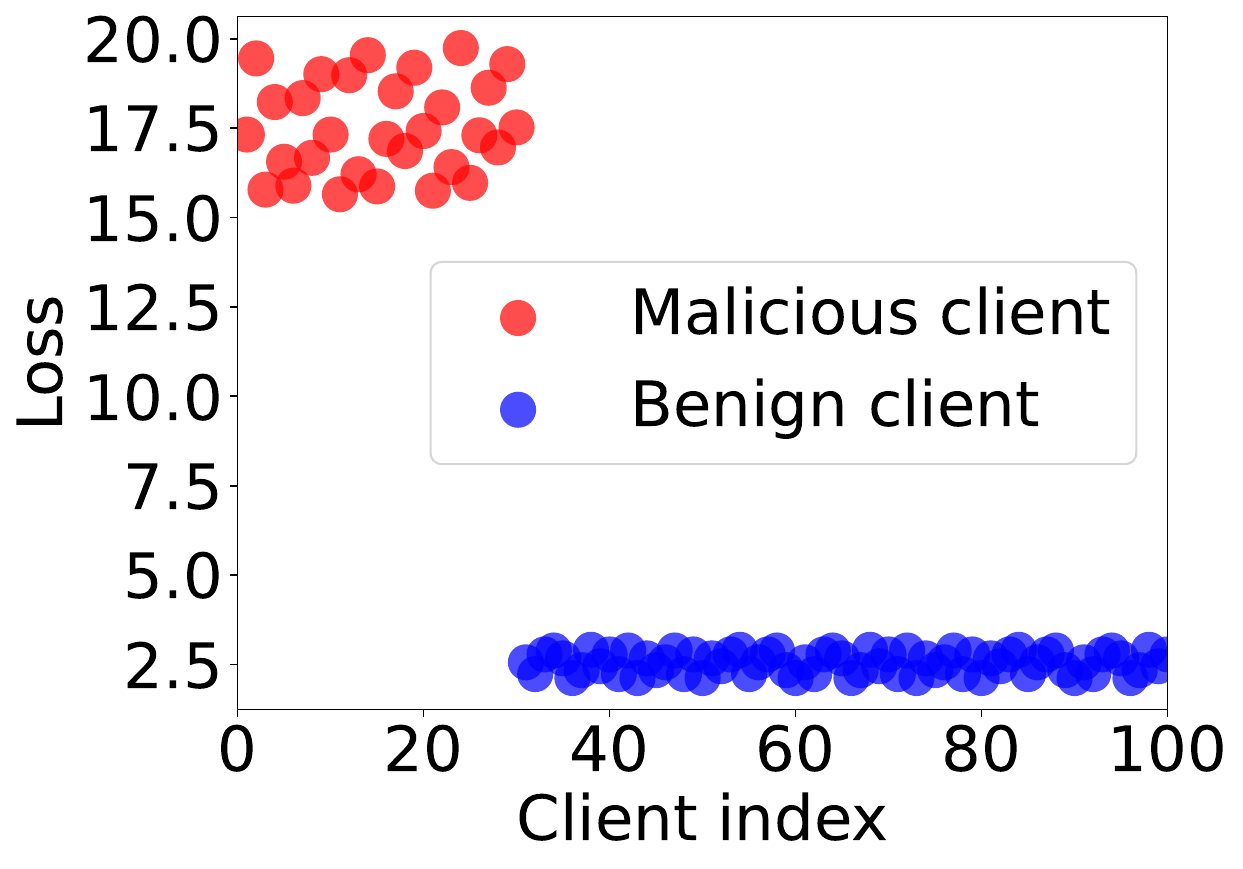}
    \caption{Adaptive attack}
  \end{subfigure}
  \caption{\textcolor{black}{The loss values of benign and malicious clients’ local models computed on the synthetic dataset, using \algSecond with the STL-10 dataset.}}
  \label{exp:safefl-cl-STL10}
\end{figure*}

\begin{figure*}[!t]
  \centering
    \begin{subfigure}{0.163\textwidth}
    \includegraphics[width=\textwidth]{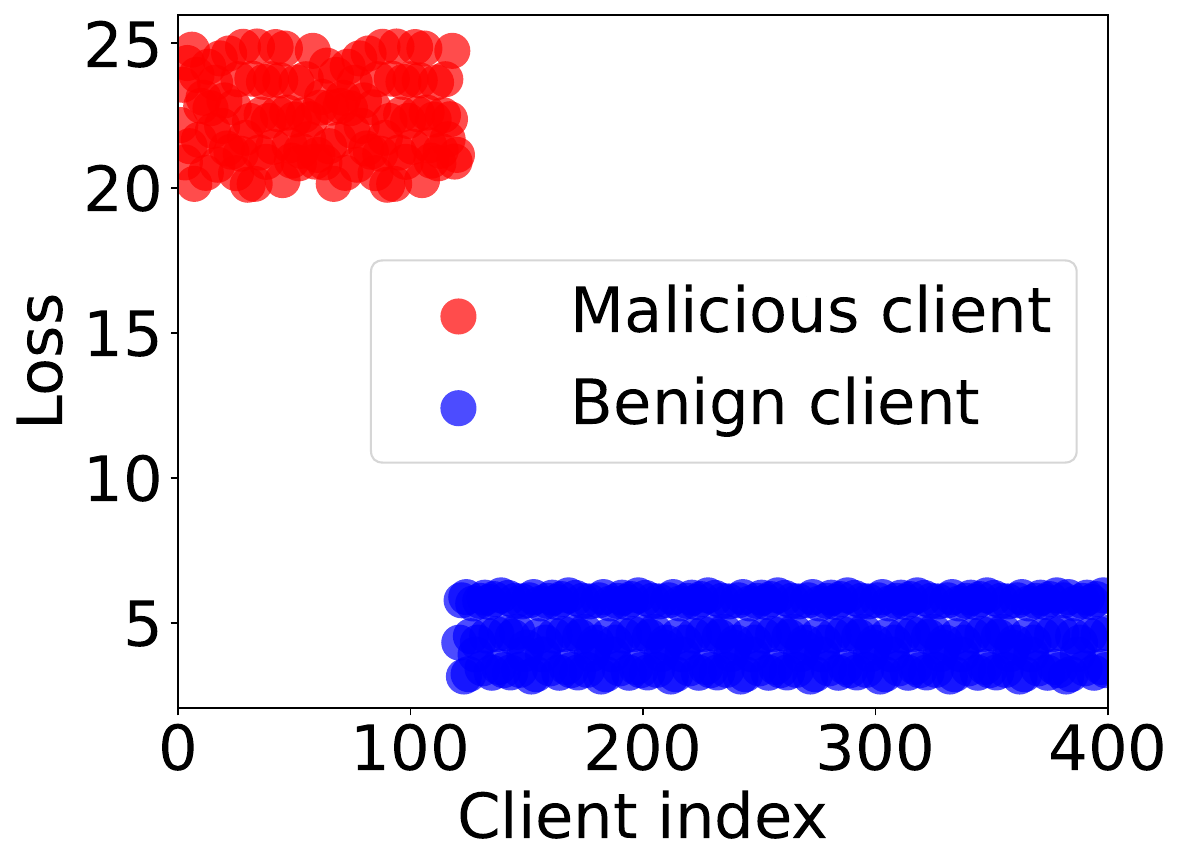}
    \caption{Trim attack}
  \end{subfigure}
    \begin{subfigure}{0.163\textwidth}
    \includegraphics[width=\textwidth]{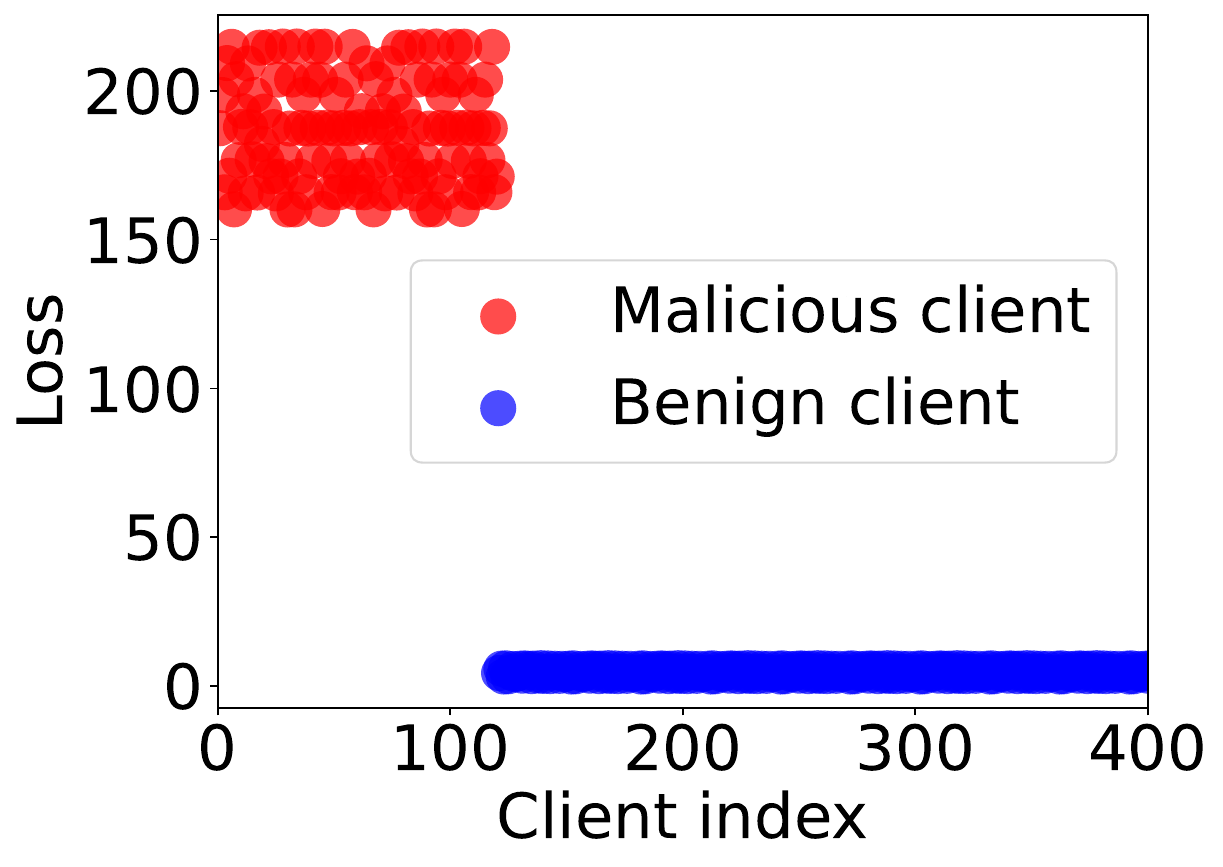}
    \caption{Scaling attack}
  \end{subfigure}
  \begin{subfigure}{0.163\textwidth}
    \includegraphics[width=\textwidth]{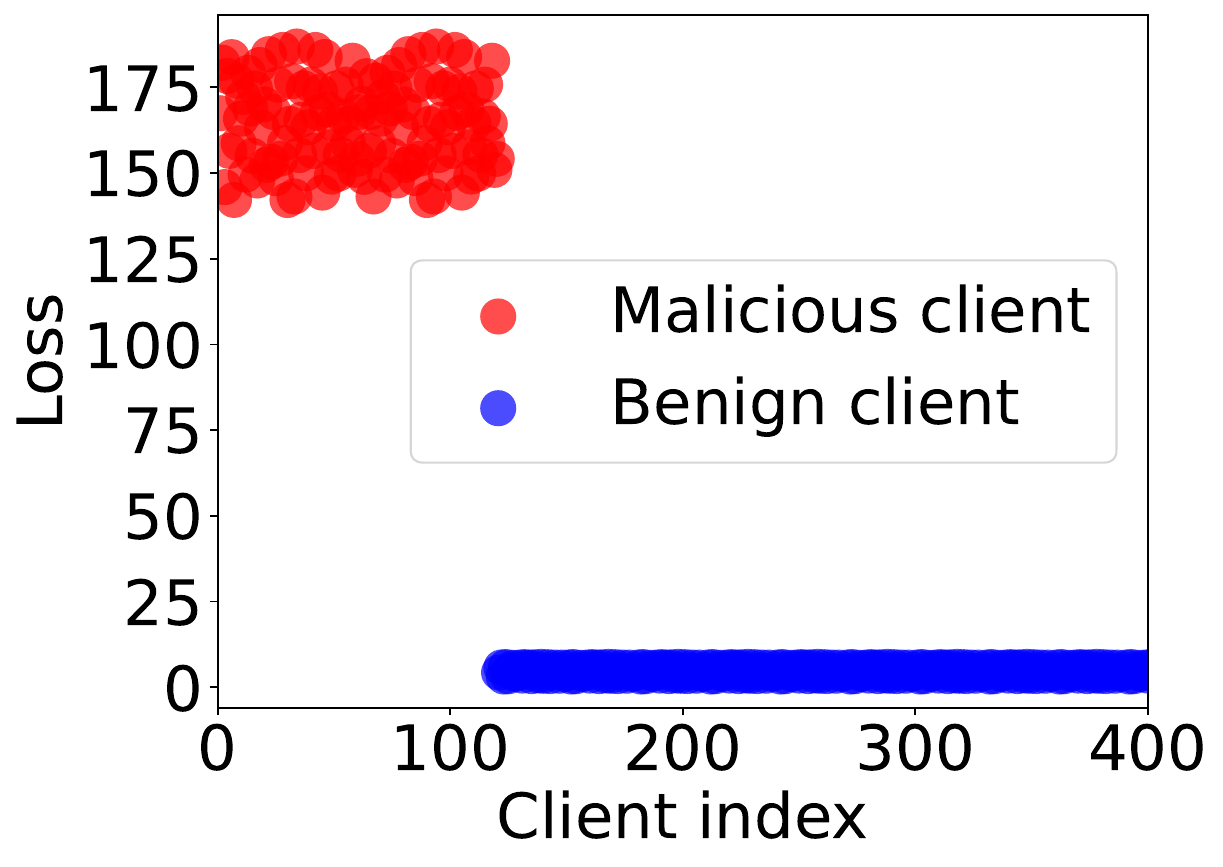}
    \caption{DBA attack}
  \end{subfigure}
  \begin{subfigure}{0.163\textwidth}
    \includegraphics[width=\textwidth]{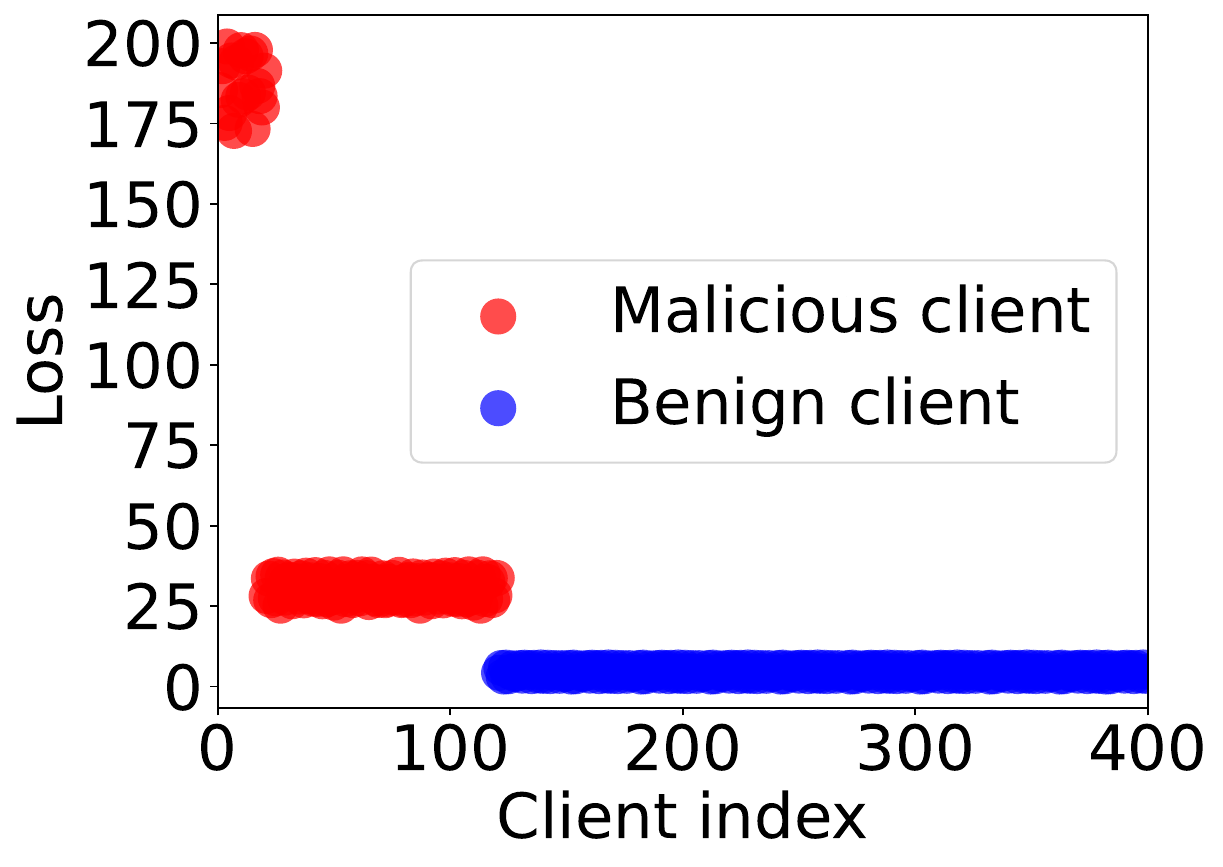}
    \caption{Trim+DBA attack}
  \end{subfigure}
    \begin{subfigure}{0.163\textwidth}
    \includegraphics[width=\textwidth]{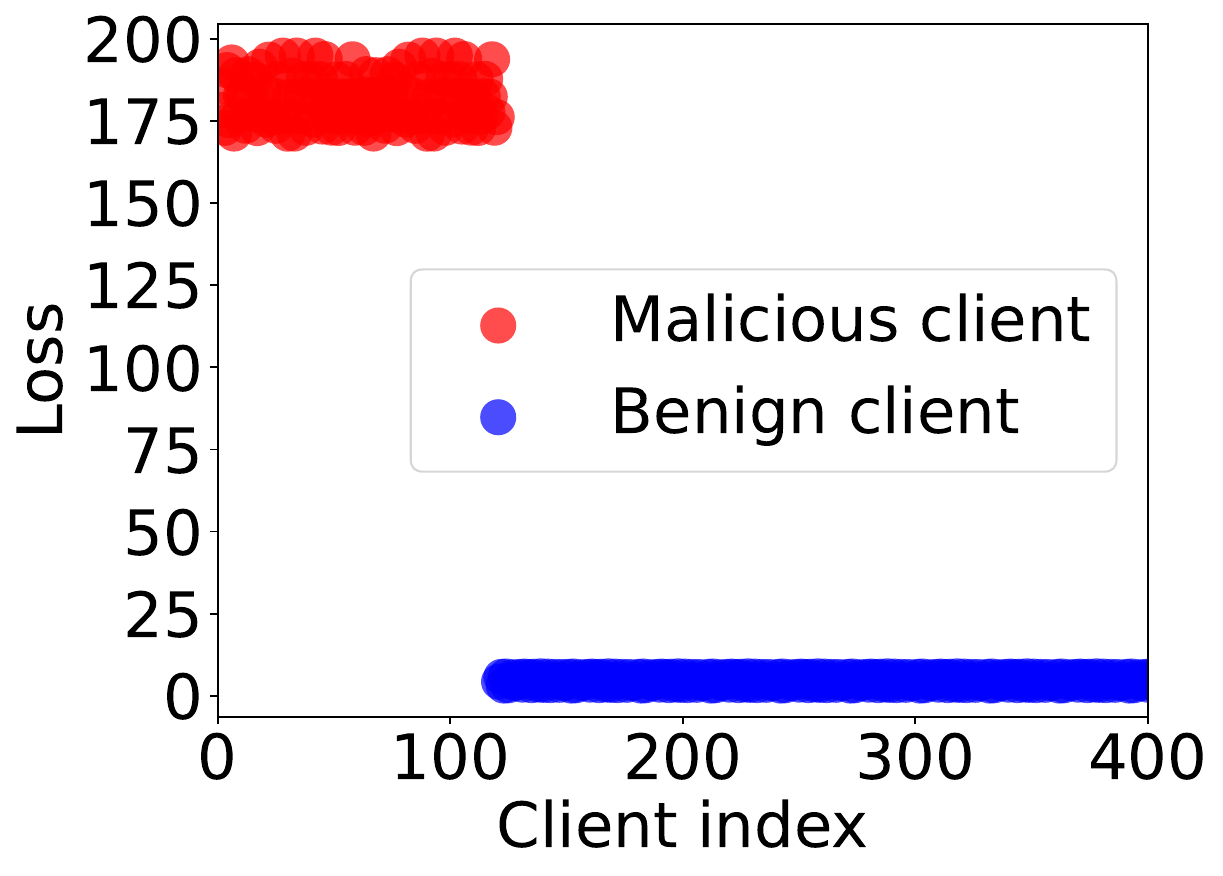}
    \caption{Scaling+DBA attack}
  \end{subfigure}
    \begin{subfigure}{0.163\textwidth}
    \includegraphics[width=\textwidth]{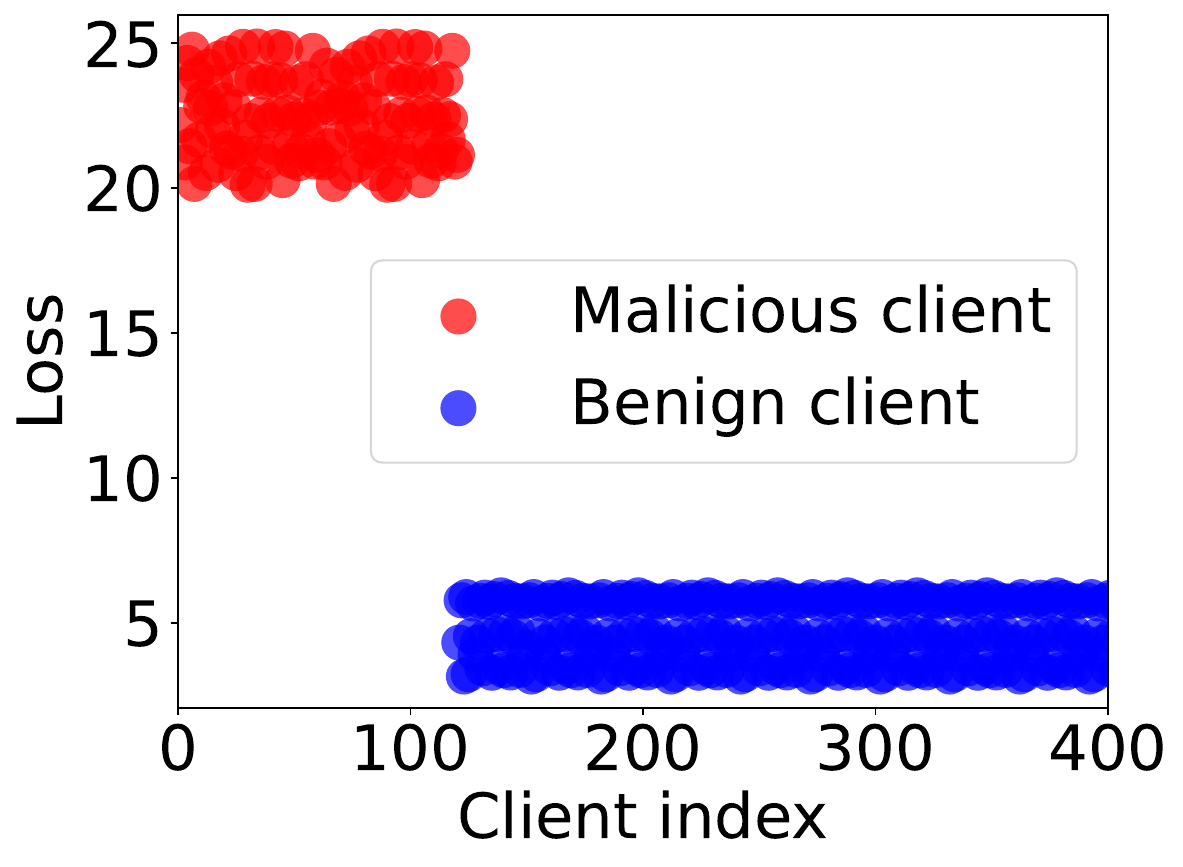}
    \caption{Adaptive attack}
  \end{subfigure}
  \caption{\textcolor{black}{The loss values of benign and malicious clients’ local models computed on the synthetic dataset, using \algSecond with the Tiny-ImageNet dataset.}}
  \label{exp:safefl-cl-STL10}
\end{figure*}

\begin{figure*}[t]
    \centering
    \subfloat[CIFAR-10]{\includegraphics[width=0.33 \textwidth]{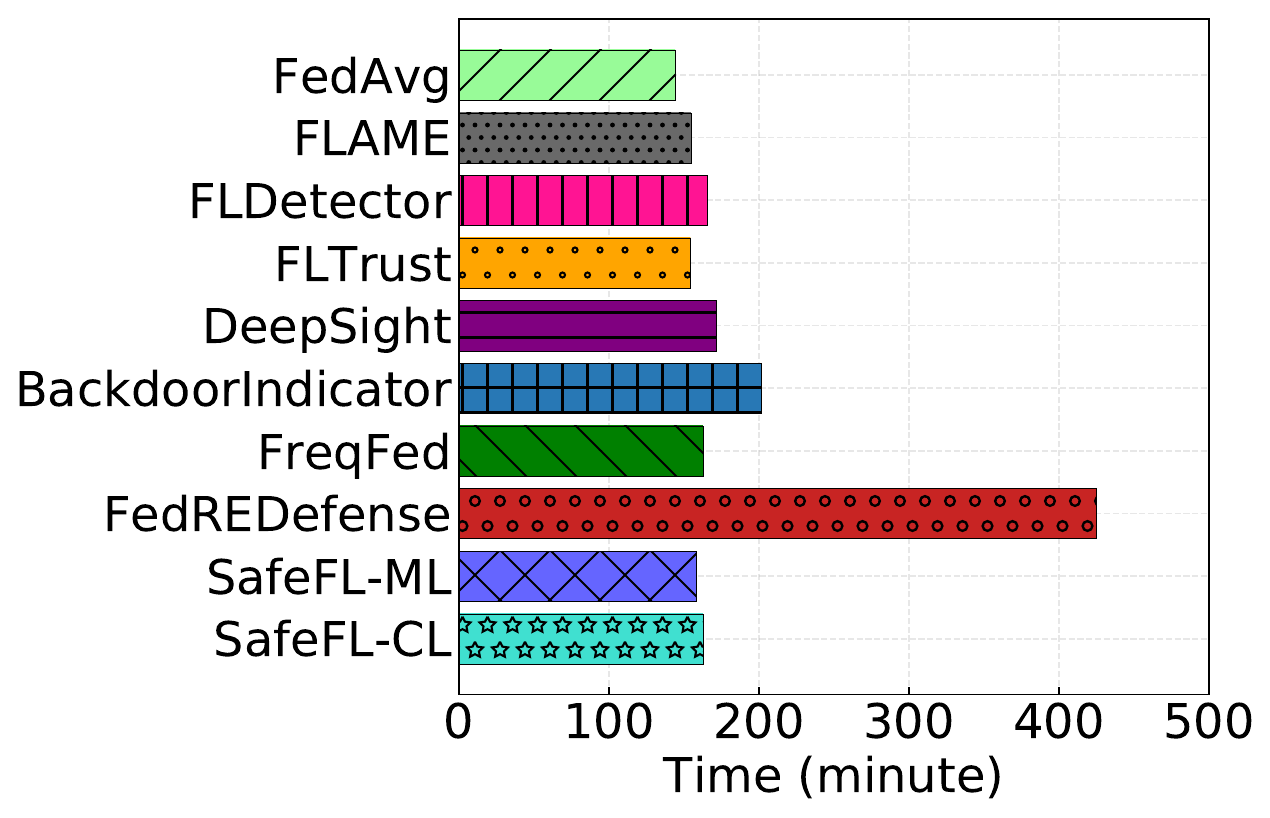}\label{fig:CIFAR10_time}}
    \subfloat[MNIST]{\includegraphics[width=0.33 \textwidth]{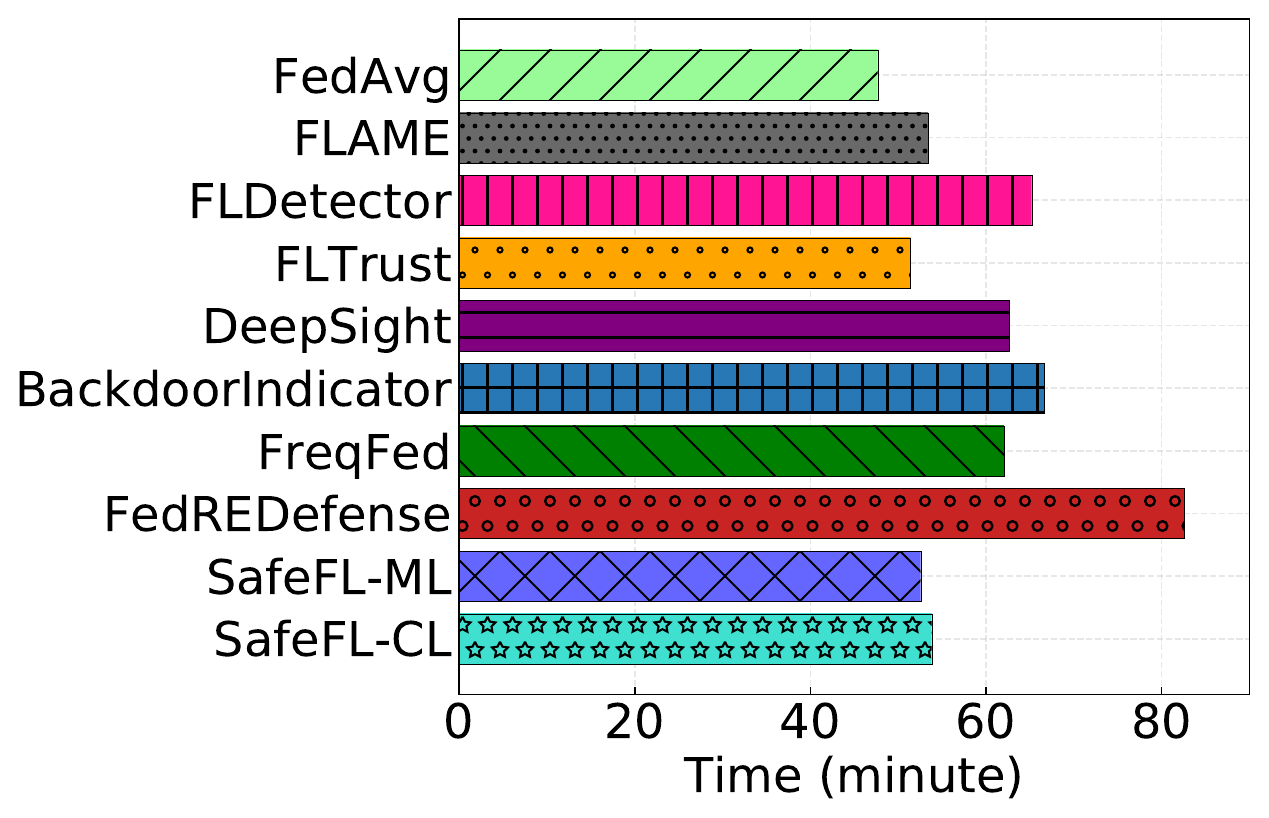}\label{fig:MNIST_time}}
    \\
	 \subfloat[FEMNIST]{\includegraphics[width=0.33 \textwidth]{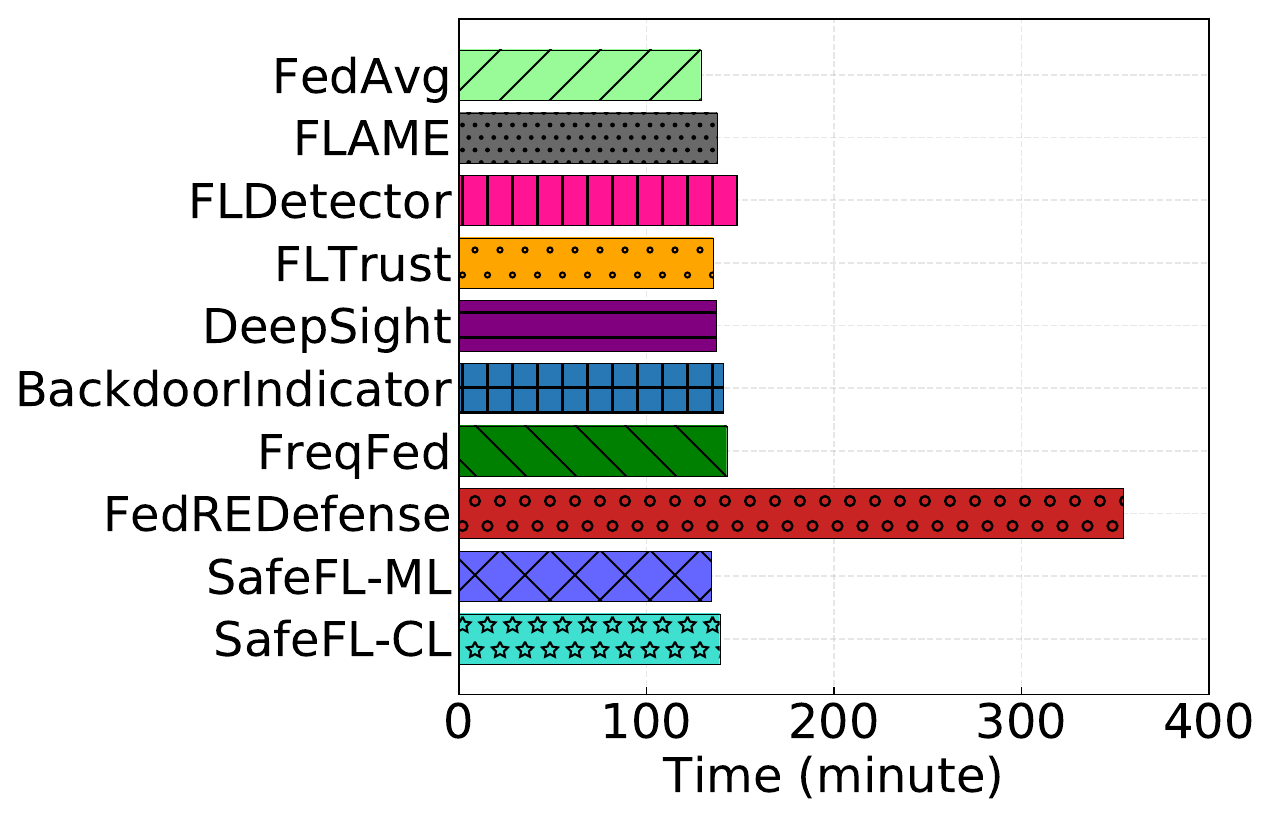}\label{fig:FEMNIST_time}}
	\subfloat[STL-10]{\includegraphics[width=0.33 \textwidth]{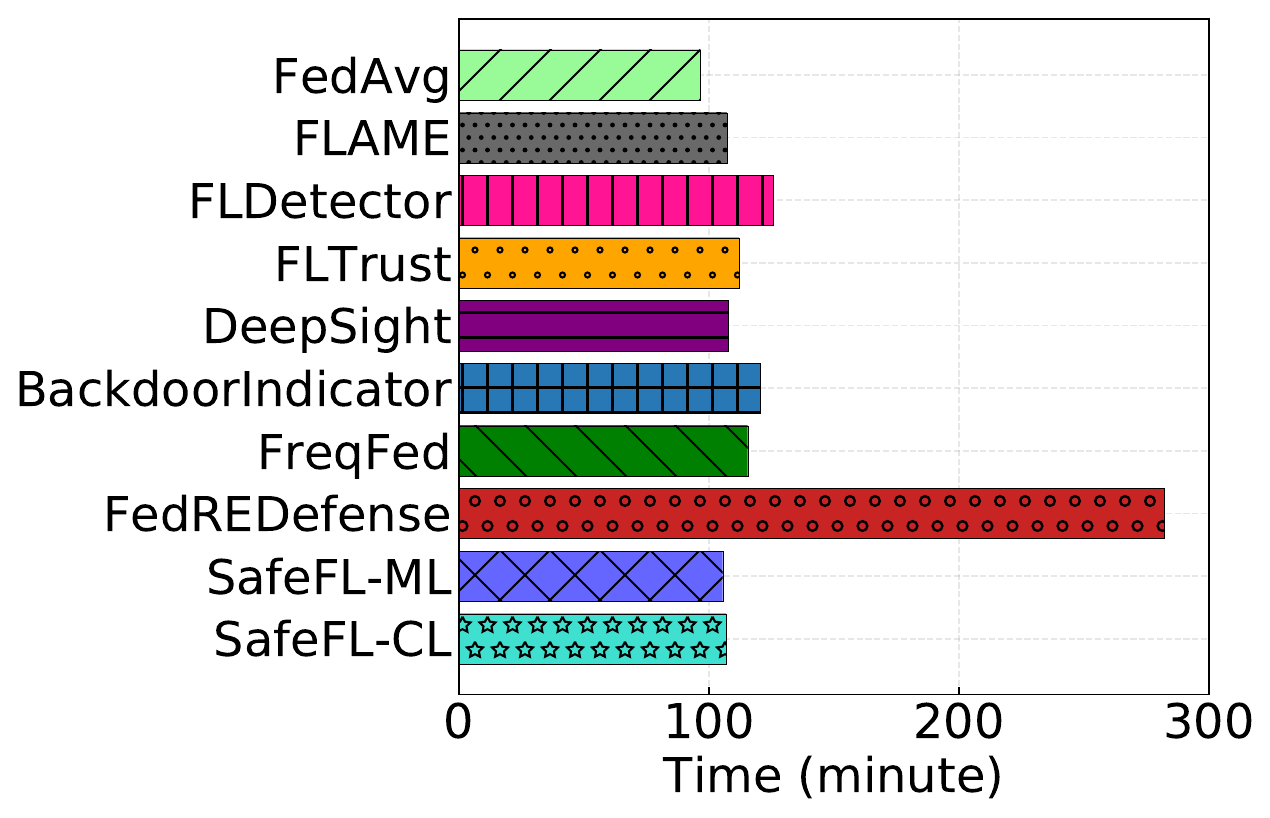}\label{fig:STL10_time}}
    \subfloat[Tiny-ImageNet]{\includegraphics[width=0.33 \textwidth]{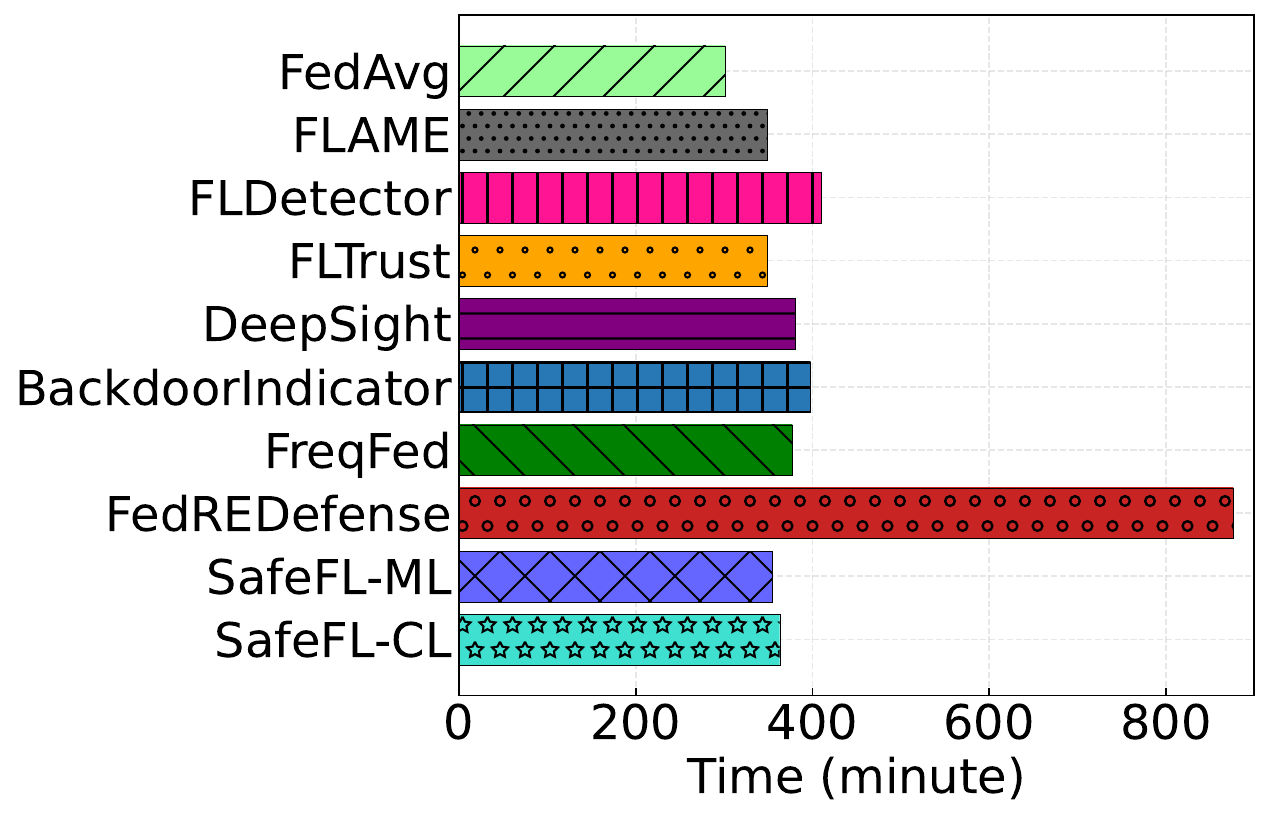}\label{fig:tiny_time}}
    \caption{\textcolor{black}{Computation costs of different methods.}}
    \label{time_comsume_fig}
\end{figure*}

\end{document}